\documentclass[10pt,oneside,a4paper,onecolumn]{elsarticle}
\usepackage{hyperref}

\journal{Computers \& Mathematics with Applications}

\usepackage{fullpage}
\usepackage{siunitx}
\usepackage{cancel}
\usepackage[usenames,dvipsnames]{xcolor}
\usepackage{multirow}

\usepackage{gensymb}

\usepackage{alltt}
\usepackage{titlesec}
\usepackage{upquote}
\usepackage{float}
\usepackage{array}
\usepackage{pgfplots}
\usetikzlibrary{positioning}
\usepackage{graphicx}
\usepackage[toc,page]{appendix}
\usepackage{amsmath,mathtools}
\usetikzlibrary{calc,decorations.markings,fit,shapes,chains,arrows}
\usepackage{amsthm}
\usepackage{amsfonts}
\usepackage{enumitem}
\usepackage{subcaption}
\usepackage{caption}
\usepackage{booktabs}
\usepackage{xcolor}
\usepackage{scalerel}
\usepackage{algorithm}
\usepackage{multicol}
\usepackage{mdframed}
\usepackage{mathtools}
\usepackage{enumerate}
\theoremstyle{remark}
\newtheorem{remark}{Remark}
\usepackage{tikzit}
\usetikzlibrary{patterns}

\tikzstyle{black dot}=[fill=black, draw=black, shape=circle]
\tikzstyle{green dot}=[fill=green, draw=black, shape=circle]

\tikzstyle{red solid line}=[-, fill=none, draw=red]
\tikzstyle{black dashed line}=[-, fill=none, draw=black, dashed]
\tikzstyle{black solid line}=[-, fill=none, draw=black]
\tikzstyle{black arrow solid}=[->, fill=none]
\tikzstyle{black arrow dashed}=[fill=none, ->, dashed, draw=black]
\tikzstyle{black dashed double sided arrow}=[<->, dashed]
\tikzstyle{red dashed}=[-, draw=red, fill=none, dashed]
\tikzstyle{red dashed arrow}=[draw=red, fill=none, dashed, ->]
\tikzstyle{blue solid}=[-, fill=none, draw=blue]
\tikzstyle{blue dashed}=[-, fill=none, draw=blue, dashed]

\usepackage{natbib}

\pgfdeclarepatternformonly{crossHatchDots}{\pgfqpoint{-1pt}{-1pt}}{\pgfqpoint{5pt}{5pt}}{\pgfqpoint{6pt}{6pt}}%
{
    \pgfpathcircle{\pgfqpoint{0pt}{0pt}}{.5pt}
    \pgfpathcircle{\pgfqpoint{3pt}{3pt}}{.5pt}
    \pgfusepath{fill}
}

\tikzset{
  block/.style={rectangle, draw, rounded corners, text centered,text width = 16em, minimum height = 2em},
  line/.style={draw, -latex'}
  }
\tikzset{
  block2/.style={text centered,text width = 22em, minimum height = 2em},
  line/.style={draw, -latex'}
  }
\tikzset{
  block3/.style={rectangle, draw, rounded corners, text centered,text width = 19em, minimum height = 1em},
  line/.style={draw, -latex'}
  }
\tikzset{
  block4/.style={rectangle, draw, rounded corners, text centered,text width = 15em, minimum height = 1em},
  line/.style={draw, -latex'}
  }
\tikzset{
  blockNS/.style={rectangle, draw, fill=black!20, rounded corners, text centered,text width = 20em, minimum height = 2em, label={center:Navier-Stokes solver}},
  line/.style={draw, -latex'}
  }
\tikzset{
  blockAC/.style={rectangle, draw, fill=black!20, rounded corners, text centered,text width = 20em, minimum height = 2em, label={center:Allen-Cahn solver}},
  line/.style={draw, -latex'}
  }
\tikzset{  
  decision/.style = {diamond, draw, minimum width=4cm, minimum height=0.2cm},
  line/.style={draw, -latex'}
  }
  
\def\@author#1{\g@addto@macro\elsauthors{\normalsize%
    \def\baselinestretch{1}%
    \upshape\authorsep#1\unskip\textsuperscript{%
      \ifx\@fnmark\@empty\else\unskip\sep\@fnmark\let\sep=,\fi
      \ifx\@corref\@empty\else\unskip\sep\@corref\let\sep=,\fi
      }%
    \def\authorsep{\unskip,\space}%
    \global\let\@fnmark\@empty
    \global\let\@corref\@empty
    \global\let\sep\@empty}%
    \@eadauthor={#1}
}

\graphicspath{{figures/}}

\begin{document}
\begin{frontmatter}
\title{A robust and accurate finite element framework for cavitating flows with fluid-structure interaction}
\author[ubc]{Suraj R. Kashyap}
\ead{suraj.kashyap@ubc.ca}

\author[ubc]{Rajeev K. Jaiman\corref{cor1}}
\ead{rjaiman@mech.ubc.ca}
\cortext[cor1]{Corresponding author}
\address[ubc]{Department of Mechanical Engineering, The University of British Columbia, Vancouver, BC V6T 1Z4}

\begin{abstract}
 In the current work, we present a unified  variational mechanics framework for the cavitating turbulent flow and the structural motion via a stabilized finite element formulation. To model the finite mass transfer rate in cavitation phenomena, we employ the homogenous mixture-based approach via phenomenological scalar transport differential equations given by the linear and nonlinear mass transfer functions. Stable linearizations of the finite mass transfer terms for the mass continuity equation and the reaction term of the scalar transport equations are derived for the robust and accurate implementation. The linearized matrices for the cavitation equation are imparted a positivity-preserving property to address numerical oscillations arising from high-density gradients typical of two-phase cavitating flows. The proposed formulation is strongly coupled in a partitioned manner with an incompressible 3D Navier-Stokes finite element solver, and the unsteady problem is advanced in time using a fully-implicit generalized-$\alpha$ time integration scheme. We first verify the implementation on the benchmark case of Rayleigh bubble collapse. We demonstrate the accuracy and convergence of the cavitation solver by comparing the numerical solutions with the analytical solutions of the Rayleigh-Plesset equation for bubble dynamics. We find our solver to be robust for large time steps and the absence of spurious oscillations/spikes in the pressure field. The cavitating flow solver is coupled with a hybrid URANS-LES turbulence model with a turbulence viscosity corrected for the presence of vapor. We validate the coupled solver for a very high Reynolds number turbulent cavitating flow over a NACA0012 hydrofoil section. Finally, the proposed method is solved in an Arbitrary Lagrangian-Eulerian framework to study turbulent cavitating flow over a pitching hydrofoil section and the coupled FSI results are explored for the characteristic features of cavitating flows such as re-entrant jet and periodic cavity shedding.
\end{abstract}

\begin{keyword}
Cavitation \sep Fluid-Structure Interaction \sep Homogeneous-Mixture  \sep Stabilized Finite Element \sep Partitioned iterative \sep Pitching Hydrofoil
\end{keyword}

\end{frontmatter}

	
\section{Introduction}\label{sec:introduction}
\label{intro}
    Cavitation is ubiquitous in natural and industrial industrial systems such as hydrofoils, nozzles, pumps, underwater vehicles and marine propellers. While cavitation can be a major source of noise, vibration and material erosion in these systems as unwanted effects \cite{carlton2018marine,kerr1940problems}, useful applications of cavitation have been developed in underwater cleaning \cite{song2004cleaning, chahine1983cleaning}, ultrasonic cleaning \cite{chahine2016cleaning, zijlstra2011acoustic}, biomedical procedures such as lithotripsy \cite{bailey2003cavitation, pishchalnikov2003cavitation} and enhanced drug delivery \cite{stride2019nucleation, husseini2005role}, etc.
    The phenomenon of cavitation involves the phase change of liquid into vapor and a highly complex interaction between the vapor and the liquid phases.  Most liquids have inherent points of weaknesses in the form of entrained microscopic gas bubbles, suspended particles, crevices along the shared boundaries with solid structures, and ephemeral voids created by the thermal motions of the liquid. When the flowing liquid encounters a region of low pressure, it has a propensity to rupture at these locations of weaknesses due to the tensile forces \cite{brennen_2013}. This rupturing creates cavities in the fluid, which can be filled with the original entrained gas or by vapor generated by evaporation at the cavity interface. These cavities can then be convected by the flow until they encounter a region of high pressure, where they can undergo violent collapse. 
     Cavitation is often encountered in marine propeller operations where fluid acceleration over propeller blades generates low-pressure regions near the blade surface. At these locations, nuclei present in seawater are prone to rupturing and forming vapor/gas-filled cavities. The cavities can then remain attached to or be shed from the blade surface in the form of cavitating structures with varying energy content \cite{ross1989mechanics}. The onset of cavitation influences the fluid-structure dynamics of propeller operation and can have several undesirable effects including performance degradation, vibration, material erosion and noise emission \cite{brennen_2013}. Traditional approaches to model the coupled fluid-structure dynamics in cavitating flows have often focused on a one-way coupling between a flow solver for the cavitation hydrodynamics and a separate finite element solver for structural deformation.
    
\begin{figure}[!h]
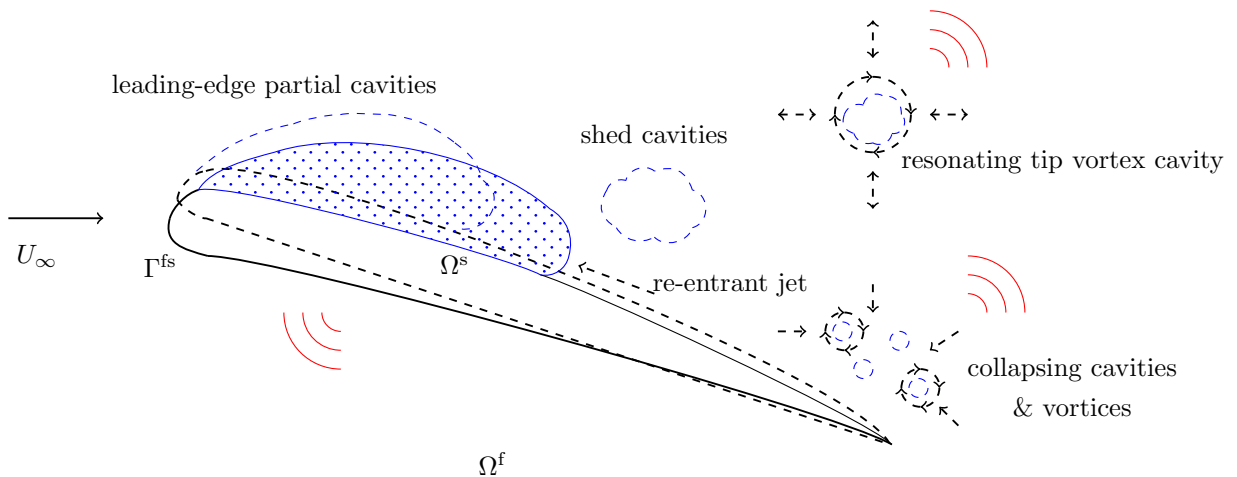

\centering
\ctikzfig{noiseSchematic}
\caption{Representative schematic demonstrating some of the prominent noise sources from cavitating blades. $\Omega^{\mathrm{f}}$ and $\Omega^{\mathrm{s}}$ are the fluid and solid domains, $\Gamma^{\mathrm{fs}}$ represents the interface between them. The emitted noise can be decomposed into two main components: (i) tonal noise originating from blade vibration and resonance in tip vortex cavities, and (ii) broadband noise resulting from violently collapsing cavities and vortices.}
\label{fig:noiseSchematic}
\end{figure}

 The reduction of noise emission in marine vessels is of interest both from an industrial and a marine-environmental perspective. For example, \cite{van1974calculation} showed that in the cavitating regime, propeller noise dominates all other sources of self-noise from ships, including electrical noise, machinery noise and boundary layer noise. Recently, \cite{carlton2018marine} provided an excellent review of noise from cavitating propellers and identified two broad categories: (i) a broadband noise component resulting from the sudden collapse of cavities and vortices, and (ii) tonal noise components from periodic fluctuations in the cavity volumes. In a classical work, \cite{kerr1940problems} identified that tonal noise emission in propellers was a result of blade vibration due to irregular cavitation and vortex-shedding dynamics. Several experimental studies \cite{carlton2018marine} have since confirmed this. In addition, \cite{arakeri1988model,maines1997case,arndt2015singing} showed that resonance in cavity-filled vortices shed from the blade tip can also emit intense tonal noise. Thus, a numerical study of propeller cavitation noise needs to consider this complex hydrodynamic interplay between cavitation and vortex shedding, and their FSI  effects with the propeller blades.	Figure \ref{fig:noiseSchematic} demonstrates some of these predominant noise-generating mechanisms in hydrodynamic cavitation, with $\Omega^{\mathrm{f}}$ representing the fluid domain, $\Omega^{\mathrm{s}}$ the solid structural domain and $\Gamma^{\mathrm{fs}}$ the fluid-structure interface.

\subsection{Transport-equation based modeling of cavitating flows}
 Cavitation in propellers manifests in the form of cavitating structures that exist across multiple orders of spatial and temporal scales \cite{arndt2015singing,brennen_2013}, making the study of marine cavitation a challenging task. Thus, each of the different approaches developed for the numerical modeling of cavitating flows is generally computationally feasible only for a select set of flow configurations. A popular approach is to represent the fluid as a homogeneous mixture of liquid and vapor. The homogeneous mixture-based approaches differ in the estimation of the density field and can be broadly classified into two categories. 
    
The first category assumes equilibrium flow theory and the density is calculated using equations of state (EoS) \cite{schnerr2008numerical,sezal2009compressible}. For isothermal flows, a barotropic equation of state is used to represent the density field as a function of the pressure \cite{ deshpande1994cavity,chen1996modeling}. The equilibrium flow approach has the advantage of easier implementation. It also does not require the use of empirical coefficients for modeling and relies on well-established equations of state. However, these models are generally used with the compressible Euler equations and solved using density-based solvers. In \cite{goncalves2010comparison}, the authors compared a compressible density-based method and an incompressible pressure-correction method to study cavitating flows with the barotropic EoS and reported better correlations with the compressible approach. Thus, this approach requires very small time-steps to capture the pressure wave propagations in the compressible fluid \cite{ghahramani2019comparative}. Another limitation of the equilibrium flow models based on barotropic EoS is that the gradients of the density and the pressure fields are parallel. Thus, the baroclinic torque which is proportional to the cross product $\left( \nabla \rho \times \nabla p\right)$ of these gradients is zero. This can result in inaccurate estimates of the vorticity production, which is an important feature of cavitating flows particularly in the closure region of attached cavities \cite{gopalan2000flow}. Hence this approach may not be appropriate for modeling the physics of cavitating flows over hydrofoils.
    
The second category of homogeneous mixture-based approaches assumes the pure liquid and vapor phases to be incompressible. The two-phase mixture density is interpolated based on a phase indicator. This phase indicator is generally in the form of the local phase fraction of either the liquid or the vapor phase. The phase indicator in the computational domain is obtained as the solution of a scalar transport equation. We refer to these transport-equation based models as TEMs in the rest of the article. The TEMs employ a source term, which is indicative of a finite mass transfer rate between the two phases by the process of cavitation. TEMs generally vary in the formulation of the source term based on phenomenological arguments. Merkle et al. \cite{merkle1998computational} used dimensional arguments for bubble clusters to relate the source term to the local pressure and phase fraction of liquid. Kunz et al. \cite{kunz1999multi} considered a similar source term as \cite{merkle1998computational}  to model the evaporation process, but modified the condensation term employing a simplified form of the Ginzburg-Landau potential. Later models by Schnerr-Sauer \cite{schnerr2001physical}, Zwart et al. \cite{zwart2004two} and Singhal et al. \cite{singhal2002mathematical} assumed the cavities to be present in the form of clusters of spherical bubbles and used directly a simplified form of the Rayleigh-Plesset equation \cite{brennen_2013} for spherical bubble dynamics to model the mass transfer rate. These models differ in multiplier terms that were derived using different phenomenological arguments for the underling bubble-bubble interaction. TEMs have been applied to the study of several cavitating flow configurations, including hydrodynamic cavitation over hydrofoils \cite{senocak2001numerical,gnanaskandan2015numerical,ji2015large}. 

Despite wide applicability, one limitation of the TEM approach is the use of semi-empirical coefficients in the source terms that need to be tuned for particular flow configurations. In addition, since the density is interpolated using the phase indicator, large spatial gradients in the density field exist across the cavity interface. This can result in unphysical oscillations/spikes in the pressure in the vicinity of the interface. For incompressible flow simulations, these pressure spikes can propagate rapidly throughout the computational domain, leading to numerical instability. These numerical artifacts have been reported in several works such as  \cite{ghahramani2019comparative}, and special treatments are required for discontinuity capturing across the interface. \cite{senocak2002pressure} suggested the ability to handle these spurious pressure spikes as one of the conditions for a robust solver for cavitating flows. 
    
\subsection{Review of numerical studies of two-phase FSI in marine propellers}
The last decade has seen advances in the numerical study of FSI effects in non-cavitating marine propeller operations. Simpler approaches have used methods based on inviscid potential flow theories. \citep{lee2014hydro} and \citep{ maljaars2018boundary} used a coupled potential theory-based boundary element method (BEM) and finite element method (FEM) approach for studying the hydro-elastic response of flexible marine propellers. \citep{jiang2018ship} presented an optimization methodology for propellers considering FSI of the fluid and propeller blades using the panel method (PM), which is derived from BEM. A loosely coupled PM-FEM approach is used for the fluid-structure interaction between the hull wake and the propeller. While inviscid models have been demonstrated to be effective for making general design decisions for propellers because of low computational cost, they are unable to capture the vortex-shedding process which is an important feature of cavitating flows. In addition, multiple modeling decisions have to be made on a case-by-case basis for different propeller geometry and inflow conditions. Within the purview of viscous flow modeling, \citep{lee2017fluid} used a tightly coupled CFD-FEM solver to study high Reynolds number flow over a flexible blade undergoing vortex-induced vibration. The authors presented the requirement of tightly-coupled FSI solvers for studying large-amplitude 3D vibrations at high Reynolds numbers. 
	
Although hydrodynamic studies of cavitating flows over propellers abound in literature, only limited examples of studies that consider FSI have been found.   \citep{ huang2013physical} used a one-way coupled commercial computational fluid dynamics (CFD) solver to determine hydrodynamic loads on a cavitating hydrofoil undergoing prescribed pitching motion.  \citep{akcabay2014influence} used a loose hybrid coupling to couple a commercial 2D URANS solver with a 2DOF hydrofoil model to study the effect of cavitation on the hydroelastic stability of hydrofoils. \citep{wu2018transient} used a similar approach to study the cavity shedding dynamics and flow-induced vibration over a hydrofoil section. Reasonable agreements were reported with experimental measurements. However, there is a need for a robust and accurate unified framework for the strongly-coupled FSI studies of cavitating flows over propellers.
    
The finite element method lends suitably to this purpose of modeling the caviating flows with fluid-structure interaction. It has long been staple for the study of structural deformations, and has also been successfully applied to fluid flow studies \citep{joshi2017variationally,joshi2018positivity}. However, not much work has yet been done on the modeling of cavitation using the finite element method. \citep{bayram2020variational} presented a variational framework for cavitating flows applied to prescribed motions of hydrokinetic turbines. A monolithic approach was taken for the coupling of the cavitation TEM and flow equations.  However, not much discussion has been made on the cavity collapse pressures, shedding dynamics or the numerical stability at high liquid-vapor density ratios seen in marine flows. 
Recently,  \citet*{joshi2017variationally} presented a variational finite element framework to study the 3D hydroelastic response of marine riser in turbulent marine flows, undergoing high-amplitude vortex-induced vibrations. A strongly-coupled partitioned approach was taken to solve the governing equations for the fluid flow, the structural deformation and the transport of the eddy viscosity. Further, \citet*{joshi2018positivity} presented a positivity preserving variational (PPV) scheme for the numerical solution of two-phase flow of immiscible fluids. A discrete upwind operator was used locally in the interface region demonstrating oscillations. This acted in the form of added diffusion, ensuring positivity of the underlying element-level matrices. The scheme was demonstrated to be effective in reducing spurious pressure oscillations in fluids with high-density ratios. Numerical solutions of cavitating flows using TEMs can benefit from similar treatment. The current work extends this framework by introducing the physics involved in the modeling of cavitating flows. 

\subsection{Current work and contributions}
In the current work, we propose a novel finite element formulation for the numerical modeling of cavitating flows. The objective is to integrate numerical modeling of cavitation into the framework of the stabilized finite element methods. Two cavitation TEMs based on homogeneous flow theory are used to model finite mass transfer rates between the liquid and vapor phases by the process of cavitation. The cavitation TEMs used are by Merkle et al. \cite{merkle1998computational} and Schnerr-Sauer \cite{schnerr2001physical}. We present stable linearizations of the two models for our proposed variational formulation. The linearized formulations are implemented in a variational finite element framework and the elemental matrices are imparted a positivity-preserving property for numerical stability in regions dominated by large density ratios. The fluid flow and cavitation solvers are coupled in a staggered partitioned manner for versatility, and predictor-corrector iterations are used for convergence stability and accuracy. A fully-implicit generalized-$\alpha$  time integration scheme \cite{jansen2000generalized} with user-controlled high-frequency damping is used to advance the solution in time, allowing for numerical stability at relatively coarse spatial and temporal discretizations. Consistent with the suggestions in \cite{senocak2002pressure}, the efficacy of the proposed method is demonstrated with respect to the requirements for a robust computational method for cavitating flows. The requirements can be summarized as (i) the accurate prediction of the pressure field, (ii) the ability to handle large density ratios of the order of 100-1000, and (iii) absence of spurious pressure spikes across the cavity interface.
	
In the sections that follow, the governing equations and the proposed formulation are first presented. Next, a benchmark test of spherical vaporous bubble collapse is used to verify the numerical implementation. The implementation is then validated on a turbulent cavitating flow configuration over a hydrofoil section. Before concluding, the case of a pitching hydrofoil is considered to explore the compatibility of the formulation with FSI and its ability to predict flow features characteristic of cavitating flows. Finally, we close our paper with some concluding remarks.

\section{Numerical formulation and implementation}
\label{sec:methodology}
\noindent In this section, we present a positivity preserving variational finite element implementation for the numerical solution of TEMs applied to cavitation modeling. In particular, we consider two types of TEM-based homogeneous mixture models classified as follows:
\begin{itemize}
    \item Model A: Cavitation model with nonlinear mass transfer rate by \textit{Schnerr and Sauer} \cite{schnerr2001physical}
    \item Model B: Cavitation model with linear mass transfer rate by \textit{Merkle et al.} \cite{merkle1998computational}
\end{itemize}
\noindent In the rest of manuscript, we shall refer to the models as A and B respectively. We design stable linearizations of models A and B for the numerical implementation within the proposed variational finite element framework.
\subsection{Governing equations}
\label{sec:GE}
\noindent The strong forms of the governing equations are presented before introducing the variational formulation. We consider the fluid physical domain $\Omega^{\mathrm{f}}(\boldsymbol{x}^{\mathrm{f}},t)$ with an associated fluid boundary $\Gamma^{\mathrm{f}}(t)$, where $\boldsymbol{x}^{\mathrm{f}}$ and $t$ represent the spatial and temporal coordinates. The working fluid, consisting of the liquid and vapor phases, is assumed to be present in the form of a continuous homogeneous mixture. The phase indicator $\phi^{\mathrm{f}}(\boldsymbol{x}^{\mathrm{f}},t)$ is used to represent the phase fraction of the liquid phase at any coordinate $(\boldsymbol{x}^{\mathrm{f}},t)$ in the homogeneous two-phase liquid-vapor mixture. The fluid density ($\rho^{\mathrm{f}}$) and dynamic viscosity ($\mu^{\mathrm{f}}$) are taken as linear combinations of $\phi^{\mathrm{f}}$
\begin{align} 
	\rho^{\mathrm{f}} 
	&=\rho_l\boldsymbol{\phi}^{\mathrm{f}} + \rho_v\left( 1-\boldsymbol{\phi}^{\mathrm{f}} \right),\label{densInterp}\\
	\mu^{\mathrm{f}}
	&=\mu_l\boldsymbol{\phi}^{\mathrm{f}} + \mu_v\left( 1-\boldsymbol{\phi}^{\mathrm{f}} \right),\label{viscInterp}
\end{align} 
\noindent where $\rho_l$ and $\rho_v$ are the densities of the pure liquid and vapor phases, respectively. $\mu_l$ and $\mu_v$ are the dynamic viscosities of the liquid and the vapor phases. 
\subsubsection{Cavitation TEM}
\noindent In the TEM approach, $\phi^{\mathrm{f}}$ is obtained as the solution of a scalar transport equation, which can be written in the conservative form in the ALE framework as: 
\begin{align} \label{TEM}
  \left. \frac{\partial \boldsymbol{\phi}^{\mathrm{f}}}{\partial t}\right|_{\boldsymbol{\chi}}
	+ \boldsymbol{\phi}^{\mathrm{f}}\nabla \cdot \boldsymbol{u}^{\mathrm{f}} 
	+ \left(\boldsymbol{u}^{\mathrm{f}}-\boldsymbol{u}^{\mathrm{m}}\right)\cdot\nabla\boldsymbol{\phi}^{\mathrm{f}} 
    = \dfrac{\dot{m}}{\rho_{l}},&&\mathrm{on}\ (\boldsymbol{x}^{\mathrm{f}},t)\in \Omega^{\mathrm{f}}
\end{align}
\noindent where $\boldsymbol{\chi}$ is the referential coordinate system, ${\boldsymbol{u}^{\mathrm{f}}} = {\boldsymbol{u}^{\mathrm{f}}}(\boldsymbol{x}^{\mathrm{f}},t)$ is the fluid velocity at each spatial point $\boldsymbol{x}^{\mathrm{f}} \in \Omega^{\mathrm{f}}$ and $\boldsymbol{u}^{\mathrm{m}}$ is the relative velocity of the spatial coordinates $\boldsymbol{x}^{\mathrm{f}}$ with respect to the referential coordinate system $\boldsymbol{\chi}$. The source term $\dot{m}$ is representative of a finite mass transfer rate that governs the rates of destruction and production of liquid by the process of cavitation. Cavitation TEMs vary in the way $\dot{m}$ is modeled.
\subsubsection*{Cavitation Model A}
\noindent For model A by \cite{schnerr2001physical}, $\dot{m}$ is given as a non-linear function of $\boldsymbol{\phi}^{\mathrm{f}}$ and $p^{\mathrm{f}}$
\begin{align}
        \dot{m}_{A}  
        =  \frac{3 \rho_{l} \rho_{v}}{\rho^{\mathrm{f}} R_{B}} \sqrt{\frac{2}{3 \rho_{l}\left|p^{\mathrm{f}}-p_{v}\right|}}
         \bigg[ C_{c} \boldsymbol{\phi}^{\mathrm{f}}&(1-\boldsymbol{\phi}^{\mathrm{f}}) \operatorname{max}\left(p^{\mathrm{f}}-p_{v}, 0\right) \nonumber \\
        &+ C_{v} \boldsymbol{\phi}^{\mathrm{f}}(1 + \phi_{nuc} -\boldsymbol{\phi}^{\mathrm{f}}) \operatorname{min}\left(p^{\mathrm{f}}-p_{v}, 0\right)
        \bigg] \label{eq:schnerrSauer}
\end{align}
\noindent Model A attempts to relate the finite mass transfer rate to the rate of growth/collapse of an equivalent spherical bubble under an external pressure field, using a simplification of the Rayleigh-Plesset equation. Cavitation is assumed to initiate from nucleation sites present in the flow by a heterogeneous nucleation process \citep{brennen_2013}. The initial concentration of nuclei per unit volume ($n_0$) with an associated nuclei diameter($d_{nuc}$) and is assumed to be a constant. It is also assumed that only vaporous cavitation occurs, and the effect of non-condensable gases is not considered. $R_B(\boldsymbol{x}^{\mathrm{f}},t)$ in Eq.~(\ref{eq:schnerrSauer}) is representative of the equivalent radius of the vapor volume at the coordinates $(\boldsymbol{x}^{\mathrm{f}},t)$, while $\phi_{nuc}$ is the phase fraction of the initial nucleation sites in an unit volume. These are calculated as
\begin{equation}
    R_B = \left( \frac{3}{4\pi n_0} \frac{1+\phi_{nuc}-\phi^{\mathrm{f}}}{\phi^{\mathrm{f}}} \right)^{1/3} \quad\mathrm{and}\quad 
    \phi_{nuc} = \frac{\dfrac{\pi n_0 d^3_{nuc}}{6}}{1+\dfrac{\pi n_0 d^3_{nuc}}{6}}
\end{equation}
  The vapor phase at any spatial location is assumed to be present in the form of a concentration of bubbles with identical radii. The model requires as input the condensation coefficient $C_c$ and the evaporation coefficient $C_v$, which require calibration for specific flow configurations. It has been applied to the study of different cavitating flow configurations, including the collapse of vaporous bubbles \cite{ghahramani2019comparative} and cavitating flow over hydrofoils \cite{ji2015large}. 
  
  We see from Eq.~(\ref{eq:schnerrSauer}) that $\dot{m}_{A}$ depends on the local pressure $p^{\mathrm{f}}$ as
\begin{align}
    \dot{m}_{A}
    \propto \dfrac{p^{\mathrm{f}}-p_{v}}{\sqrt{\left| p^{\mathrm{f}}-p_{v} \right|}} &&\ p^{\mathrm{f}}(\boldsymbol{x}^{\mathrm{f}},t)\in {\rm I\!R}-\{0\}
\end{align}
    It is observed that $\dot{m}_{A}$ is not defined when $p^{\mathrm{f}}=p_{v}$ and can lead to numerical instability. In the current work, we take 
\begin{equation}
    \left.\dot{m}_{A}\right|_{p^{\mathrm{f}}=p_{v}} = \lim_{p^{\mathrm{f}}\to p_{v}}\frac{p^{\mathrm{f}}-p_{v}}{\sqrt{\left|p^{\mathrm{f}}-p_{v}\right|}} = 0
\end{equation}
\subsubsection*{Cavitation model B}
\noindent For model B, $\dot{m}$ is a linear function of both $\boldsymbol{\phi}^{\mathrm{f}}$ and the local pressure $p^{\mathrm{f}}$
\begin{align}
        \dot{m}_{B}  
        =  C_{dest}\frac{\rho_l^2}{\rho_v}\frac{\boldsymbol{\phi}^{\mathrm{f}}}{\frac{1}{2}\rho_lU_{\infty}^2t_{\infty}}&\operatorname{min}\left(p^{\mathrm{f}}-p_{v}, 0\right) \nonumber \\
         &+  C_{prod}\rho_l\frac{1- \boldsymbol{\phi}^{\mathrm{f}}}{\frac{1}{2}\rho_lU_{\infty}^2t_{\infty}}\operatorname{max}\left(p^{\mathrm{f}}-p_{v}, 0\right) \label{eq:merkle}
\end{align}
\noindent where $p_{v}$ is the saturation vapor pressure, $U_{\infty}$ is the free-stream velocity and $t_{\infty}$ is the mean flow time-scale. For hydrofoils, $t_{\infty}$ is taken as $t_{\infty}$ = $C/U_{\infty}$, where $C$ is the chord length. $C_{dest}$ and $C_{prod}$ are semi-empirical coefficients influencing the rates of destruction and production of liquid by the process of cavitation. In the current work, we assume $C_{dest}=1$ and $C_{prod}=80$ \cite{senocak2004interfacial,huang2013physical} . Model B has been derived using dimensional arguments for clusters of large bubbles and has been applied to macro-scale cavitation over hydrofoils \cite{huang2013physical,bayram2020variational}.

\subsubsection{Fluid momentum and mass conservation}
\noindent The unsteady Navier-Stokes equations for the fluid momentum and mass conservation can be written in an ALE framework as
\begin{align} 
	\left.\rho^{\mathrm{f}} \frac{\partial \boldsymbol{u}^{\mathrm{f}}}{\partial t}\right|_{\boldsymbol{\chi}}
	+\rho^{\mathrm{f}}\left(\boldsymbol{u}^{\mathrm{f}}-\boldsymbol{u}^{\mathrm{m}}\right) \cdot \nabla \boldsymbol{u}^{\mathrm{f}}
	-\nabla \cdot \boldsymbol{\sigma}
	=\boldsymbol{f}^{\mathrm{f}},&&\mathrm{on}\ (\boldsymbol{x}^{\mathrm{f}},t)\in \Omega^{\mathrm{f}}, \label{NS_mom}\\
	\left. \frac{\partial \rho^{\mathrm{f}}}{\partial t}\right|_{\boldsymbol{\chi}}+ \rho^{\mathrm{f}}\nabla \cdot \boldsymbol{u}^{\mathrm{f}} + \left(\boldsymbol{u}^{\mathrm{f}}-\boldsymbol{u}^{\mathrm{m}}\right)\cdot\nabla\rho^{\mathrm{f}} = 0,&&\mathrm{on}\ (\boldsymbol{x}^{\mathrm{f}},t)\in \Omega^{\mathrm{f}}, \label{NS_mass}
\end{align} 
\noindent \noindent where $\boldsymbol{f}^{\mathrm{f}}$ is the body force applied on the fluid and
\begin{equation}
    \boldsymbol{\sigma} = \boldsymbol{\sigma}^{\mathrm{f}} + \boldsymbol{\sigma}^{\mathrm{des}}
\end{equation}
where ${\boldsymbol{\sigma}^{\mathrm{f}}}$ and $\boldsymbol{\sigma}^{\mathrm{des}}$ are the Cauchy stress tensor for a Newtonian fluid and the turbulent stress tensor respectively, given by
\begin{align}
	{\boldsymbol{\sigma}^{\mathrm{f}}} 
	&= -{p^{\mathrm{f}}}\boldsymbol{I} + \mu^{\mathrm{f}}( \nabla{\boldsymbol{u}^{\mathrm{f}}}+ (\nabla{\boldsymbol{u}^{\mathrm{f}}})^T),\\
	{\boldsymbol{\sigma}^{\mathrm{des}}} 
	&= \mu_T( \nabla{\boldsymbol{u}^{\mathrm{f}}}+ (\nabla{\boldsymbol{u}^{\mathrm{f}}})^T),
\end{align}
where ${p^{\mathrm{f}}}$ denotes the fluid pressure and $\mu_T$ is the turbulent viscosity. $\boldsymbol{\sigma}^{\mathrm{des}}$ is modeled using the Boussinesq approximation and in the current work a hybrid URANS-LES turbulence model is applied. The details of the turbulence model implementation can be found in \citet*{joshi2017variationally}. 

\subsubsection{Convective form of TEM and local fluid compressibility}
\noindent In the present work, the conservative form of the transport equation is re-arranged in the form a convection-reaction equation. Taking the material derivative of Eq.~(\ref{densInterp}) in the ALE framework, we obtain
\begin{equation}\label{densMaterialDerivative}
  \left. \frac{\partial \rho^{\mathrm{f}}}{\partial t}\right|_{\boldsymbol{\chi}}+\left(\boldsymbol{u}^{\mathrm{f}}-\boldsymbol{u}^{\mathrm{m}}\right) \cdot \nabla \rho^{\mathrm{f}} = \left( \rho_l - \rho_v \right) \left( \left. \frac{\partial \boldsymbol{\phi}^{\mathrm{f}}}{\partial t}\right|_{\boldsymbol{\chi}}+\left(\boldsymbol{u}^{\mathrm{f}}-\boldsymbol{u}^{\mathrm{m}}\right) \cdot \nabla \boldsymbol{\phi}^{\mathrm{f}} \right)
\end{equation} 
\noindent Combining equations (\ref{TEM}), (\ref{NS_mass}) and (\ref{densMaterialDerivative}), the following forms of the mass continuity equation and the phase indicator transport equation are obtained, which are used in the current implementation.
\begin{align} 
	\nabla \cdot \boldsymbol{u}^{\mathrm{f}} 
	= \left(\frac{1}{\rho_{l}} - \frac{1}{\rho_{v}} \right) \dot{m},
	&&\mathrm{on}\ (\boldsymbol{x}^{\mathrm{f}},t)\in \Omega^{\mathrm{f}}, \label{NS_massMod}\\
	\left. \frac{\partial \boldsymbol{\phi}^{\mathrm{f}}}{\partial t}\right|_{\boldsymbol{\chi}}
	+\left(\boldsymbol{u}^{\mathrm{f}}-\boldsymbol{u}^{\mathrm{m}}\right) \cdot \nabla \boldsymbol{\phi}^{\mathrm{f}} 
	= \frac{\rho^{\mathrm{f}}}{\rho_{l}\rho_{v}}\dot{m},
	&&\mathrm{on}\ (\boldsymbol{x}^{\mathrm{f}},t)\in \Omega^{\mathrm{f}}, \label{phiMod}
\end{align} 
It is observed that the divergence of the velocity field is no longer zero, and local dilation effects are introduced that are governed by the finite mass transfer rate. This local compressibility exists only within the two-phase mixture and the pure phases are incompressible, since for the cavitation models chosen no mass transfer occurs when $\boldsymbol{\phi}^{\mathrm{f}}$ equals $0$ or $1$.
\subsubsection{Turbulence viscosity modification}
\noindent The URANS implementations were developed for incompressible turbulent flows, and are not fully capable of accounting for the local compressibility induced by the presence of vapor. For cavitating flows over hydrofoils, this leads to an over-prediction of the turbulent stresses at the cavity closure and has been reported by \cite{coutier2003numerical,huang2013physical}. This reduces the momentum of the characteristic re-entrant jet in cavitating flows and does not allow it to penetrate the cavity. Thus, the periodic cavity breakup and shedding is undermined and the cavity remains attached to the surface in a quasi-steady state. This has also been observed by the authors in preliminary investigations, although not presented here. Hence, in the current work the turbulent viscosity is modified in the presence of vapor. An approach similar to that adopted in \cite{coutier2003numerical} is taken. The turbulent viscosity $\mu_T$ is modified as
\begin{align}
    \mu_{T_{mod}}&=\frac{f(\rho^{\mathrm{f}})}{\rho^{\mathrm{f}}}\mu_T\\
    f(\rho^{\mathrm{f}})&=\rho_{v}+\left(\frac{\rho_{v}-\rho^{\mathrm{f}}}{\rho_{v}-\rho_{l}}\right)^{n}\left(\rho_{l}-\rho_{v}\right), \quad n \gg 1
\end{align}
\noindent A value of $n=10$ is used in the current work, which has been shown to agree well with experimental observations of cavity shedding \cite{coutier2003numerical}.

\subsubsection{FSI boundary conditions and fluid mesh deformation}
In Section~\ref{sec:explore}, we study cavitating flow over a pitching hydrofoil. We briefly review the FSI boundary conditions and the ALE mesh motion in the continuum setting. The modeling of FSI requires the satisfaction of the velocity continuity and traction equilibrium at the fluid-structure boundary $\Gamma^{\mathrm{fs}}$. 
Let us consider a structural domain $\Omega^{s} \subset \mathbb{R}^{d}$ with an associated structural boundary $\Gamma^{\mathrm{s}}(0)$ at time $t=0$. Let the mapping $\varphi^{\mathrm{s}}\left(x^{\mathrm{s}}, t\right)$ map the deformation of the structure from its initial configuration $\Omega^{s}$ to a deformed configuration $\Omega^{s}(t)$ at time $t$, where $x^{\mathrm{s}}$ denote the material coordinates. We denote the initial fluid-structure interface at $t=0$ by $\Gamma^{\mathrm{fs}}(0)=\Gamma^{\mathrm{f}}(0) \cap \Gamma^{\mathrm{s}}(0)$. At time $t$ the interface will then be deformed as $\Gamma^{\mathrm{fs}}(t)=\varphi^{\mathrm{s}}\left(\Gamma^{\mathrm{fs}}, t\right)$. The following kinematic and dynamic conditions are satisfied on $\Gamma^{\mathrm{fs}}$
\begin{align}
\boldsymbol{u}^{\mathrm{f}}\left(\boldsymbol{\varphi}^{\mathrm{s}}\left(\boldsymbol{x}^{\mathrm{s}}, t\right), t\right) &=\boldsymbol{u}^{\mathrm{s}}\left(\boldsymbol{x}^{\mathrm{s}}, t\right), & & \forall \boldsymbol{x}^{\mathrm{s}} \in \Gamma^{\mathrm{fs}} \\
\int_{\boldsymbol{\varphi}^{\mathrm{s}}(\gamma, t)} \boldsymbol{\sigma}^{\mathrm{f}} \cdot \mathbf{n}^{\mathrm{f}} d \Gamma+\int_{\gamma} \boldsymbol{\sigma}^{\mathrm{s}} \cdot \mathbf{n}^{\mathrm{s}} d \Gamma &=0, && \forall \gamma \subset \Gamma^{\mathrm{fs}}
\end{align}
where $u^{s}$ is the velocity of the structural domain, $\mathbf{n}^{\mathrm{f}}$ and $\mathbf{n}^{\mathrm{s}}$ are the unit normals to the deformed fluid elements $\boldsymbol{\varphi}^{\mathrm{s}}(\gamma, t)$ and their corresponding structural elements $\gamma$ on the interface $\Gamma^{\mathrm{fs}}$ respectively. The structural stress tensor $\sigma^{\mathrm{s}}$ is modeled depending on the type of material.

The fluid spatial coordinates are updated to conform to the structural deformation. The motion of the coordinates which are not at $\Gamma^{\mathrm{fs}}$ is modeled as an elastic material in equilibrium and the mesh equation is solved as
\begin{align}
\nabla \cdot \boldsymbol{\sigma}^{\mathrm{m}} &=\mathbf{0}, && \text { on } \Omega^{\mathrm{f}}, \label{eq:meshUpdate}\\
\boldsymbol{\eta}^{\mathrm{f}} &=\boldsymbol{\eta}_{D}^{\mathrm{f}}, && \forall \boldsymbol{x}^{\mathrm{f}} \in \Gamma_{D}^{\mathrm{m}}
\end{align}
where $\boldsymbol{\sigma}^{\mathrm{m}}=\left(1+k_{m}\right)\left[\nabla \boldsymbol{\eta}^{\mathrm{f}}+\left(\nabla \boldsymbol{\eta}^{\mathrm{f}}\right)^{T}+\left(\nabla \cdot \boldsymbol{\eta}^{\mathrm{f}}\right) \boldsymbol{I}\right]$ is the stress experienced at the fluid spatial coordinates due to the
strain induced by the deformation of the interface, $\boldsymbol{\eta}^{\mathrm{f}}$ is the displacement of the fluid spatial coordinates. The amount of deformation of the spatial coordinates is controlled using the local stiffness parameter $k_{m}$. Dirichlet conditions for the fluid mesh displacement $\boldsymbol{\eta}_{D}^{\mathrm{f}}$ are satisfied on the boundary $\Gamma_{D}^{\mathrm{m}}$.

\subsection{Temporal Discretization}
\noindent Both the fluid mass and momentum equations and the phase indicator transport equation are discretized in time using a generalized-$\alpha$ predictor-corrector time integration method \cite{chung1993time,jansen2000generalized}. For linear problems, the generalized-$\alpha$ method can be second-order accurate and unconditionally stable. This also enables the use of a single parameter called the spectral radius $\rho_{\infty}$, to achieve user-controlled high-frequency damping.
\subsubsection{Cavitation TEM}
\noindent Let $\partial_t{\boldsymbol{\phi}}^{{\mathrm{f}},\mathrm{n+\alpha_m}}$ be the temporal derivative of $\boldsymbol{\phi}^{\mathrm{f}}$ at time $t^\mathrm{n+\alpha_m}$. Using the generalized-$\alpha$ method, $\phi^{\mathrm{f}}$ is solved for at the $\mathrm{n}+\alpha$ time-level and integrated in time $t \in\left[t^{\mathrm{n}}, t^{\mathrm{n}+1}\right]$ according to the rules
\begin{align}
	\phi^{\mathrm{f}, \mathrm{n+1}} &= \phi^{\mathrm{f}, \mathrm{n}} + \Delta t\partial_t{\phi}^{\mathrm{f}, \mathrm{n}} + \gamma\Delta t(\partial_t{\phi}^{\mathrm{f}, \mathrm{n+1}} - \partial_t{\phi}^{\mathrm{f}, \mathrm{n}}), \nonumber \\
	\partial_t{\phi}^{\mathrm{f}, \mathrm{n}+\alpha_\mathrm{m}} &= \partial_t{\phi}^{\mathrm{f}, \mathrm{n}} + \alpha_\mathrm{m} (\partial_t{\phi}^{\mathrm{f}, \mathrm{n+1}} - \partial_t{\phi}^{\mathrm{f}, \mathrm{n}}),\label{eq:genAlpha} \\
	\phi^{\mathrm{f}, \mathrm{n}+\alpha} &= \phi^{\mathrm{f}, \mathrm{n}} + \alpha (\phi^{\mathrm{f}, \mathrm{n+1}} - \phi^{\mathrm{f}, \mathrm{n}}), \nonumber
\end{align}
where $\Delta t$ is the time step size and $\partial_t{\phi}^{\mathrm{f}, \mathrm{n}+\alpha_\mathrm{m}}$ is the temporal derivative of $\phi^{\mathrm{f}}$ at the $\mathrm{n}+\alpha_\mathrm{m}$ time level. $\alpha_\mathrm{m}$, $\alpha$ and $\gamma$ are generalized-$\alpha$ parameters based on $\rho_{\infty}$ \cite{chung1993time}.

\noindent The transport equation is arranged in the following convection-reaction form for numerical solving
\begin{align}
	G(\partial_t{\phi}^{\mathrm{f}, \mathrm{n}+\alpha_\mathrm{m}}, \phi^{\mathrm{f}, \mathrm{n}+\alpha}) 
	= \partial_t{\phi}^{\mathrm{f}, \mathrm{n}+\alpha_\mathrm{m}} 
	+ \left(\boldsymbol{u}^{\mathrm{f}}-\boldsymbol{u}^{\mathrm{m}}\right)\cdot\nabla\phi^{\mathrm{f}, \mathrm{n}+\alpha} 
	+ s\phi^{\mathrm{f}, \mathrm{n}+\alpha} 
	- f 
	= 0,
\end{align}\label{eq:TEM_CDR}\\
\noindent where $\boldsymbol{u}^{\mathrm{f}}-\boldsymbol{u}^{\mathrm{m}}$ is the convection velocity.
The reaction coefficient $s$ and the source term $f$ for model A are given by
\begin{align}
	s_{A} &= -\frac{3 }{ R_{B}} \sqrt{\frac{2}{ 3\rho_{l}\left|p^{\mathrm{f}}-p_{v}\right|}} 
        \bigg[ C_{c} (1-\boldsymbol{\phi}^{\mathrm{f}}) \operatorname{max}\left(p^{\mathrm{f}}-p_{v}, 0\right) 
        + C_{v} (1 + \phi_{nuc} -\boldsymbol{\phi}^{\mathrm{f}}) \operatorname{min}\left(p^{\mathrm{f}}-p_{v}, 0\right)
        \bigg],\\
	f_{A} &= 0.
\end{align}
while for model B, we have
\begin{align}
	s_{B} &= \frac{\rho^{\mathrm{f}}}{\rho_{v}}\bigg[ -\frac{\rho_{l}}{\rho_{v}}\frac{C_{dest}}{\frac{1}{2}\rho_lU_{\infty}^2t_{\infty}}\operatorname{min}\left(p-p_{v}, 0\right) 
         +  \frac{C_{prod}}{\frac{1}{2}\rho_lU_{\infty}^2t_{\infty}}\operatorname{max}\left(p-p_{v}, 0\right) \bigg],\\
	f_{B} &= \frac{\rho^{\mathrm{f}}}{\rho_{v}}\bigg[\frac{C_{prod}}{\frac{1}{2}\rho_lU_{\infty}^2t_{\infty}}\operatorname{max}\left(p-p_{v}, 0\right) \bigg].
\end{align}

\subsubsection{Navier-Stokes equations}
\noindent The Navier-Stokes equations are also discretized using the generalized-$\alpha$ time integration for consistency with the phase indicator transport equation.  The variational formulation employs the following relations for time integration during $t \in\left[t^{\mathrm{n}}, t^{\mathrm{n}+1}\right]$
\begin{align}
	\boldsymbol{u}^{\mathrm{f}, \mathrm{n+1}} 
	&= \boldsymbol{u}^{\mathrm{f}, \mathrm{n}} 
	+ \Delta t\partial_t\boldsymbol{u}^{\mathrm{f}, \mathrm{n}} 
	+ \gamma \Delta t( \partial_t\boldsymbol{u}^{\mathrm{f}, \mathrm{n+1}} - \partial_t\boldsymbol{u}^{\mathrm{f}, \mathrm{n}}), \\
	\boldsymbol{u}^{\mathrm{f}, \mathrm{n+\alpha}} 
	&= \boldsymbol{u}^{\mathrm{f}, \mathrm{n}} 
	+ \alpha(\boldsymbol{u}^{\mathrm{f}, \mathrm{n+1}} - \boldsymbol{u}^{\mathrm{f}, \mathrm{n}}), \\
	\partial_t\boldsymbol{u}^{\mathrm{f}, \mathrm{n+\alpha_m}} 
	&= \partial_t\boldsymbol{u}^{\mathrm{f}, \mathrm{n}} 
	+ \alpha_\mathrm{m}( \partial_t\boldsymbol{u}^{\mathrm{f}, \mathrm{n+1}} - \partial_t\boldsymbol{u}^{\mathrm{f}, \mathrm{n}}).
\end{align}
The semi-discrete forms of the Navier-Stokes equations are written as
\begin{align} 
	\left.\rho^{\mathrm{f}} \partial_{t} \boldsymbol{u}^{\mathrm{f}, \mathrm{n}+\alpha_{\mathrm{m}}}\right|_{\chi}
	+\rho^{\mathrm{f}}\left(\boldsymbol{u}^{\mathrm{f}, \mathrm{n}+\alpha}-\boldsymbol{u}^{\mathrm{m}, \mathrm{n}+\alpha}\right) \cdot \nabla \boldsymbol{u}^{\mathrm{f}, \mathrm{n}+\alpha}
	-\nabla \cdot \boldsymbol{\sigma}^{\mathrm{n}+\alpha} 
	- \boldsymbol{f}^{\mathrm{n}+\alpha} 
	&= 0,\label{NS_momSemiDisc}\\
	\nabla \cdot \boldsymbol{u}^{\mathrm{f},\mathrm{n}+\alpha}
	-\left(\frac{1}{\rho_{l}} - \frac{1}{\rho_{v}} \right)\dot{m}
	&=0,
\end{align}

\subsection{Spatial discretization and variational statement}
 Next, we present the stabilized variational statements of the governing equations.  We first present the positivity preserving variational form of the cavitation model, followed by the Navier-Stokes equations. The stabilized variational form for the incompressible Navier-Stokes has been discussed in detail in other work \cite{joshi2017variationally,joshi2017positivity}, which we shall not reproduce here. However, the presence of the non-zero divergence of the fluid velocity introduces additional terms to the formulation, which merits a brief discussion.

\subsubsection{Cavitation TEM}
The fluid computational domain $\Omega^{\mathrm{f}}$ is spatially discretized into $\mathrm{n_{el}}$ number of elements such that $\Omega^{\mathrm{f}} = \cup_\mathrm{e=1}^\mathrm{n_{el}} \Omega^{{\mathrm{f}},\mathrm{e}}$ and $\emptyset = \cap_\mathrm{e=1}^\mathrm{n_{el}} \Omega^{{\mathrm{f}},\mathrm{e}}$. The trial solutions are taken from from the space $\mathcal{S}^\mathrm{h}$, which equal the given Dirichlet boundary condition at the boundary $\Gamma_D$. The test functions are taken from the space $\mathcal{V}^\mathrm{h}$, which vanish on the Dirichlet boundary. 
\noindent We state the variational form of the phase indicator transport equation to find $\phi^{\mathrm{f}}_\mathrm{h}(\boldsymbol{x}^{\mathrm{f}},t^{\mathrm{n}+\alpha}) \in \mathcal{S}^\mathrm{h}$ such that $\forall w_\mathrm{h} \in \mathcal{V}^\mathrm{h}$,
\begin{align} \label{eq:TEMvariational}
	&\int_{\Omega^{\mathrm{f}}} \bigg( w_\mathrm{h}\partial_t{\phi}^{\mathrm{f}, \mathrm{n+\alpha_m}}_\mathrm{h} 
	+ w_\mathrm{h}\left(\boldsymbol{u}^{\mathrm{f}, \mathrm{n}+\alpha}-\boldsymbol{u}^{\mathrm{m}, \mathrm{n}+\alpha}\right)\cdot\nabla\phi^{\mathrm{f}, \mathrm{n}+\alpha}_\mathrm{h} 
	+ w_\mathrm{h}s\phi^{\mathrm{f}, \mathrm{n}+\alpha}_\mathrm{h} 
	- w_\mathrm{h}f \bigg) \mathrm{d}\Omega^{\mathrm{f}} \nonumber \\
	&+ \displaystyle\sum_\mathrm{e=1}^\mathrm{n_{el}}\int_{\Omega^{{\mathrm{f}},\mathrm{e}}}\bigg( \big(\left(\boldsymbol{u}^{\mathrm{f}, \mathrm{n}+\alpha}-\boldsymbol{u}^{\mathrm{m}}\right)\cdot\nabla w_\mathrm{h} \big)\tau_\phi \big( \partial_t{\phi^{\mathrm{f}, \mathrm{n}+\alpha_m}}_\mathrm{h} + \left(\boldsymbol{u}^{\mathrm{f}, \mathrm{n}+\alpha}-\boldsymbol{u}^{\mathrm{m}}\right)\cdot\nabla\phi^{\mathrm{f}, \mathrm{n}+\alpha}_\mathrm{h} + s\phi^{\mathrm{f}, \mathrm{n}+\alpha}_\mathrm{h} -f \big) \bigg) \mathrm{d}\Omega^{{\mathrm{f}},\mathrm{e}}  \nonumber \\
	&+ \displaystyle\sum_\mathrm{e=1}^\mathrm{n_{el}}\int_{\Omega^{{\mathrm{f}},\mathrm{e}}} \chi \frac{|\boldsymbol{\mathcal{R}}_\phi|}{|\nabla\phi^{\mathrm{f}}_\mathrm{h}|}k_s^\mathrm{add} \nabla w_\mathrm{h}\cdot \bigg( \frac{\left(\boldsymbol{u}^{\mathrm{f}, \mathrm{n}+\alpha}-\boldsymbol{u}^{\mathrm{m}}\right)\otimes \left(\boldsymbol{u}^{\mathrm{f}, \mathrm{n}+\alpha}-\boldsymbol{u}^{\mathrm{m}}\right)}{|\left(\boldsymbol{u}^{\mathrm{f}, \mathrm{n}+\alpha}-\boldsymbol{u}^{\mathrm{m}}\right)|^2} \bigg) \cdot \nabla\phi^{\mathrm{f}, \mathrm{n}+\alpha}_\mathrm{h} \mathrm{d}\Omega^{{\mathrm{f}},\mathrm{e}}\\
	&+ \sum_\mathrm{e=1}^\mathrm{n_{el}} \int_{\Omega^{{\mathrm{f}},\mathrm{e}}}\chi \frac{|\boldsymbol{\mathcal{R}}_\phi|}{|\nabla \phi^{\mathrm{f}, \mathrm{n}+\alpha}_\mathrm{h}|} k^\mathrm{add}_{c} \nabla w_\mathrm{h} \cdot \bigg( \mathbf{I} - \frac{\left(\boldsymbol{u}^{\mathrm{f}, \mathrm{n}+\alpha}-\boldsymbol{u}^{\mathrm{m}}\right)\otimes \left(\boldsymbol{u}^{\mathrm{f}, \mathrm{n}+\alpha}-\boldsymbol{u}^{\mathrm{m}}\right)}{|\left(\boldsymbol{u}^{\mathrm{f}, \mathrm{n}+\alpha}-\boldsymbol{u}^{\mathrm{m}}\right)|^2} \bigg) \cdot \nabla\phi^{\mathrm{f}, \mathrm{n}+\alpha}_\mathrm{h} \mathrm{d}\Omega^{{\mathrm{f}},\mathrm{e}} \nonumber \\
	&= 0,\nonumber 	
\end{align}
where the first line represents the standard Galerkin finite element terms. The second line consists of linear stabilization terms with the stabilization parameter $\tau_\phi$ given by \cite{shakib1991new}
\begin{align}
	\tau_\phi &= \bigg[ \bigg( \frac{2}{\Delta t}\bigg)^2 +\left(\boldsymbol{u}^{\mathrm{f}}-\boldsymbol{u}^{\mathrm{m}}\right)\cdot \boldsymbol{G}\left(\boldsymbol{u}^{\mathrm{f}}-\boldsymbol{u}^{\mathrm{m}}\right) + s^2\bigg] ^{-1/2},
\end{align}
where $\boldsymbol{G}$ is the element contravariant metric tensor, which is defined as
\begin{align}
	\boldsymbol{G} = \frac{\partial \boldsymbol{\xi}^T}{\partial \boldsymbol{x}^{\mathrm{f}}}\frac{\partial \boldsymbol{\xi}}{\partial \boldsymbol{x}^{\mathrm{f}}},
\end{align}
where $\boldsymbol{x}^{\mathrm{f}}$ and $\boldsymbol{\xi}$ are the physical and parametric coordinates respectively. $\boldsymbol{\mathcal{R}}_\phi$ is the residual of the phase indicator transport equation given as
\begin{align}
	\boldsymbol{\mathcal{R}}_\phi = \partial_t\phi^{\mathrm{f}, \mathrm{n}+\alpha}_\mathrm{h} + \left(\boldsymbol{u}^{\mathrm{f}, \mathrm{n}+\alpha}-\boldsymbol{u}^{\mathrm{m}}\right)\cdot\nabla\phi^{\mathrm{f}, \mathrm{n}+\alpha}_\mathrm{h} + s\phi^{\mathrm{f}, \mathrm{n}+\alpha}_\mathrm{h} -f. 	
\end{align}
The linear stabilization terms are used to address spurious oscillations in the solution when convection and reaction effects are dominant. However, separate treatment is required to address oscillations in the solution near the regions of high gradients \cite{joshi2018positivity}. Across the cavity interface, the transition from the liquid to the vapor phase takes place over the span of a few elements, and the spatial gradient of $\phi^{\mathrm{f}}$ is very high. As seen in Eq.~(\ref{densInterp}) and Eq.~(\ref{viscInterp}), physical properties of the fluid such as the density and the viscosity are obtained as weighted linear interpolations of $\phi^{\mathrm{f}}$. Unbounded oscillations in $\phi^{\mathrm{f}}$ can result in negative values of $\rho^{\mathrm{f}}$ and $\mu^{\mathrm{f}}$, which are unphysical and can induce numerical instability. To address this, we introduce additional non-linear stabilization terms that impart a positivity property to the underlying element-level matrices. The third and fourth lines of Eq.~(\ref{eq:TEMvariational}) contain the positivity preserving nonlinear stabilization terms in the streamwise and crosswind directions respectively\cite{joshi2018positivity}. These essentially act as added diffusion in the region of high gradients and ensure that the element-level matrix is an $M$-matrix. 

The PPV parameters $\chi$, $k_s^\mathrm{add}$ and $k_c^\mathrm{add}$ for the phase indicator transport equation are obtained as:
\begin{align}
	\chi &= \frac{2}{|s|h + 2|\left(\boldsymbol{u}^{\mathrm{f}}-\boldsymbol{u}^{\mathrm{m}}\right)|},\\
	k_s^\mathrm{add} &= \mathrm{max} \bigg\{ \frac{||\left(\boldsymbol{u}^{\mathrm{f}}-\boldsymbol{u}^{\mathrm{m}}\right)| - \tau_\phi|\left(\boldsymbol{u}^{\mathrm{f}}-\boldsymbol{u}^{\mathrm{m}}\right)|s|h}{2} - \tau_\phi|\left(\boldsymbol{u}^{\mathrm{f}}-\boldsymbol{u}^{\mathrm{m}}\right)|^2 + \frac{sh^2}{6}, 0 \bigg\},\\
	k_c^\mathrm{add} &= \mathrm{max} \bigg\{ \frac{|\left(\boldsymbol{u}^{\mathrm{f}}-\boldsymbol{u}^{\mathrm{m}}\right)|h}{2} + \frac{sh^2}{6}, 0 \bigg\},
\end{align} 
where $|\left(\boldsymbol{u}^{\mathrm{f}}-\boldsymbol{u}^{\mathrm{m}}\right)|$ is the magnitude of the convection velocity and $h$ is the characteristic element length \cite{joshi2017positivity}. We next present the Navier-Stokes equations and its variational form for the two-phase solver.

\subsubsection{Navier-Stokes equations}
\noindent The trial solution is taken from the function space $\mathcal{S}^\mathrm{h}$, the values of which satisfy the Dirichlet boundary condition at the Dirichlet boundary $\Gamma_D$. Let $\mathcal{V}^\mathrm{h}$ be the space of test functions which vanish on $\Gamma_D$. We formulate the variational statement for the fluid flow equations to find $[ {\boldsymbol{u}}_\mathrm{h}^\mathrm{n+\alpha}, {p}_\mathrm{h}^\mathrm{n+1}] \in\mathcal{S}^\mathrm{h}$ such that $\forall [\boldsymbol{\psi}_\mathrm{h}, q_\mathrm{h}] \in \mathcal{V}^\mathrm{h}$,
\begin{align}
	&\int_{\Omega^{\mathrm{f}}} \boldsymbol{\psi}_{h}^{\mathrm{f}} \cdot\left(\left.\rho^{\mathrm{f}} \partial_{t} \boldsymbol{u}_{h}^{\mathrm{f}, \mathrm{n}+\alpha_{\mathrm{m}}}\right|_{\boldsymbol{\chi}}+\rho^{\mathrm{f}}\left(\boldsymbol{u}_{h}^{\mathrm{f}, \mathrm{n}+\alpha}-\boldsymbol{u}^{\mathrm{m}}\right) \cdot \nabla \boldsymbol{u}_{h}^{\mathrm{f}, \mathrm{n}+\alpha}\right) d \Omega\nonumber \\
	&+ \int_{\Omega^{\mathrm{f}}} \nabla \boldsymbol{\psi}_{h}^{\mathrm{f}}: \boldsymbol{\sigma}_{h}^{\mathrm{n}+\alpha} d \Omega 
	+ \int_{\Omega^{\mathrm{f}}\left(t^{\mathrm{n}+1}\right)} q_{h}\left(\nabla \cdot \boldsymbol{u}_{h}^{\mathrm{f}, \mathrm{n}+\alpha}\right) d \Omega\nonumber \\
	&+ \sum_{e=1}^{\mathrm{n}_{\mathrm{el}}} \int_{\Omega^{\mathrm{f}}} \frac{\tau_{m}}{\rho^{\mathrm{f}}}\left(\rho^{\mathrm{f}}\left(\boldsymbol{u}_{h}^{\mathrm{f}, \mathrm{n}+\alpha}-\boldsymbol{u}^{\mathrm{m}}\right) \cdot \nabla \boldsymbol{\psi}_{h}^{\mathrm{f}}+\nabla q_{h}\right) \cdot \boldsymbol{\mathcal{R}}_{m} d \Omega^{\mathrm{e}}
	+ \sum_{e=1}^{\mathrm{n}_{\mathrm{el}}} \int_{\Omega^{e}} \nabla \cdot \boldsymbol{\psi}_{h}^{\mathrm{f}} \tau_{c} \rho^{\mathrm{f}} \boldsymbol{\mathcal{R}}_{c} d \Omega^{e}\\
	&-\displaystyle\sum_\mathrm{e=1}^\mathrm{n_{el}}\int_{\Omega^\mathrm{e}} \tau_\mathrm{m} \boldsymbol{\psi}_\mathrm{h}\cdot (\boldsymbol{\mathcal{R}}_\mathrm{m} \cdot \nabla {\boldsymbol{u}}_\mathrm{h}^\mathrm{n+\alpha}) \mathrm{d\Omega^e} 
	-\displaystyle\sum_\mathrm{e=1}^\mathrm{n_{el}}\int_{\Omega^\mathrm{e}} \frac{\nabla \boldsymbol{\psi}_\mathrm{h}}{\rho(\phi)}:(\tau_\mathrm{m}\boldsymbol{\mathcal{R}}_\mathrm{m} \otimes \tau_\mathrm{m}\boldsymbol{\mathcal{R}}_\mathrm{m}) \mathrm{d\Omega^e}\nonumber \\
	&= \int_{\Omega^{\mathrm{f}}} \boldsymbol{\psi}_{h}^{\mathrm{f}} \cdot \boldsymbol{f}^{\mathrm{f}, \mathrm{n}+\alpha} \; d \Omega
	+\int_{\Gamma_{N}^{\mathrm{f}}} \boldsymbol{\psi}_{h}^{\mathrm{f}} \cdot \boldsymbol{h}^{\mathrm{f}, \mathrm{n}+\alpha} \; d \Gamma
	+ \underbrace{\int_{\Omega^{\mathrm{f}}} q_{h}\left(\frac{1}{\rho_{l}} - \frac{1}{\rho_{v}} \right) \dot{m}^{\mathrm{f}, \mathrm{n}+\alpha} \; d \Omega\nonumber} 
\end{align}
\noindent where the first and second lines contain the standard Galerkin finite element terms of the momentum and the continuity equation. The third line contains the Galerkin Least Squares stabilization terms for the momentum and mass continuity equations. The fourth line contains stabilization terms based on the multi-scale argument \cite{hughes2005conservation, hsu2010improving}. The fifth line contains the Galerkin terms for the body force and the Neumann boundary in the momentum equation. The term in under-braces corresponds to the Galerkin projection of the term in the mass conservation equation dependent on $p^{\mathrm{f}}$, and has been introduced in this formulation. The element-wise residual of the momentum and the continuity equations denoted by $\boldsymbol{\mathcal{R}}_\mathrm{m}$ and $\boldsymbol{\mathcal{R}}_\mathrm{c}$ respectively are given by
\begin{align}
	\boldsymbol{\mathcal{R}}_\mathrm{m}({\boldsymbol{u}^{\mathrm{f}}},{p}^{\mathrm{f}}) 
	&= \rho^{\mathrm{f}}\partial_t{\boldsymbol{u}}_\mathrm{h}^\mathrm{n+\alpha_m} 
	+ \rho^{\mathrm{f}}\left(\boldsymbol{u}^{\mathrm{f}, \mathrm{n}+\alpha}-\boldsymbol{u}^{\mathrm{m}}\right) \cdot \nabla{\boldsymbol{u}}_\mathrm{h}^{\mathrm{f}, \mathrm{n}+\alpha} - \nabla \cdot {\boldsymbol{\sigma}}_\mathrm{h}^{\mathrm{f}, \mathrm{n}+\alpha} - \boldsymbol{f}^{\mathrm{f}, \mathrm{n}+\alpha}_\mathrm{h}, \\
	\boldsymbol{\mathcal{R}}_\mathrm{c}({\boldsymbol{u}^{\mathrm{f}}},{p}^{\mathrm{f}}, \phi^{\mathrm{f}}) &= \nabla\cdot\boldsymbol{u}^{\mathrm{f}, \mathrm{n}+\alpha}_\mathrm{h}-\left(\frac{1}{\rho_{l}} - \frac{1}{\rho_{v}} \right) \dot{m}^{\mathrm{f}, \mathrm{n}+\alpha}.
\end{align}
$\tau_\mathrm{m}$ and $\tau_\mathrm{c}$ are stabilization parameters \cite{brooks1982streamline, tezduyar1992incompressible, franca1992stabilized} defined as 
\begin{align}
	\tau_\mathrm{m} &= \bigg[ \bigg( \frac{2}{\Delta t}\bigg)^2 + {\boldsymbol{u}}_\mathrm{h}\cdot \boldsymbol{G}{\boldsymbol{u}}_\mathrm{h} + C_I \bigg(\frac{\mu(\phi)}{\rho(\phi)}\bigg)^2 \boldsymbol{G}:\boldsymbol{G}\bigg] ^{-1/2},\\
	\tau_\mathrm{c} &= \frac{1}{\mathrm{tr}(\boldsymbol{G})\tau_\mathrm{m}}, \label{tau_c}
\end{align}
where $C_I$ is a constant derived from the element-wise inverse estimate \cite{harari1992c} and $\mathrm{tr(\boldsymbol{G})}$  denotes the trace of $\boldsymbol{G}$. 

\subsection{Finite element matrix form}   

\noindent The Navier-Stokes equations are coupled with the phase fraction transport equation in a partitioned manner as shown in Eq.~(\ref{eq:NSmatrices}) and Eq.~(\ref{eq:phimatrices}). The equations are linearized with a Newton-Raphson technique to solve for the incremental velocity, pressure and phase indicator which are advanced in time using the Generalized-$\alpha$ time integration method. The linearized matrices for the fluid flow and the phase indicator transport equations are arranged in the form
        \begin{equation}
            \left[\begin{array}{cc}
                \boldsymbol{K}_{\Omega^{\mathrm{f}}} & \boldsymbol{G}_{\Omega^{\mathrm{f}}} \\
                -\boldsymbol{G}_{\Omega^{\mathrm{f}}}^{T} & \boldsymbol{C}_{\Omega^{\mathrm{f}}}
                \end{array}\right]\left\{\begin{array}{c}
                \Delta \boldsymbol{u}^{\mathrm{f}, n+\alpha} \\
                \Delta p^{\mathrm{f}, n+1}
                \end{array}\right\}=\left\{\begin{array}{l}
                {\boldsymbol{\mathcal { R }}}_{\mathrm{m}} \\
                {\boldsymbol{\mathcal { R }}}_{\mathrm{c}}
            \end{array}\right\} \label{eq:NSmatrices}
        \end{equation}
        \begin{equation}
            \left[\boldsymbol{K}_{\phi}\right]\{\Delta \phi^{\mathrm{f}, n+\alpha}\}=\{{\boldsymbol{\mathcal { R }}}_{\phi}\} \label{eq:phimatrices}
        \end{equation}

\noindent where $\boldsymbol{K}_{\Omega^{\mathrm{f}}}$ is the stiffness matrix of the momentum equation, $\boldsymbol{G}_{\Omega^{\mathrm{f}}}$ is the discrete gradient operator and $\boldsymbol{G}^T_{\Omega^{\mathrm{f}}}$ is the divergence operator. $\boldsymbol{C}_{\Omega^{\mathrm{f}}}$ consists of the pressure-pressure stabilization term and the terms in the mass continuity equation having dependency on the pressure. $\boldsymbol{K}_{\phi}$ is the stiffness matrix for the phase indicator transport equation, consisting of the transient, convection, reaction, linear stabilization terms and the non-linear PPV stabilization terms. $\Delta \boldsymbol{u}^{\mathrm{f}, n+\alpha}$, $\Delta p^{\mathrm{f}, n+1}$ and $\Delta \phi^{\mathrm{f}, n+\alpha}$ are the increments in velocity, pressure and the phase indicator respectively. \\
\noindent Linearization of the phase indicator transport equation requires the derivative of the terms in Eq.~(\ref{eq:TEM_CDR}) with respect to $\phi^{\mathrm{f}, n+\alpha}$. 
\begin{align}
    \dfrac{\partial G}{\partial \phi^{\mathrm{f}, n+\alpha}}
    &= \dfrac{\partial \left( \partial_t{\phi}^{\mathrm{f}, \mathrm{n}+\alpha_\mathrm{m}} \right)}{\partial \phi^{\mathrm{f}, n+\alpha}}
    + \cancelto{0}{\dfrac{\partial \left( \left(\boldsymbol{u}^{\mathrm{f}}-\boldsymbol{u}^{\mathrm{m}}\right)\cdot\nabla\phi^{\mathrm{f}, \mathrm{n}+\alpha} \right)}{\partial \phi^{\mathrm{f}, n+\alpha}}}
    + \dfrac{\partial \left( s\phi^{\mathrm{f}, \mathrm{n}+\alpha} \right)}{\partial \phi^{\mathrm{f}, n+\alpha}}
    - \cancelto{0}{\dfrac{\partial f}{\partial \phi^{\mathrm{f}, n+\alpha}}}\\
    &= \dfrac{\alpha_\mathrm{m}}{\alpha \gamma \Delta t} + \dfrac{\partial \left( s\phi^{\mathrm{f}, \mathrm{n}+\alpha} \right)}{\partial \phi^{\mathrm{f}, n+\alpha}}
\end{align}
where the derivative of the transient term reduces to a constant obtained from the generalized-$\alpha$ relations in Eq.~(\ref{eq:genAlpha}). Owing to the symmetrical property of second order cross-derivatives, the convective term also reduces to zero. The derivative of the reaction term $\dfrac{\partial \left(s\phi^{\mathrm{f}, \mathrm{n}+\alpha}\right)}{\partial \phi^{\mathrm{f}, \mathrm{n}+\alpha}}$ needs to be calculated. Similarly, in the linearization of the mass continuity equation we encounter the term $\dfrac{\partial \dot{m}^{\mathrm{f}, \mathrm{n}+\alpha}}{\partial p^{\mathrm{f}, \mathrm{n}+1}}$. 
\noindent Care must be taken in forming the terms $\dfrac{\partial \dot{m}}{\partial p^{\mathrm{f}, \mathrm{n}+1}}$ and $\dfrac{\partial \left(s\phi^{\mathrm{f}, \mathrm{n}+\alpha}\right)}{\partial \phi^{\mathrm{f}, \mathrm{n}+\alpha}}$ for a stable linearization of the matrices  $\boldsymbol{C}_{\Omega^{\mathrm{f}}}$ and $\boldsymbol{K}_{\phi}$ respectively. In model A, $\dot{m}$ varies non-linearly with both $p$ and $\phi^{\mathrm{f}}$, along with a discontinuity at $p=p_v$. Model B presents a simpler linearization with $\dot{m}$ being a linear function of both $p$ and $\phi^{\mathrm{f}}$. We propose linearlizations for the said terms in Tables.~(\ref{tab:mDot}) and (\ref{tab:tableSphi}) respectively. The presented linearlizations have been tested to be stable over a range of temporal and spatial discretizations on two different cavitating flow configurations.\\
\begingroup
\setlength{\tabcolsep}{25pt} 
\renewcommand{\arraystretch}{1.75} 
\begin{table}[]
\centering
\resizebox{1\textwidth}{!}{%
\tikz\node[draw=red,thick,double,inner sep=1pt]{
\begin{tabular}{ccc}
\toprule
                    & \textbf{Model A}                                                                                                                    & \textbf{Model B}                                                                                                                                                                                                                                   \\ \midrule
\textit{\textbf{$p^{\mathrm{n}+1}_h>p_v$}} & $C_{c} \boldsymbol{\phi}^{\mathrm{f},\mathrm{n}+\alpha}_h(1-\boldsymbol{\phi}^{\mathrm{f},\mathrm{n}+\alpha}_h)\dfrac{ \rho_{l} \rho_{v}}{\rho^{\mathrm{f}} R_{B}} \sqrt{\dfrac{3}{2 \rho_{l}\left(p^{\mathrm{n}+1}_h-p_{v}\right)}}$               & $C_{prod}\rho_l\dfrac{1- \boldsymbol{\phi}^{\mathrm{f},\mathrm{n}+\alpha}_h}{\dfrac{1}{2}\rho_lU_{\infty}^2t_{\infty}}$                \\ 
\textit{\textbf{$p^{\mathrm{n}+1}_h<p_v$}} & $ C_{v} \boldsymbol{\phi}^{\mathrm{f},\mathrm{n}+\alpha}_h(1 + \phi_{nuc} -\boldsymbol{\phi}^{\mathrm{f},\mathrm{n}+\alpha}_h)\dfrac{ \rho_{l} \rho_{v}}{\rho^{\mathrm{f}} R_{B}} \sqrt{\dfrac{3}{2 \rho_{l}\left(p_{v}-p^{\mathrm{n}+1}_h\right)}}$ & $C_{dest}\dfrac{\rho_l^2}{\rho_v}\dfrac{\boldsymbol{\phi}^{\mathrm{f},\mathrm{n}+\alpha}_h}{\dfrac{1}{2}\rho_lU_{\infty}^2t_{\infty}}$ \\ 
\textit{\textbf{$p^{\mathrm{n}+1}_h=p_v$}} & 0, using $\lim_{p^{\mathrm{n}+1}_h\to p_{v}}\dfrac{p^{\mathrm{n}+1}_h-p_{v}}{\sqrt{\left|p^{\mathrm{n}+1}_h-p_{v}\right|}} = 0$                                                                                                                                   & 0  \\  \bottomrule                                                                                                 
\end{tabular}};%
}
\caption{$\dfrac{\partial \dot{m}^{\mathrm{f}, \mathrm{n}+\alpha}}{\partial p^{\mathrm{f}, \mathrm{n}+1}}$ in linearized matrix $\boldsymbol{C}_{\Omega^{\mathrm{f}}}$}
\label{tab:mDot}
\end{table}
\endgroup

\begingroup
\setlength{\tabcolsep}{25pt} 
\renewcommand{\arraystretch}{1.75} 
\begin{table}[]
\centering
\resizebox{1\textwidth}{!}{%
\tikz\node[draw=red,thick,double,inner sep=1pt]{
\begin{tabular}{ccc}
\toprule
                    & \textbf{Model A}                                                                                                                             & \textbf{Model B}                                                                                                                                                 \\ \midrule
\textit{\textbf{$p^{\mathrm{n}+1}_h>p_v$}} & $-C_{c} (1-2\boldsymbol{\phi}^{\mathrm{f},\mathrm{n}+\alpha}_h)\dfrac{3 }{ R_{B}} \sqrt{\dfrac{2\left(p^{\mathrm{n}+1}_h-p_{v}\right)}{3 \rho_{l}}}$            & $\dfrac{\rho^{\mathrm{f}}}{\rho_{v}}\dfrac{C_{prod}}{\dfrac{1}{2}\rho_lU_{\infty}^2t_{\infty}}\left(p^{\mathrm{n}+1}_h-p_{v}\right)$                \\ 
\textit{\textbf{$p^{\mathrm{n}+1}_h<p_v$}} & $C_{v} (1 + \phi_{nuc} -2\boldsymbol{\phi}^{\mathrm{f},\mathrm{n}+\alpha}_h)\dfrac{3 }{ R_{B}} \sqrt{\dfrac{2\left(p_{v}-p^{\mathrm{n}+1}_h\right)}{3 \rho_{l}}}$ & $-\dfrac{\rho^{\mathrm{f}}\rho_{l}}{\rho^2_{v}}\dfrac{C_{dest}}{\dfrac{1}{2}\rho_lU_{\infty}^2t_{\infty}}\left(p^{\mathrm{n}+1}_h-p_{v}\right)$ \\ 
\textit{\textbf{$p^{\mathrm{n}+1}_h=p_v$}} & 0, using $\lim_{p^{\mathrm{n}+1}_h\to p_{v}}\dfrac{p^{\mathrm{n}+1}_h-p_{v}}{\sqrt{\left|p^{\mathrm{n}+1}_h-p_{v}\right|}} = 0$                                                                                                                                            & 0                   \\  \bottomrule
\end{tabular}};%
}
\caption{$\dfrac{\partial \left(s\phi^{\mathrm{f}, \mathrm{n}+\alpha}\right)}{\partial \phi^{\mathrm{f}, \mathrm{n}+\alpha}}$ in linearized matrix $\boldsymbol{K}_{\phi}$}
\label{tab:tableSphi}
\end{table}
\endgroup
\subsection{Implementation details}
Algorithm~\ref{algo:solverAlgo} shows the algorithm for the staggered partitioned coupling of the implicit Navier-Stokes and cavitation solvers. This provides two unique advantages compared to monolithic couplings - (i) simplicity of linearization and implementation, and (ii) the flexibility to couple additional governing equations as demanded by the case being studied. For example, in Sections \ref{sec:validation} and \ref{sec:explore}, the additional equations for turbulence modeling and the ALE mesh update are similarly coupled using a staggered partitioned approach. For stable and accurate convergence, non-linear predictor-multicorrector iterations are performed within each time step. Let us consider the time level $t^\mathrm{n}$ and the associated velocity $\boldsymbol{u}^{\mathrm{f}}(\boldsymbol{x},t^\mathrm{n})$, pressure $p^{\mathrm{f}}(\boldsymbol{x},t^\mathrm{n})$ and phase indicator $\phi^{\mathrm{f}}(\boldsymbol{x},t^\mathrm{n})$ fields. Each non-linear predictor-corrector iteration consists of one pass through the steps \textbf{[A]}, \textbf{[B]}, \textbf{[C]} and \textbf{[D]}. In step \textbf{[A]} of the predictor-corrector iteration $\mathrm{k}$, a predictor fluid velocity $\boldsymbol{u}^{\mathrm{f}, \mathrm{n+1}}_\mathrm{(k+1)}$ and pressure $p^{\mathrm{f}, \mathrm{n+1}}_\mathrm{(k+1)}$ is obtained from the solution of the Navier-Stokes equations(Eq.~(\ref{eq:NSmatrices})). These are passed to the cavitation solver in step \textbf{[B]}. An updated phase indicator $\phi^{\mathrm{f}, \mathrm{n+1}}_\mathrm{(k+1)}$ is obtained in step \textbf{[C]} by solving Eq.~(\ref{eq:phimatrices}), which is then passed to the Navier-Stokes solver in step \textbf{[D]}. This updated phase indicator value is used to interpolate the fluid density and dynamic viscosity, and prepare the Navier-Stokes matrices for the next iteration $\mathrm{k+1}$. This cyclic process is continued till the solvers have achieved the convergence criteria. Then the coupled solver is advanced to the next time level $t^\mathrm{n+1}$.
\begin{figure}[H]
\centering
\floatstyle{ruled}
\newfloat{algorithm}{H}{loa}
\floatname{algorithm}{Algorithm}
\begin{algorithm}
\caption{Partitioned coupling of implicit Navier-Stokes and cavitation solvers}
\label{algo:solverAlgo}
\textbf{Input:} $\boldsymbol{u}^{\mathrm{f}, 0}$, $p^{\mathrm{f}, 0}$, $\phi^{\mathrm{f}, 0}$ \\
\qquad    \textbf{for} n $\leftarrow$ 0 to $n_{last}$ \textbf{do}
\begin{tabbing}
\qquad Predict solution\\
\qquad \qquad $\bigg[ \boldsymbol{u}^{\mathrm{f}, \mathrm{n+1}}_{(0)} \;\;
                p^{\mathrm{f}, \mathrm{n+1}}_{(0)} \;\;
                \phi^{\mathrm{f}, \mathrm{n+1}}_{(0)}
                \bigg]
                \leftarrow
                \bigg[
                \boldsymbol{u}^{\mathrm{f}, \mathrm{n}} \;\;
                p^{\mathrm{f}, \mathrm{n}} \;\;
                \phi^{\mathrm{f}, \mathrm{n}}
                \bigg]$\\ \\
\qquad Interpolate density and viscosity fields\\                
\qquad \qquad $\rho^{\mathrm{f}}(\phi^{\mathrm{f}, \mathrm{n}+1}_{(0)})$, $\mu^{\mathrm{f}}(\phi^{\mathrm{f}, \mathrm{n}+1}_{(0)})$\\ \\
\qquad \textbf{for} k $\leftarrow$ 0 to convergence/$k_{max}$ \textbf{do} \\ \\
\qquad \qquad \begin{tikzpicture}
	\begin{pgfonlayer}{nodelayer}
		\node [style=none] (0) at (-3, 5) {};
		\node [style=none] (1) at (-3, -4) {};
		\node [style=none] (2) at (-9, -4) {};
		\node [style=none] (3) at (-9, 5) {};
		\node [style=none] (4) at (-0.5, 5) {};
		\node [style=none] (5) at (5.5, 5) {};
		\node [style=none] (6) at (5.5, -4) {};
		\node [style=none] (7) at (-0.5, -4) {};
		\node [style=none] (8) at (-0.75, 3) {};
		\node [style=none] (9) at (-2.75, 3) {};
		\node [style=none] (10) at (-2.75, -3) {};
		\node [style=none] (11) at (-0.75, -3) {};
		\node [style=none] (12) at (-6, 4.25) {\textbf{[A] Navier-Stokes Solver}};
		\node [style=none] (13) at (2.5, 4.25) {\textbf{[C] Cavitation Solver}};
		\node [style=none] (14) at (-1.75, 0) {\textbf{Non-linear}};
		\node [style=none] (15) at (-1.75, 3.5) {\textbf{[D]}};
		\node [style=none] (16) at (-1.75, -3.5) {$\boldsymbol{u}^\mathrm{n+1}_\mathrm{(k+1)}, p^\mathrm{n+1}_\mathrm{(k+1)}$};
		\node [style=none] (17)[anchor=west] at (-8.75, 3.5) {1. Interpolate solution};
		\node [style=none] (18)[anchor=west] at (-8.75, 1.5) {2. Solve};
		\node [style=none] (19)[anchor=west] at (-8.75, 0.25) {3. Correct solution};
		\node [style=none] (20)[anchor=west] at (-8.75, -1.75) {4. Update solution};
		\node [style=none] (21)[anchor=west] at (-0.25, 3.5) {1. Interpolate solution};
		\node [style=none] (22)[anchor=west] at (-0.25, 1.5) {2. Solve};
		\node [style=none] (23)[anchor=west] at (-0.25, -0.5) {3. Correct solution};
		\node [style=none] (24)[anchor=west] at (-0.25, -2.5) {4. Update solution};
		\node [style=none] (25)[anchor=west] at (-8.25, 3) {$\boldsymbol{u}^{\mathrm{f}, \mathrm{n+\alpha}}_\mathrm{(k+1)} \leftarrow \boldsymbol{u}^{\mathrm{f}, \mathrm{n}} + \alpha(\boldsymbol{u}^{\mathrm{f}, \mathrm{n+1}}_\mathrm{(k)} - \boldsymbol{u}^{\mathrm{f}, \mathrm{n}})$};
		\node [style=none] (26)[anchor=west] at (-8.25, 2.25) {$p^{\mathrm{f}, \mathrm{n+1}}_\mathrm{(k+1)} \leftarrow p^{\mathrm{f}, \mathrm{n+1}}_\mathrm{(k)}$};
		\node [style=none] (27)[anchor=west] at (-8.25, 1) {$\Delta \boldsymbol{u}^{\mathrm{f}, \mathrm{n+\alpha}}$ and $\Delta p^{\mathrm{f}, \mathrm{n+1}}$ in Eq.~({\ref{eq:NSmatrices}})};
		\node [style=none] (28)[anchor=west] at (-8.25, -0.25) {$\boldsymbol{u}^{\mathrm{f}, \mathrm{n+\alpha}}_\mathrm{(k+1)} \leftarrow \boldsymbol{u}^{\mathrm{f}, \mathrm{n+\alpha}}_\mathrm{(k+1)} + \Delta\boldsymbol{u}^{\mathrm{f}, \mathrm{n+\alpha}}$};
		\node [style=none] (29)[anchor=west] at (-8.25, -1) {$p^{\mathrm{f}, \mathrm{n+1}}_\mathrm{(k+1)} \leftarrow p^{\mathrm{f}, \mathrm{n+1}}_\mathrm{(k+1)} + \Delta p^{\mathrm{f}, \mathrm{n+1}}$};
		\node [style=none] (30)[anchor=west] at (-8.25, -2.25) {$\boldsymbol{u}^{\mathrm{f}, \mathrm{n+1}}_\mathrm{(k+1)} \leftarrow \boldsymbol{u}^{\mathrm{f}, \mathrm{n}} + \dfrac{1}{\alpha}(\boldsymbol{u}^{\mathrm{f}, \mathrm{n+\alpha}}_\mathrm{(k+1)} - \boldsymbol{u}^{\mathrm{f}, \mathrm{n}})$};
		\node [style=none] (31)[anchor=west] at (-8.25, -3) {$p^{\mathrm{f}, \mathrm{n+1}}_\mathrm{(k+1)} \leftarrow p^{\mathrm{f}, \mathrm{n+1}}_\mathrm{(k+1)} + \Delta p^{\mathrm{f}, \mathrm{n+1}}$};
		\node [style=none] (32)[anchor=west] at (0.25, 3) {$\phi^{\mathrm{f}, \mathrm{n+\alpha}}_\mathrm{(k+1)} \leftarrow \phi^{\mathrm{f}, \mathrm{n}} + \alpha(\phi^{\mathrm{f}, \mathrm{n+1}}_\mathrm{(k)} - \phi^{\mathrm{f}, \mathrm{n}})$};
		\node [style=none] (33)[anchor=west] at (0.25, 1) {$\Delta \phi^{\mathrm{f}, \mathrm{n+\alpha}}$ in Eq.~(\ref{eq:phimatrices})};
		\node [style=none] (34)[anchor=west] at (0.25, -1) {$\phi^{\mathrm{f}, \mathrm{n+\alpha}}_\mathrm{(k+1)} \leftarrow \phi^{\mathrm{f}, \mathrm{n+\alpha}}_\mathrm{(k+1)} + \Delta\phi^{\mathrm{f}, \mathrm{n+\alpha}}$};
		\node [style=none] (35)[anchor=west] at (0.25, -3) {$\phi^{\mathrm{f}, \mathrm{n+1}}_\mathrm{(k+1)} \leftarrow \phi^{\mathrm{f}, \mathrm{n}} + \dfrac{1}{\alpha}(\phi^{\mathrm{f}, \mathrm{n+\alpha}}_\mathrm{(k+1)} - \phi^{\mathrm{f}, \mathrm{n}})$};
		\node [style=none] (36) at (-1.75, -0.7) {\textbf{iterations}};
		\node [style=none] (37) at (-1.75, 2.75) {$\phi^\mathrm{n+1}_\mathrm{(k+1)}$};
		\node [style=none] (38) at (-1.75, -2.75) {\textbf{[B]}};
	\end{pgfonlayer}
	\begin{pgfonlayer}{edgelayer}
		\draw [style=black dashed line] (3.center) to (0.center);
		\draw [style=black dashed line] (0.center) to (1.center);
		\draw [style=black dashed line] (1.center) to (2.center);
		\draw [style=black dashed line] (2.center) to (3.center);
		\draw [style=black dashed line] (4.center) to (5.center);
		\draw [style=black dashed line] (5.center) to (6.center);
		\draw [style=black dashed line] (6.center) to (7.center);
		\draw [style=black dashed line] (7.center) to (4.center);
		\draw [style=black arrow solid, in=15, out=525] (8.center) to (9.center);
		\draw [style=black arrow solid, bend right=15] (10.center) to (11.center);
	\end{pgfonlayer}
\end{tikzpicture}\\ \\
\qquad \textbf{end do}
\end{tabbing}
\textbf{end do}
\end{algorithm}
\vspace{-0.5cm}
\end{figure}

A similar partitioned iterative coupling is used between the ALE fluid-cavitation and the structural displacement. In the current study, the structural displacement is dictated by the prescribed motion. Alternatively, for fully-coupled FSI studies the structural displacement can be obtained by solving the governing equations for the structural mechanics as presented in \citet*{joshi2017variationally}). At time $t^{\mathrm{n}}$, let us consider a predictor structural displacement $\boldsymbol{\eta}^{\mathrm{s}}\left(\boldsymbol{x}^{\mathrm{s}}, t^{\mathrm{n}}\right)$ obtained from the structural update in the Lagrangian reference frame. These structural displacements are then passed to the fluid solvers, while satisfying kinematic conditions at the fluid-structure interface $\Gamma^{\mathrm{fs}}$ as follows. At the time $t^{\mathrm{n+1}}$, the displacement $\boldsymbol{\eta}^{\mathrm{f}}$ of the fluid spatial coordinates (i,e., Eulerian mesh coordinates) at the wetted boundary $\Gamma^{\mathrm{fs}}$ are equated to the structural displacement $\boldsymbol{\eta}^{\mathrm{s}}$. 
\begin{align}
    \boldsymbol{\eta}^{\mathrm{f}, \mathrm{n}+1}=\boldsymbol{\eta}^{\mathrm{s}},  && \text { on } \Gamma^{\mathrm{fs}}
\end{align}
In addition, the velocity continuity is satisfied at the interface is satisfied at the time $t^{\mathrm{n}+\alpha}$ as
\begin{align}
    \boldsymbol{u}^{\mathrm{f}, \mathrm{n}+\alpha}=\boldsymbol{u}^{\mathrm{m}, \mathrm{n}+\alpha},  && \text { on } \Gamma^{\mathrm{fs}}
\end{align}
with the mesh velocity at the interface evaluated as
\begin{align}
   \boldsymbol{u}^{\mathrm{m}, \mathrm{n}+\alpha}=\frac{\boldsymbol{\eta}^{\mathrm{f}, \mathrm{n}+1}-\boldsymbol{\eta}^{\mathrm{f}, \mathrm{n}}}{\Delta t},  && \text { on } \Gamma^{\mathrm{fs}}
\end{align}
Away from the interface $\Gamma^{\mathrm{fs}}$ and any Dirichlet conditions on $\boldsymbol{\eta}^{\mathrm{f}}$ on the Dirichlet boundary $\Gamma_{D}^{\mathrm{m}}$, Eq.~(\ref{eq:meshUpdate}) is solved for updating the fluid spatial coordinates. The flow-cavitation equations are then solved with the updated kinematic boundary conditions and the displaced fluid spatial coordinates. The updated hydrodynamic forces are passed to the structural solver to correct the deformations. This sequence is repeated iteratively till converged solutions are obtained.

For the finite element discretization of the variables $\boldsymbol{u}^{\mathrm{f}}$, $p^{\mathrm{f}}$ and $\phi^{\mathrm{f}}$, we consider equal-order interpolations. The Harwell-Boeing sparse matrix format is used to form and store the matrices for the linear system of equations. A Generalized Minimal RESidual (GMRES)\cite{saad1986gmres} algorithm is used to solve the linear system. We observe that 2-3 non-linear iterations(as described in Algorithm~(\ref{algo:solverAlgo})) are sufficient to obtain a converged solution at each time-step. The solver uses communication protocols based on standard message passing interface \cite{MPI} for parallel computing on distributed memory clusters. 

\section{Numerical verification and convergence}  
\noindent A stabilized numerical method for the solution of cavitating flows has been presented. In this section we verify the implementation and discuss its convergence and efficacy. 

\subsection{Analytical solution of vaporous spherical bubble collapse}

\noindent The numerical implementation is verified by comparison with the analytical solution of the Rayleigh-Plesset equation for spherical bubble dynamics. It has influenced several works on the study of cavitation and bubble dynamics, and has seen many adaptations over the last century. We present here only salient features relevant to the current study and interested readers are directed to comprehensive texts such as \cite{brennen_2013,franc2006fundamentals}. 

\begin{figure}[!h]
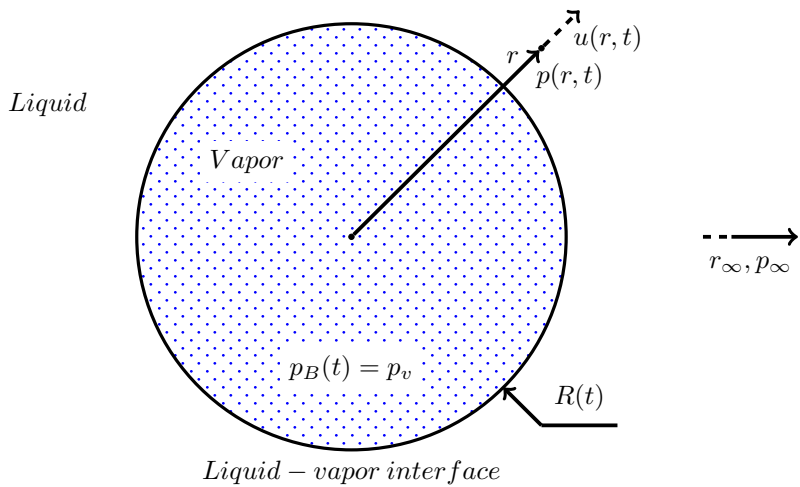

    \centering
    \ctikzfig{bubSchematic}
    \caption{Representative schematic for iso-thermal collapse of a vaporous spherical bubble}
    \label{fig:bubSchematic}
\end{figure}

We consider the case of an iso-thermal collapse of a spherical vaporous bubble. A bubble of radius $R_0$ is initialized in an infinite domain of liquid with a constant far-field pressure $p_\infty$. It is assumed that the bubble contents are homogeneous and consist only of saturated vapor and no non-condensable gases. It is also assumed that the pure liquid is incompressible. In the absence of any non-condensable gases, the uniform pressure $p_B(t)$ inside the bubble equals $p_v$. Fig.~(\ref{fig:bubSchematic}) shows the schematic of the domain. 
Spherical symmetry is assumed and a one-dimensional analysis is performed along the radial direction. 

Using mass conservation in the incompressible liquid outside the bubble, the radial outward velocity $u(r,t)$ at a radial distance $r$ from the center of the bubble can be shown to follow an inverse square law of the form
\begin{equation}
    u(r,t) \propto 1/r^2 \label{eq:uProportionality}
\end{equation}
Assuming that the liquid at the interface is in thermal equilibrium with the vapor at the saturation temperature, no mass transfer in the form of evaporation or condensation occurs at the interface. Thus, at the interface the velocity $u(R,t)=dR/dt$. Thus Eq.~(\ref{eq:uProportionality}) can be written as
\begin{equation}
    u(r,t) = \frac{R^{2}}{r^2} \frac{d R}{d t} \label{eq:mass_bub}
\end{equation}
\noindent where $R(t)$ is the bubble radius. Momentum conservation in the $r$-direction is given by
\begin{equation}
    -\frac{1}{\rho_{l}} \frac{\partial p}{\partial r}=\frac{\partial u}{\partial t}+u \frac{\partial u}{\partial r}-\nu_{l}\left[\frac{1}{r^{2}} \frac{\partial}{\partial r}\left(r^{2} \frac{\partial u}{\partial r}\right)-\frac{2 u}{r^{2}}\right] \label{eq:mom_bub}
\end{equation}
 Substituting the expression for $u$ from Eq.~(\ref{eq:mass_bub}) the following relation is obtained 
\begin{equation}
    -\frac{1}{\rho_{l}} \frac{\partial p}{\partial r}=2\left(\frac{d R}{d t} \right)^2\left[\frac{R}{r^2}-\frac{R^4}{r^5} \right] + \frac{R^2}{r^2}\frac{d^2R}{dt^2} \label{eq:presDer_bub}
\end{equation}
Assuming surface tension and viscous effects to be negligible, and the absence of any mass transfer at the interface, $p_{r=R}=p_B(t)=p_v$. Using this assumption, spatial integration of Eq.~(\ref{eq:presDer_bub}) between $r = R$ and $r \rightarrow\infty$, gives the well-known Rayleigh equation \cite{rayleigh1917viii}
\begin{equation}
    \frac{p_v-p_{\infty}}{\rho_{l}}=R \frac{d^{2} R}{d t^{2}}+\frac{3}{2}\left(\frac{d R}{d t}\right)^{2} \label{eq:rayleighEquation}
\end{equation}
Eq.~(\ref{eq:rayleighEquation}) can be integrated to obtain the interface velocity during the collapse
\begin{equation}
    \frac{d R}{d t} = -\sqrt{\frac{2}{3} \frac{p_{v}-p_{\infty}}{\rho_{l}}\left(1-\frac{R_{0}^{3}}{R^{3}}\right)} \label{eq:interfaceVelocity}
\end{equation}
The interface acceleration is obtained as 
\begin{equation}
    \frac{d^2 R}{d t^2} = \frac{p_v-p_{\infty}}{\rho_{l}}\frac{R_{0}^3}{R^{4}}
    \label{eq:interfaceAcceleration}
\end{equation}
Using the kinematic relations in Eq.~(\ref{eq:interfaceVelocity}) and Eq.~(\ref{eq:interfaceAcceleration}), Eq.~(\ref{eq:presDer_bub}) can be integrated between $r$ (in the liquid outside the bubble) and $r\rightarrow \infty$ to obtain the pressure $p(r, t)$ at the radial distance $r$ from the centre of the bubble 
\begin{equation}
    \frac{p(r, t)-p_{\infty}}{p_{\infty}-p_v} = -\frac{4}{3}\left( 1 - \frac{R_{0}^{3}}{R^{3}} \right)\left( \frac{R}{r} - \frac{1}{4}\frac{R^{4}}{r^{4}} \right) - \frac{R^3_{0}}{R^2 r}
    \label{eq:presNonDr}
\end{equation}
\noindent In the absence of thermal effects and non-condensable gas content, we encounter the special case of a vaporous bubble collapsing to zero volume. The total time $t_{tc}$ taken by such a bubble to collapse from an initial radius $R_0$ is presented by Rayleigh's relation \cite{rayleigh1917viii}:
\begin{equation}
    t_{tc} = 0.915R_0\left( \frac{\rho_l }{p_{\infty}-p_v} \right)^{1/2}
\end{equation}
\subsection{Numerical case setup}
As we discussed in Section~\ref{sec:introduction}, the numerical modeling of cavitation requires careful consideration of the cavitation model based on the spatial-temporal scales of the particular flow configuration. We use cavitation model A for the numerical study of this case of micro-scale bubble collapse because of its origins in the Rayleigh-Plesset equation. In addition, model A has been previously used to study this configuration in \cite{ghahramani2019comparative}, where an implementation based on the finite volume method was used. For comparison of the efficacy of the present variational finite element implementation, we set up our numerical case using similar geometrical parameters and model coefficients as in \cite{ghahramani2019comparative}.

\begin{figure}[!h]
\centering
\includegraphics[width=1\columnwidth]{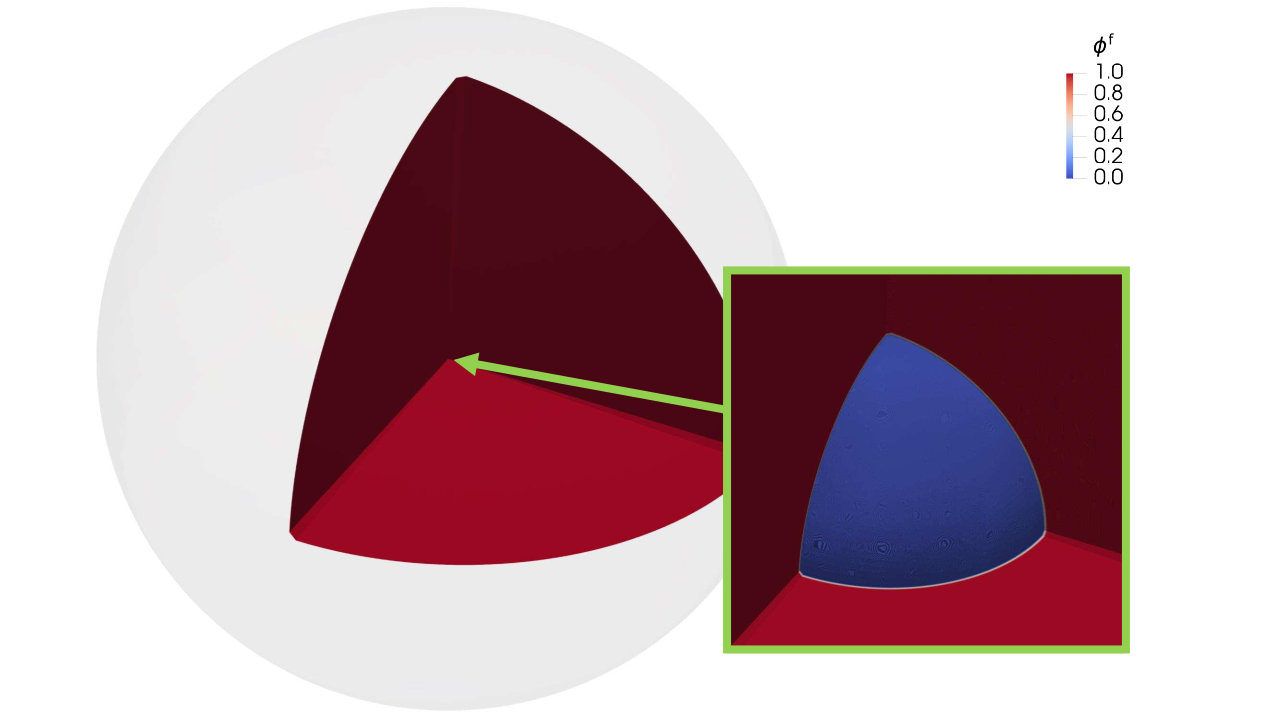}
\caption{Sliced octant of domain showing the initial phase indicator values. Inlay showing a close-up of the bubble at the centre of the domain. Inside the bubble, the pressure is initialized to the vapor pressure $p_v$, with $\phi^{\mathrm{f}}=0.01$ corresponding to the vapor phase. Outside the bubble, the initial pressure has a continuous distribution according to Eq.~(\ref{eq:presNonDr}) and a constant $\phi^{\mathrm{f}}=1$ corresponding to the liquid phase.}
\label{fig:bubInit}
\end{figure}

For the numerical study we consider a 3D spherical domain of $R_\infty = 0.5\si{m}$, consisting of quiescent liquid. A spherical vaporous bubble of radius $R_0 = 4\times 10^{-4}\si{m}$ is initialized in the centre of the domain. The spherical domain is discretized using $840264$ hexahedral elements, with 49 nodes in the radial direction resolving the initial bubble. The initial phase indicator and pressure inside the bubble are set to $0.01$ and $p_v = 2320\si{Pa}$ respectively, corresponding to the vapor phase. Outside the bubble, the phase indicator is set to $1$. The pressure in the liquid is initialized according to Eq.~(\ref{eq:presNonDr}). Fig.~(\ref{fig:bubInit}) shows the initial conditions for the phase indicator. A far-field pressure $P_\infty = 1\times 10^{5}~\si{Pa}$ is weakly enforced over the outer surface boundary boundary of the domain using a traction boundary condition. The densities of the pure liquid and vapor phases are taken as $\rho_l=1000~ \si{kg.m^{-3}}$ and $\rho_v=0.01389~\si{kg.m^{-3}}$, giving a density ratio $\dfrac{\rho_l}{\rho_v}\approx 72000$. The dynamic viscosities of the two phases are set to $\mu_l=0.001~\si{kg.m^{-1}.s^{-1}}$ and $\mu_v=10^{-5}~\si{kg.m^{-1}.s^{-1}}$. The model parameters $n_0$ and $d_{nuc}$ are assumed to be $10^8$ and $10^{-4}~\si{m}$ respectively. The condensation and evaporation coefficients are set to $C_c=C_v=100$.
\subsection{Results and discussion}
\begin{figure}[p]
    \centering
    \begin{subfigure}[b]{\textwidth}
        \centering
        \includegraphics[height=0.42\textheight]{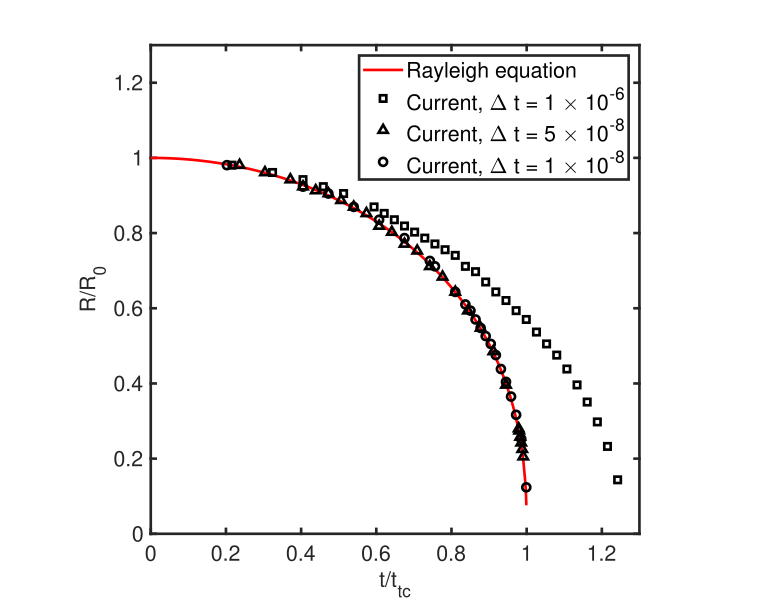}
        \caption{}
        \label{fig:bubCollapseRadius}
    \vspace{0.06\textheight}
    \end{subfigure}
    \begin{subfigure}[b]{\textwidth}
        \centering
        \includegraphics[height=0.42\textheight]{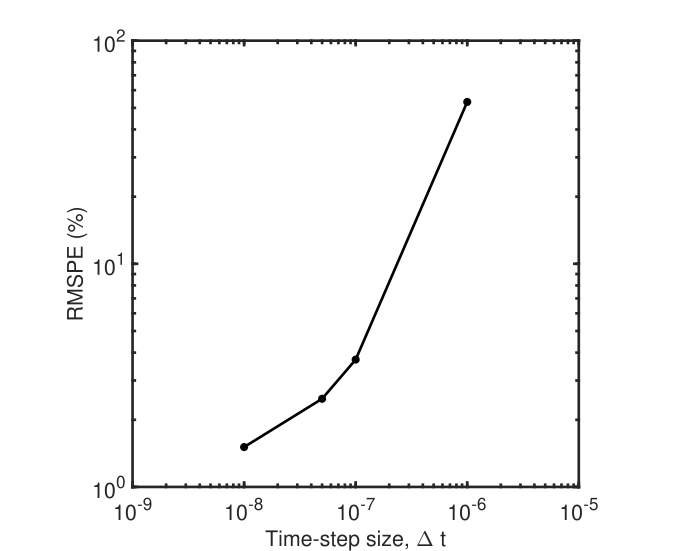}
        \caption{}
        \label{fig:RMSPE}
    \end{subfigure}
    \caption{Numerical assessment of spherical bubble collapse problem: (a) comparison of bubble radius against analytical solution of the Rayleigh equation (b) root mean square percentage error in bubble radius calculation at different values of the numerical time-step size $\Delta t$ }
    \label{fig:fpr-tpr}
\end{figure}

\begin{figure*}[t!]
    \centering
    \begin{subfigure}[b]{0.5\columnwidth}
        \centering
        \centerline{\includegraphics[height=9.5cm,clip=false]{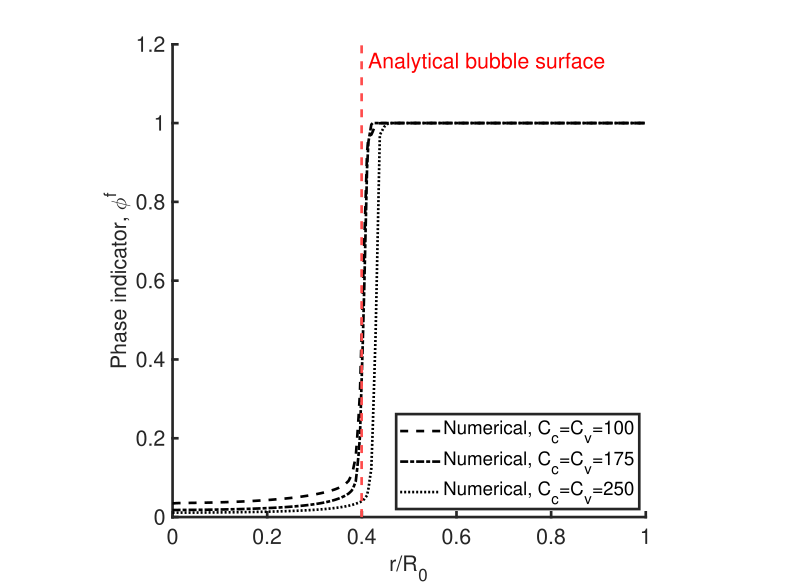}}
        \subcaption{}
        \label{fig:opradial_Cvar}
    \end{subfigure}\hspace{0.1em}%
    \begin{subfigure}[b]{0.5\columnwidth}
        \centering
        \centerline{\includegraphics[height=9.5cm,clip=false]{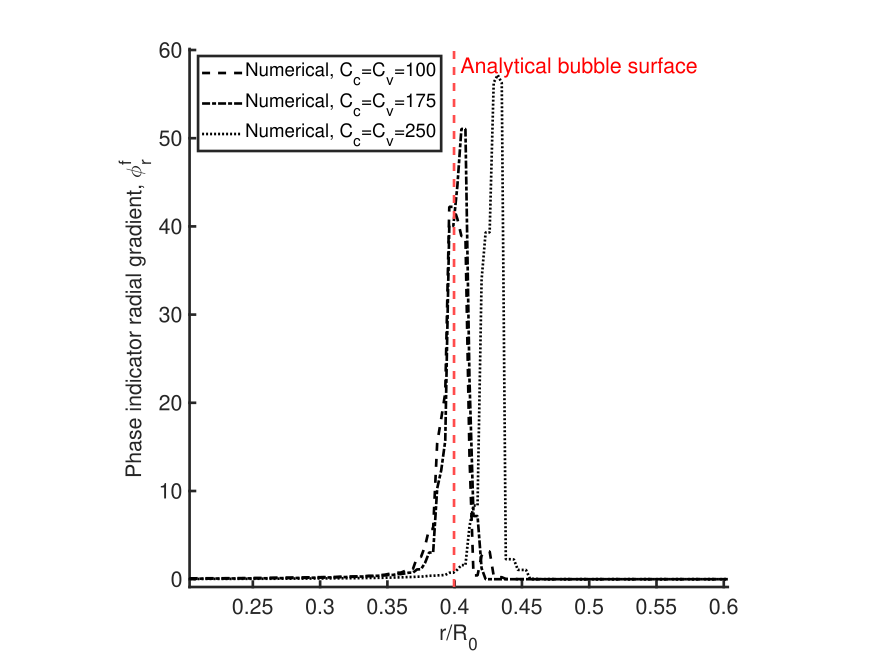}}
        \subcaption{}
        \label{fig:opradialGrad_Cvar}
    \end{subfigure}
    \caption{Spherical bubble collapse problem: Distribution of $\phi^{\mathrm{f}}$ and $\left(\dfrac{\partial \phi^{\mathrm{f}}}{\partial r}\right)$ in the vicinity of the bubble surface}\label{fig:phiProfiles}
\end{figure*}

\noindent The results of the numerical study are first compared to the exact solution of Eq.~(\ref{eq:interfaceVelocity}). Fig.~(\ref{fig:bubCollapseRadius}) shows the evolution of the bubble radius $R(t)$ with time. In the numerical study, we consider the bubble radius to be the effective radius of the volume of vapor in the domain. 
\begin{equation}
    R = \left( \dfrac{3}{4\pi}\int_{\Omega^{\mathrm{f}}} \left( 1 - \phi^{\mathrm{f}} \right) \mathrm{d}\Omega^{\mathrm{f}} \right)^{1/3} \label{eq:numBubRadius}
\end{equation}
Since no evaporation occurs outside the bubble, and no convection of vapor occurs though the boundary at $R_\infty$ $\left(\dfrac{\partial \phi^{\mathrm{f}}}{\partial r}=0\right)$, the only vapor content in the domain is in the bubble. This is also confirmed later in Fig.~(\ref{fig:opradial_Cvar}). Thus, Eq.~(\ref{eq:numBubRadius}) is seen to represent the bubble radius well. 
The bubble radius and the solution time are non-dimensionalized using the initial bubble radius $R_0$ and the Rayleigh collapse time $t_{tc}$ respectively. We consider predictions obtained using four different values of the numerical time-step $\Delta t$. It can be observed from Fig.~(\ref{fig:bubCollapseRadius}) that solutions obtained using $\Delta t \leq 1\times 10^{-7}~s$ are able to capture the evolution of the bubble radius well. To proceed, we sample the error between the numerical and exact bubble radius at four instances during the collapse process. These sampling instances correspond to the physical times $0.405t_{tc}$, $0.810t_{tc}$, $0.945t_{tc}$ and $0.972t_{tc}$. We compute the Root Mean Square Percentage Error (RMSPE) using the sampled results as 
\begin{equation}
    \mathrm{RMSPE}=\sqrt{\frac{1}{n} \cdot \sum_{i=1}^{n} \Delta R_{\mathrm{rel}, i}^{2}} \cdot 100 \%
\end{equation}
with
\begin{equation}
    \Delta R_{\mathrm{rel}, i}=\frac{R_{num,i}}{R_{exact,i}}-1
\end{equation}
where $R_{num,i}$ is the numerically obtained bubble radius and $R_{exact,i}$ is the exact solution and $i$ denotes the index of the sampling time. Figure~(\ref{fig:RMSPE}) shows the calculated RMSPE for different values of $\Delta t$. We observe that the RMSPE is below $2.5 \%$ for $\Delta t \leq 5\times 10^{-8}~\si{s}$, indicating good agreement with the exact solution. Thus, for the rest of this study we present results obtained with $\Delta t = 5\times 10^{-8}~\si{s}$.\\

\begin{remark}
The solution is observed to be convergent and stable across a range of time-step sizes, even at large values of the order of $\Delta t = 1\times 10^{-6}\si{s}$. No numerical oscillations are seen in the solution fields in the domain. \cite{ghahramani2019comparative} reported the presence of spurious pressure pulses in the domain for $\Delta t > 5\times 10^{-8}\si{s}$. We note the efficacy of the present implementation in suppressing these spurious oscillations.
\end{remark}

\begin{figure*}[t!]
    \centering
    \begin{subfigure}[b]{0.5\columnwidth}
        \centering
        \centerline{\includegraphics[height=8.5cm]{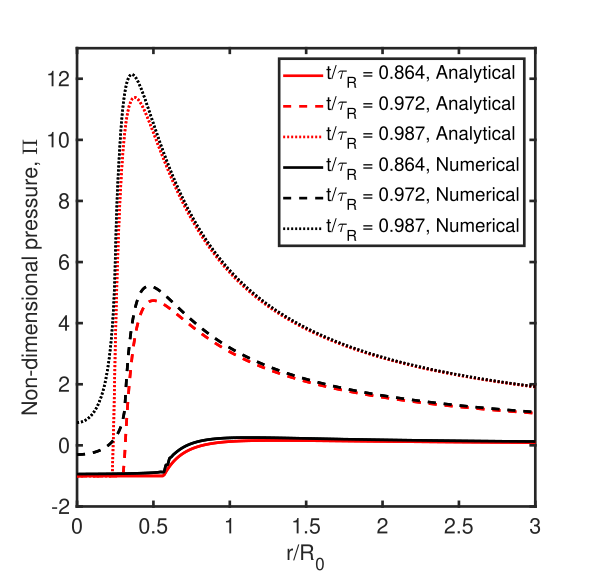}}
        \subcaption{}\label{fig:bubCollapsePres}
    \end{subfigure}\hfill%
    ~ 
    \begin{subfigure}[b]{0.5\columnwidth}
        \centering
        \centerline{\includegraphics[height=7.5cm,clip=false]{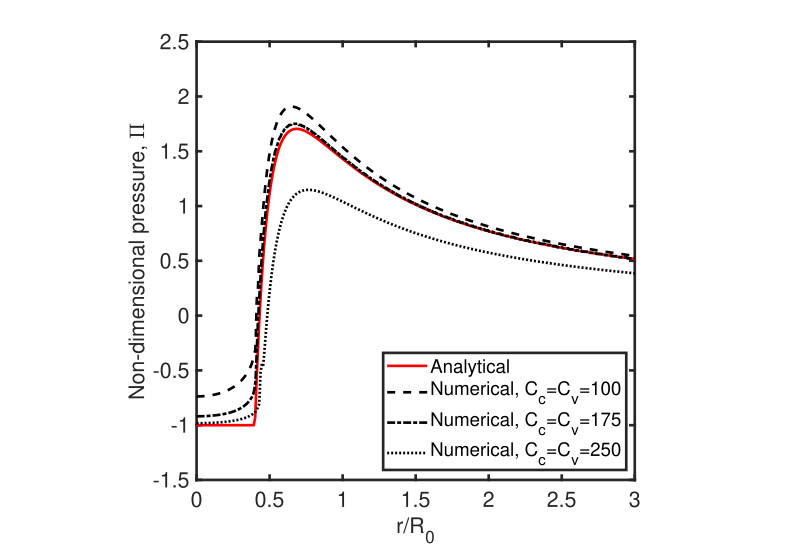}}
        \subcaption{}\label{fig:Pradial_Cvar}
    \end{subfigure}%
    ~ 
    \begin{subfigure}[b]{0.5\columnwidth}
        \centering
        \centerline{\includegraphics[height=7.5cm,clip=false]{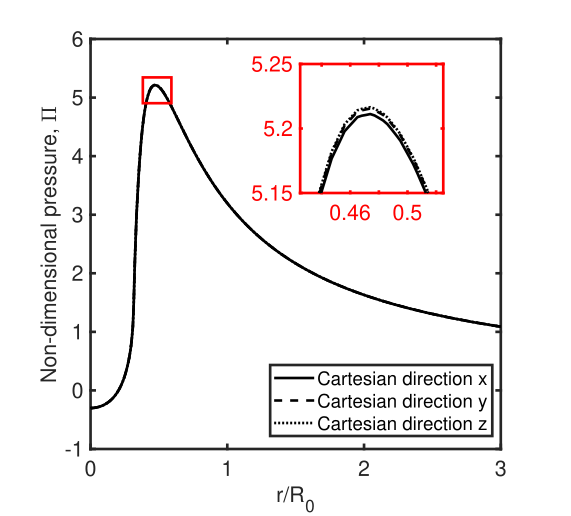}}
        \subcaption{}\label{fig:piSymmetry}
    \end{subfigure}%
    \caption{Spherical bubble collapse problem:  (a)  comparison of non-dimensional pressure $\Pi$ with the analytical solution at different time steps, (b) sensitivity of non-dimensional pressure $\Pi$ on the coefficients $C_c$ and $C_v$, and (c) assessment of spherical symmetry. 
    }
    \label{fig:pressureProfiles}
\end{figure*}

We next look at the predicted pressure field, which is another quantity of interest in cavitating flows. It is well known that the cavity collapse process can result in high pressures in the domain, several times the magnitude of the collapse driving pressure $p_\infty$. Thus, for any effective study of the fluid-structure interaction and noise effects in cavitating flows, it is important that the cavitation solver is able to accurately capture the pressure field. Figure~(\ref{fig:bubCollapsePres}) shows the non-dimensional pressure $\Pi = \dfrac{p^{\mathrm{f}}-p_\infty}{p_\infty-p_v}$ in the domain at different times during the bubble collapse, compared to the analytical solution of Eq.~(\ref{eq:presNonDr}). The pressure $p^{\mathrm{f}}$ is taken in the radial direction along the $z$-coordinate axis, with $r=0$ at the center of the bubble. The numerically obtained pressures are observed to be in good agreement with the analytical solution. The peak pressures are close to the expected values, and an improvement in accuracy is obtained compared to results presented in \cite{ghahramani2019comparative}. No spurious spikes or numerical oscillations are observed in the pressure. 

However, it is observed that at the final stages of the collapse process, the pressure inside the bubble is higher than $p_v$. This deviates from the assumption that the bubble pressure $p_B(t)$ equals $p_v$ at all times. Increasing the coefficients $C_c$ and $C_v$ appears to resolve this. Fig.~(\ref{fig:Pradial_Cvar}) compares the pressure profile at $t=0.945\tau_R$ for three values of the evaporation and condensation coefficients. At $C_c=C_v=175$, the pressure profile nearly matches the analytical solution. The pressure inside the bubble is also observed to approach closer to the theoretical value of the vapor pressure. On further increasing the $C_c=C_v$ to $250$, the vapor pressure inside the bubble is recovered. However, the peak pressure is observed to be under-predicted. Thus, a systematic tuning of the semi-empirical coefficients is required, which is one of the limitations of such phenomenological models. However, further tuning of these parameters is beyond the scope of the current work. 

Unlike in \citep{ghahramani2019comparative} where spherical symmetry was assumed in the numerical simulation, in the current study we consider the full 3D spherical domain. Numerical discretization and the weakly enforced far-field pressure boundary condition can introduce asymmetry in the solution field. To determine the ability of the solver to preserve the symmetrical nature of the solution, the non-dimensional pressure $\Pi$ is plotted along the three cartesian directions. Fig.~(\ref{fig:piSymmetry}) shows the comparison of $\Pi$ (at $t=0.972t_{tc}$, $C_c=C_v=100$) in the $x$, $y$ and $z$ directions as a function of the distance from the centre of the bubble. It is observed that the 3D solver is able to naturally preserve the symmetry of the collapsing bubble. 

\begin{remark}
 We note the ability of the present implementation to accurately predict the pressure in the domain, and the absence of spurious pressure oscillations across the bubble interface. The vapor pressure inside the cavity is also recovered, although it requires careful calibration of the model coefficients. It is possible that an adaptive calibration (e.g., deep learning models \cite{miyanawala2017efficient, bukka2021assessment})) of these coefficients can aid in generalizing the model to multiple flow configurations. This can be explored in future studies.
\end{remark}

\noindent Next, we investigate the predicted values of the phase indicator $\phi^{\mathrm{f}}$ in the domain. As discussed previously, presence of large spatial gradients of $\phi^{\mathrm{f}}$ across the bubble interface can result in unbounded numerical oscillations in the density field, and in turn the pressure field. Fig.~(\ref{fig:opradial_Cvar}) shows the distribution of $\phi^{\mathrm{f}}$ along the radial direction, while Fig.~(\ref{fig:opradialGrad_Cvar}) shows the gradient $\left(\dfrac{\partial \phi^{\mathrm{f}}}{\partial r}\right)$ of $\phi^{\mathrm{f}}$ with respect to the radial distance $r$. We observe no oscillations in the solution in the vicinity of the bubble surface. The solution is seen to be bounded within the range $\phi^{\mathrm{f}} \in \left[0, 1\right]$. The gradient $\dfrac{\partial \phi^{\mathrm{f}}}{\partial r}$ is $\geq 0$ across the interface, indicating monotone solutions. The location of the peak values of the gradient are observed to be coincident with the exact value of the bubble radius. Once again, a value of $C_c=C_v=175$ is observed to best represent the expected distribution of $\phi^{\mathrm{f}}$ in the domain.

\begin{remark}
 We note the ability of the present implementation to recover monotone and bounded solutions across the bubble interface. The absence of spurious oscillations in $\phi^{\mathrm{f}}$ in the vicinity of the interface allows the solver to be stable at large values of the density ratio $\dfrac{\rho_l}{\rho_v}$. It is worth emphasizing, in the current study, the density ratio is taken $\approx 72000$.\\
\end{remark}

\section{Turbulent cavitating flow over a hydrofoil: A validation study} \label{sec:validation}
\begin{figure}[!h]
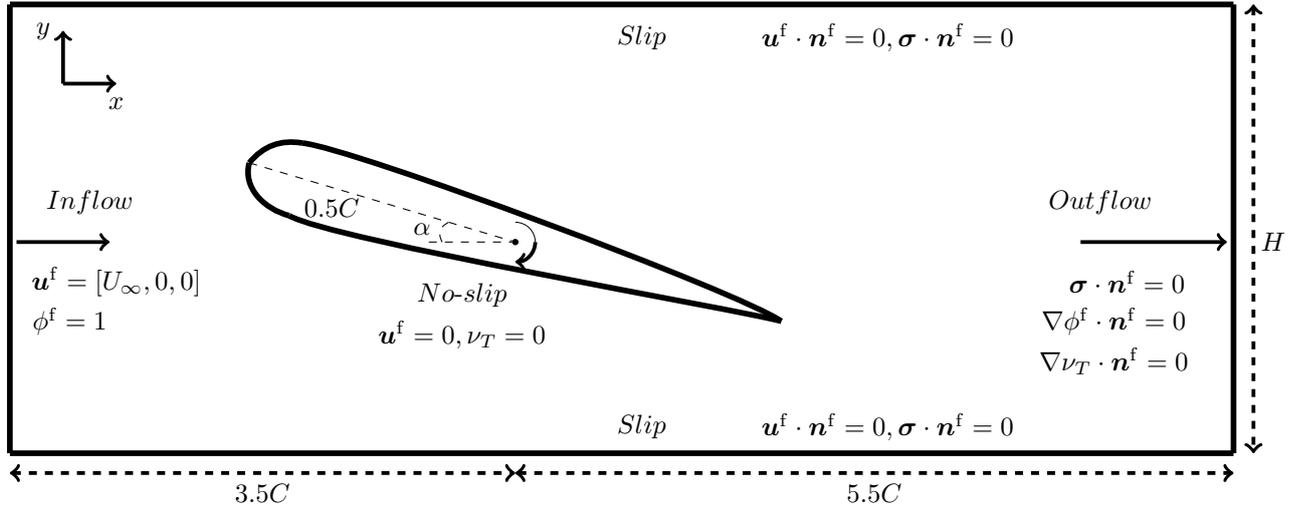

\centering
\ctikzfig{pitchingSchematic}
\caption{Representative computational domain and associated boundary conditions for cavitating flow over hydrofoil}
\label{fig:pitchingSchematic}
\end{figure}

\noindent In this section, we validate the proposed numerical method on the case of turbulent cavitating flow over a hydrofoil. This is an often-encountered scenario in marine propellers where fluid acceleration over the hydrofoil surface can result in very low pressures and cavity inception near the blade leading edge. For this study, we use cavitation model B because of its origins in flows involving large bubble clusters. It is computationally less expensive, and has been previously applied to the study of macro-scale cavitation over hydrofoils. The turbulence model is validated first on non-cavitating flow before proceeding to the case of cavitating flow. Fig.~(\ref{fig:pitchingSchematic}) shows the general schematic of the computational domain used in the sections to follow. $C$ is the hydrofoil chord length, $\alpha$ is the angle of attack of the incoming flow, $H$ is the channel height and $\nu_{T}$ is the kinematic turbulence viscosity. Specific details are given in the respective case descriptions.

\subsection{Turbulence model validation}
\begin{figure}[!h]
\centering
\includegraphics[width=\columnwidth]{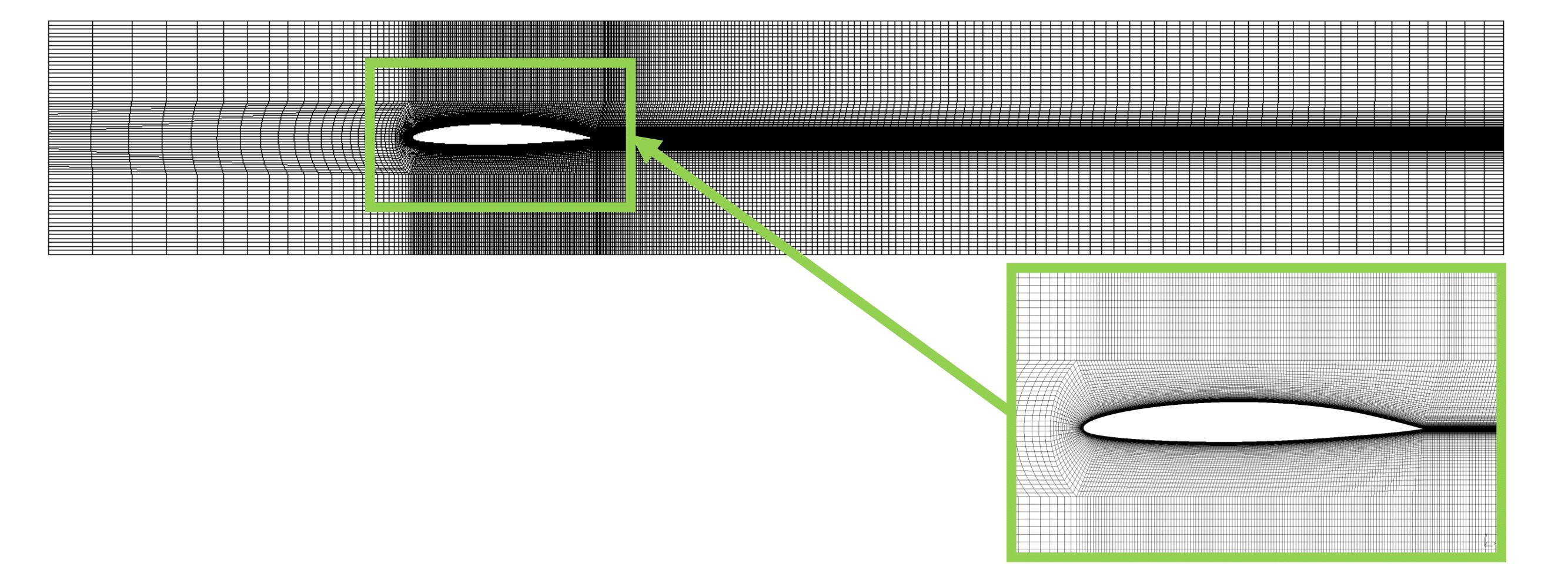}
\caption{Computational mesh for NACA66 hydrofoil. Inlay showing mesh in the vicinity of the hydrofoil}
\label{fig:NACA66grid}        
\end{figure}

\noindent In the current study, we employ a hybrid URANS-LES model to model turbulence. We validate the turbulence model on non-cavitating flow over a hydrofoil section. A NACA66 hydrofoil with chord length $C=0.15m$ and a span equal to $0.3C$ is used in the study. The height of the channel $H=1.28C$. Fig.~(\ref{fig:NACA66grid}) shows the computational grid. The grid consists of 76364 hexahedral elements (eight-node bricks) resolving the cross-section. 30 nodes resolve the spanwise direction. A target $y^{+}=y u_{\tau}/\nu=1$ was maintained in the discretization of the hydrofoil boundary layer, where $y$ is the height of the first node from the wall, $u_{\tau}$ is the friction velocity and $\nu$ is the kinematic viscosity of the single phase liquid. The flow Reynolds number is $8\times 10^5$, with a free-stream velocity $U_{\infty}=5.333~\si{m.s^{-1}}$. Liquid water at $25^{\circ}C$ is taken as the working fluid, with a single-phase density of $\rho_l=999.19~\si{kg.m^{-3}}$. A Dirichlet velocity condition equal to $U_{\infty}$ is set at the inlet with a natural traction-free outflow condition at the flow exit. A symmetric boundary condition is used on the top and bottom surfaces with periodic conditions on the spanwise surfaces. The time-step $\Delta t$ of the numerical simulation is set to $1.407 \times 10^{-4}$($T_{ref}/200$, where $T_{ref}=C/U_\infty$). 

\begin{figure*}[t!]
    \centering
    \begin{subfigure}[b]{0.5\columnwidth}
        \centering
        \centerline{\includegraphics[height=7.5cm,clip=false]{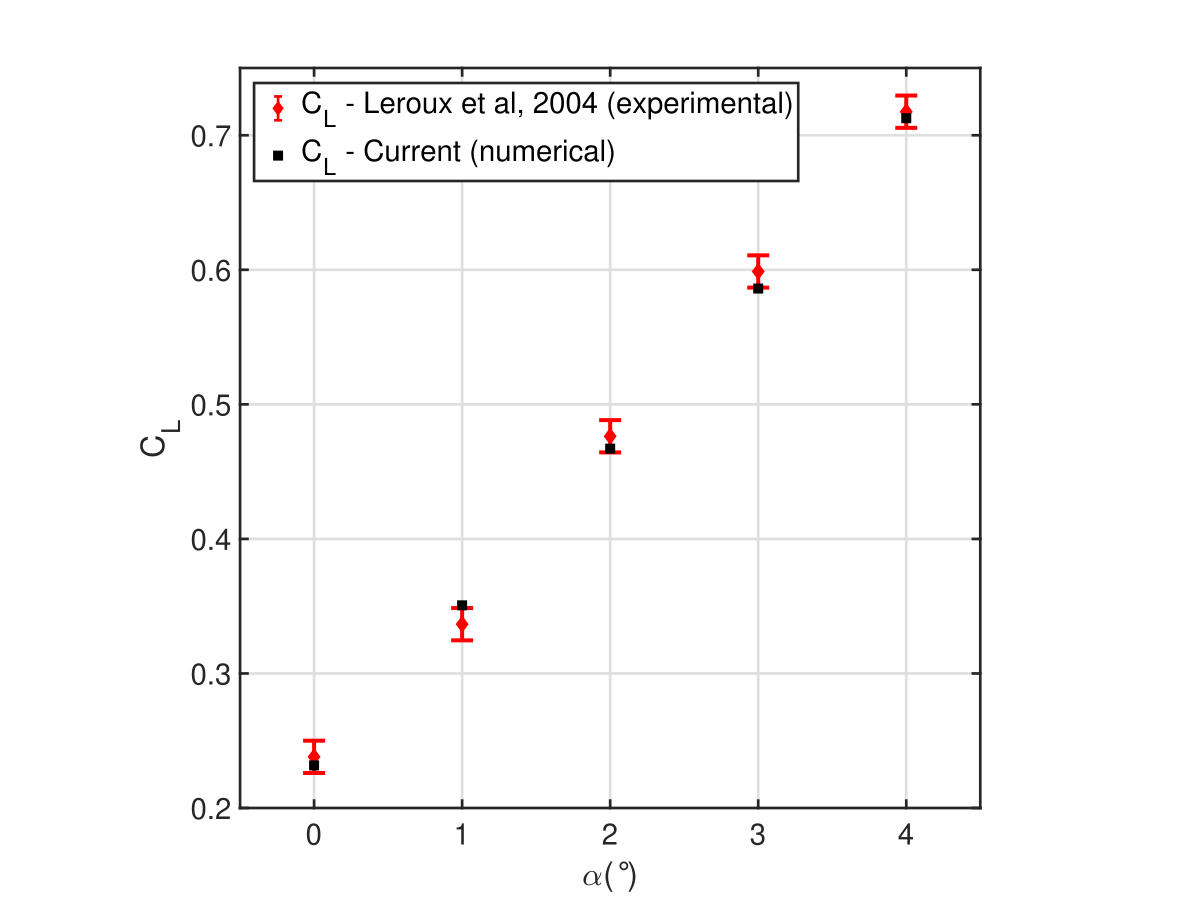}}
        \subcaption{}
    \end{subfigure}%
    ~ 
    \begin{subfigure}[b]{0.5\columnwidth}
        \centering
        \centerline{\includegraphics[height=7.5cm,clip=false]{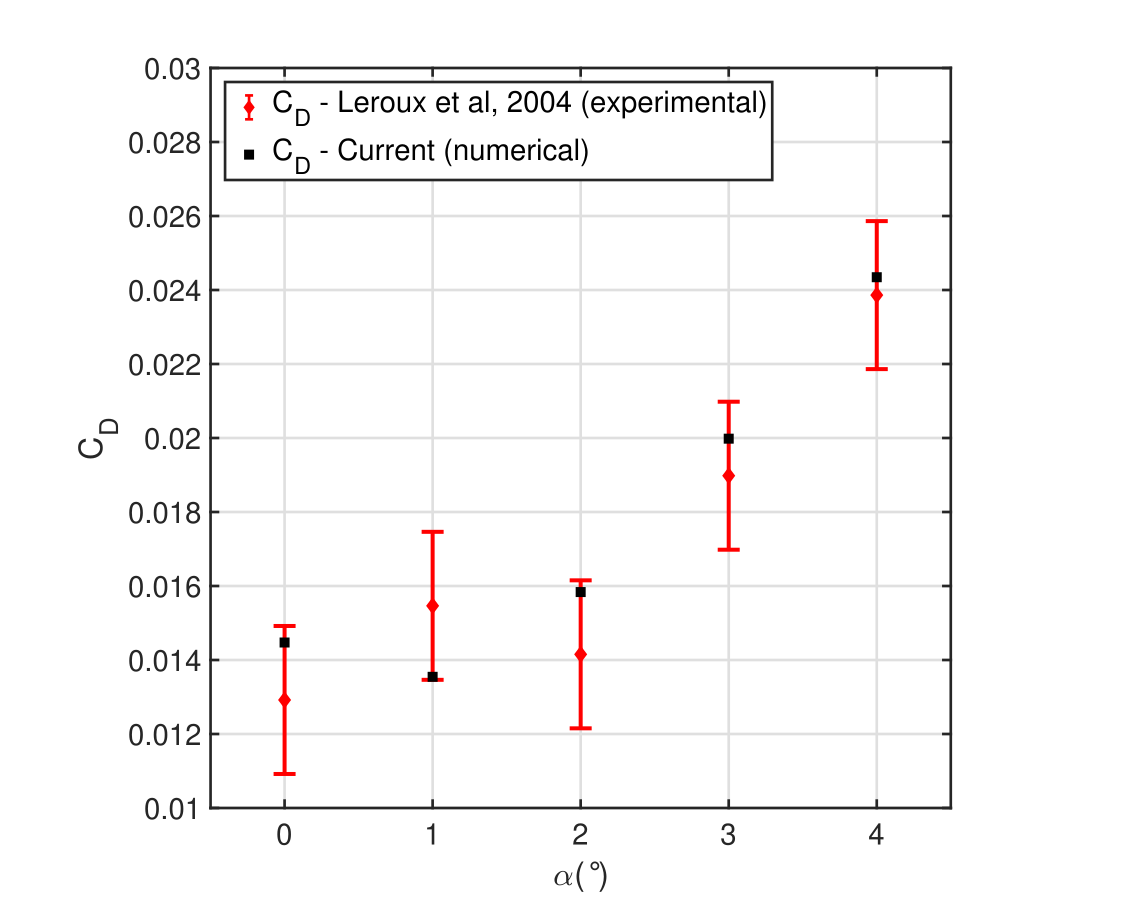}}
        \subcaption{}
    \end{subfigure}
    \caption{NACA66 hydrofoil problem: Predicted $C_L$ and $C_D$ in the non-cavitating range compared with the experimental values from \cite{leroux2004experimental}}\label{fig:turbValidation}
\end{figure*}

For our validation, the angle of attack $\alpha$ of the hydrofoil is varied between $0^{\circ}-4^{\circ}$ for the non-cavitating flow condition, and the time-averaged lift $(C_L)$ and drag $(C_D)$ coefficients were monitored.
 Fig.~(\ref{fig:turbValidation}) shows the comparison of $C_L$ and $C_D$ obtained from the numerical simulation with the experimental results in \cite{leroux2004experimental}. The predicted numerical results are observed to agree well with the experimental values in the non-cavitating regime, and are within the uncertainties for $C_L (\Delta C_L = 0.012)$ and $C_D (\Delta C_D = 0.002)$ reported in \cite{leroux2004experimental}.

\subsection{Cavitating flow over a hydrofoil} \label{section:cavFlowOverHydrofoil}
\begin{figure}[!h]
\centering
\includegraphics[width=0.77\columnwidth]{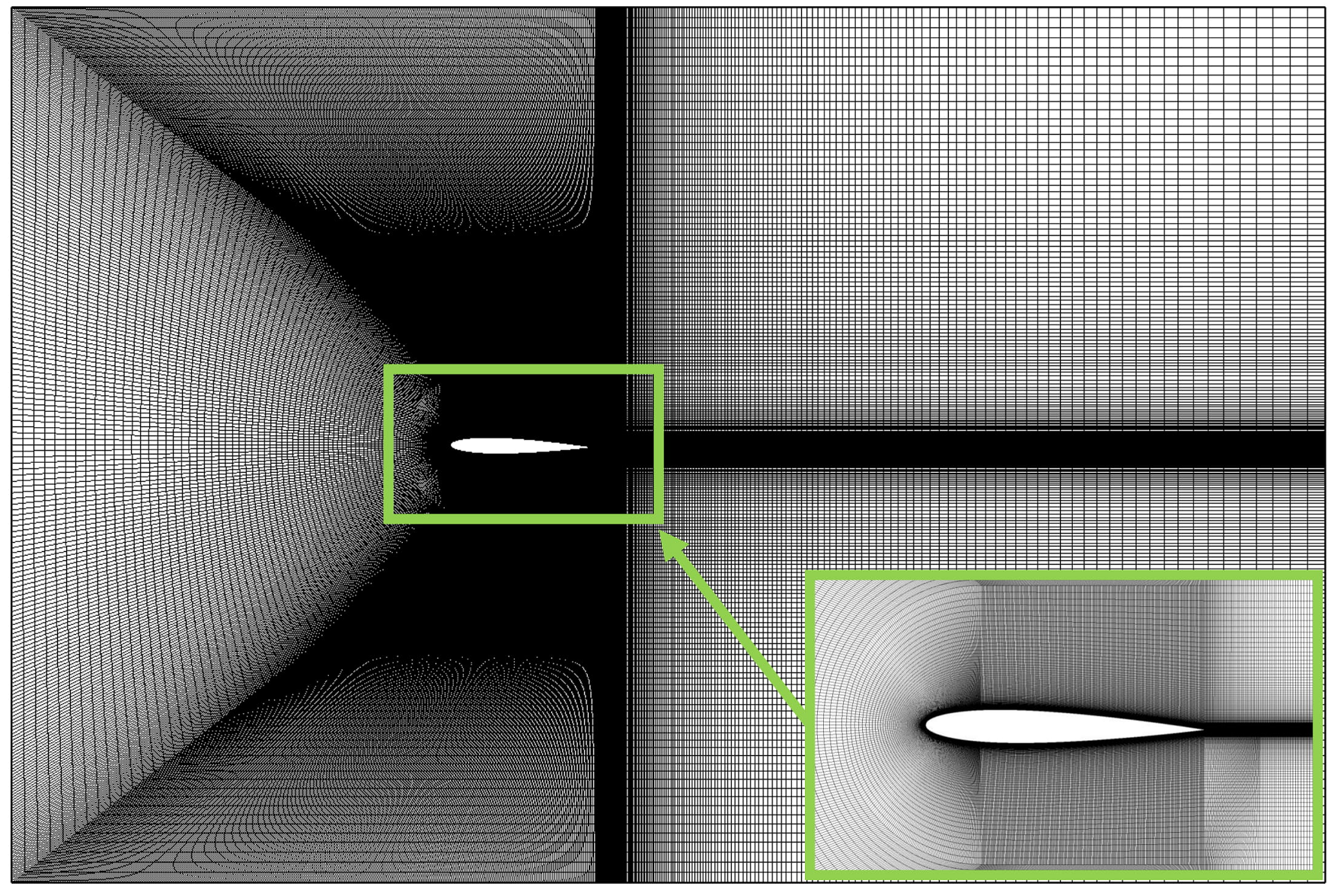}
\caption{Computational mesh for NACA0012 hydrofoil. Inlay showing mesh in the vicinity of the hydrofoil}
\label{fig:NACA0012grid}        
\end{figure}
To proceed further, the case of turbulent cavitating flow over a hydrofoil is studied using our numerical solver. A NACA0012 hydrofoil with $\alpha=1^{\circ}$ is consider with a Reynolds number of $2\times 10^6$. The chord length $C=1$ with the height of the channel $H=6C$. 
 Figure~(\ref{fig:NACA0012grid}) shows the computational grid used in the study. The domain is discretized with 90340 hexahedral elements and a 2D periodic boundary condition is applied in the spanwise direction. A target $y^+$ equal to 1 is enforced at the hydrofoil surface. The fluid domain is initialized with a liquid phase fraction of $\phi^{\mathrm{f}}=1$. A freestream velocity of $U_\infty=1\si{m.s^{-1}}$ is applied at the inlet as a dirichlet boundary condition. A traction-free outflow boundary condition is used, weakly setting $p_\infty=0$. The phase fraction is set to $1$ at the inlet, along with a Neumann boundary condition at the outflow. A density ratio ($\rho_l/\rho_v$) of 1000 is used in the study. The cavitation number of the flow is defined as $\sigma = \dfrac{p_\infty-p_v}{0.5\rho_l U_{\infty}^2}$, where $p_{\infty}$ is the free-steam hydrostatic pressure. In the current study, the cavitation number of the flow is set to $0.42$. The vapor pressure is set to meet the cavitation number of the flow. A time-step of $\Delta t=1\times 10^{-4}\si{s}$ ($t_{ref}/1000$) is utilized for the study. 

\begin{figure}[!h]
\centering
\includegraphics[width=0.85\columnwidth]{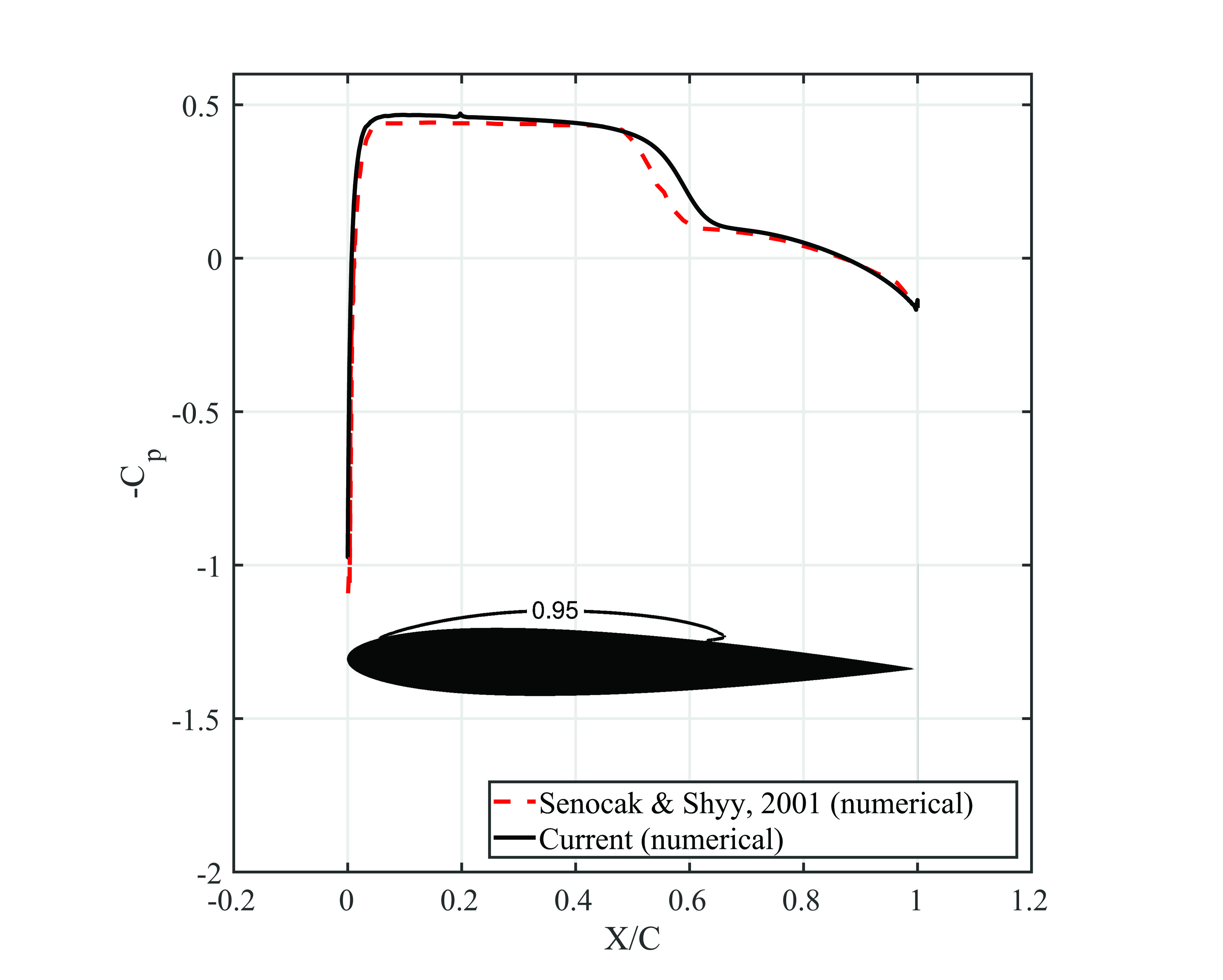}
\caption{Comparison of predicted pressure coefficient $C_p$ with results presented in \cite{senocak2001numerical}}
\label{fig:NACA0012validation}        
\end{figure}

Figure~(\ref{fig:NACA0012validation}) shows the result of the numerical study. The pressure coefficient ($C_p = \dfrac{p^{\mathrm{f}}-p_{\infty}}{0.5\rho_lU_{\infty}^2}$) on the suction surface of the hydrofoil is compared against the numerical results of \cite{senocak2001numerical} based on Kunz et al. \cite{kunz1999multi} model. A good agreement can be seen between these two numerical studies. The cavity pressure has been captured consistently, which is important for the prediction of hydrodynamic loads on the hydrofoil surface. A sharp gradient in the pressure over the hydrofoil suction surface can be observed, corresponding to the cavity closure location. Inlay shows the partial sheet cavity on the suction surface of the hydrofoil, marked by an iso-contour of $\phi^{\mathrm{f}}=0.95$. Notably, it is observed to stabilize over time to form a thin attached partial cavity on the hydrofoil surface. A slight shift in the cavity closure location is observed which is attributed to the difference in the cavitation and turbulence models used in the two studies. In addition, the model coefficients used in the current study are taken from\cite{huang2013physical} and \cite{senocak2004interfacial} where the geometries studied were different.  No cavity separation and shedding are observed, and a fully attached turbulent flow exists over the hydrofoil surface. The absence of a re-entrant jet is consistent with the observations of \cite{gopalan2000flow} for thin cavities.

\section{Application to fluid-structure interaction of caviating hydrofoil}\label{sec:explore}
\begin{figure}[!h]
\centering
\frame{\includegraphics[width=0.77\columnwidth]{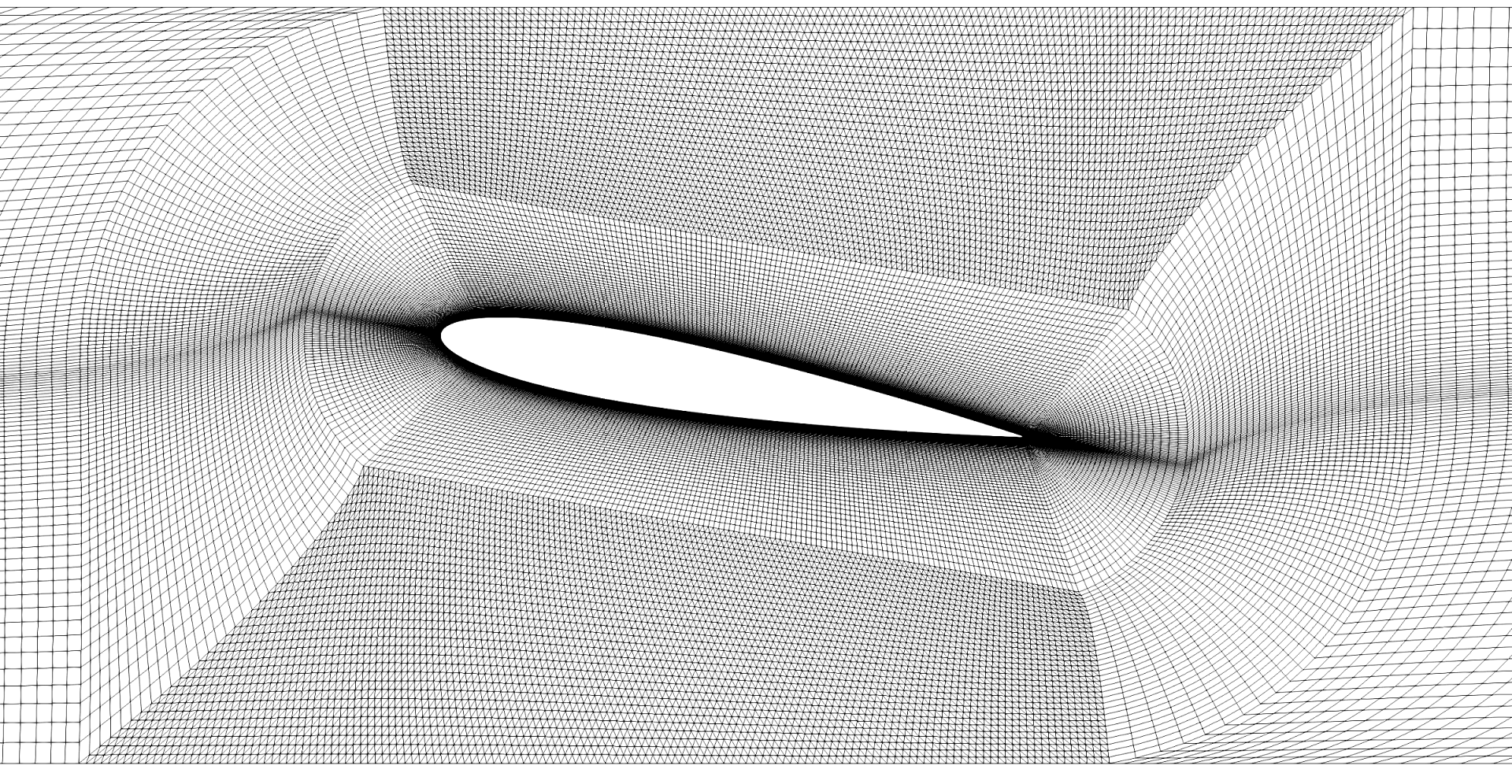}}
\caption{Hybrid computational grid for NACA0012 hydrofoil deformed to $\alpha=10^{\circ}$}
\label{fig:NACA0012hybridGrid}        
\end{figure}

\noindent In this closing section, we explore the ability of the proposed implementation to model freely moving cavitating hydrofoil and to predict some key features of cavitating flow over hydrofoils. We also test the compatibility of the solver on configurations with moving solid boundaries. Before proceeding to our fully-coupled cavitation and FSI demonstration, the primary objective of this section is to evaluate the feasibility of the implementation for turbulent cavitating flows over a stationary hydrofoil section. We consider the same NACA0012 hydrofoil geometry as in Sec.~(\ref{section:cavFlowOverHydrofoil}) with the following modifications. The height of the channel $H$ is reduced to $1.28C$. The computational grid is also modified to a hybrid grid system comprising hexahedral (8-node brick) and prism (6-node wedge) elements. This is to prevent the formation of highly skewed elements (in fully hexahedral grids) resulting from mesh deformations at high angles of attack. Figure~(\ref{fig:NACA0012hybridGrid}) shows the representative computational grid in the vicinity of the hydrofoil, demonstrating deformation to $\alpha=10^{\circ}$. Cavitation model B is used for the following studies. The values of the model coefficients and fluid properties defined in Section~(\ref{section:cavFlowOverHydrofoil}) are used. The cavitation number $\sigma$ is increased to $1.2$ and the flow Reynolds number $Re$ is set to $1\times 10^6$, with $U_\infty = 1\si{m.s^{-1}}$. A traction-free outflow is used, setting the pressure weakly to $0$ at the outflow boundary. 

\subsection{Stationary hydrofoil}\label{sec:statHydrofoil}
Leading-edge cavitation over hydrofoils can behave in different ways based on flow conditions defined by the flow Reynolds number $Re$, the cavitation number $\sigma$ and the angle of attack $\alpha$ of the incoming flow. Under certain combinations of these flow parameters, periodic cavity growth and shedding can be observed. Integral to this cavity shedding process is the formation of a periodic re-entrant jet along the hydrofoil surface. This jet periodically flows along the hydrofoil surface from the cavity closure location towards the leading edge, detaching the attached cavity. The capturing of this flow phenomena is of interest to the study of propeller vibration and noise because of two reasons. First, the periodic shedding of the cavity leads to periodic fluctuations in the hydrodynamic loading on the propeller blade. If the frequency of this loading is close to the natural frequency of the blade, it can result in structural excitation - leading to vibration and tonal noise emission. Second, the shed cavities exist in the form of clouds of vaporous bubbles. These bubbles are prone to collapsing near the trailing edge of the blade with localized high-amplitude water hammer impacts. This can contribute to broadband underwater noise emission. We investigate here the turbulent cavitating flow over a hydrofoil section at $\alpha=10^{\circ}$. 

\begin{figure}
\begin{subfigure}{.5\textwidth}
  \centering
  \includegraphics[width=.9\linewidth]{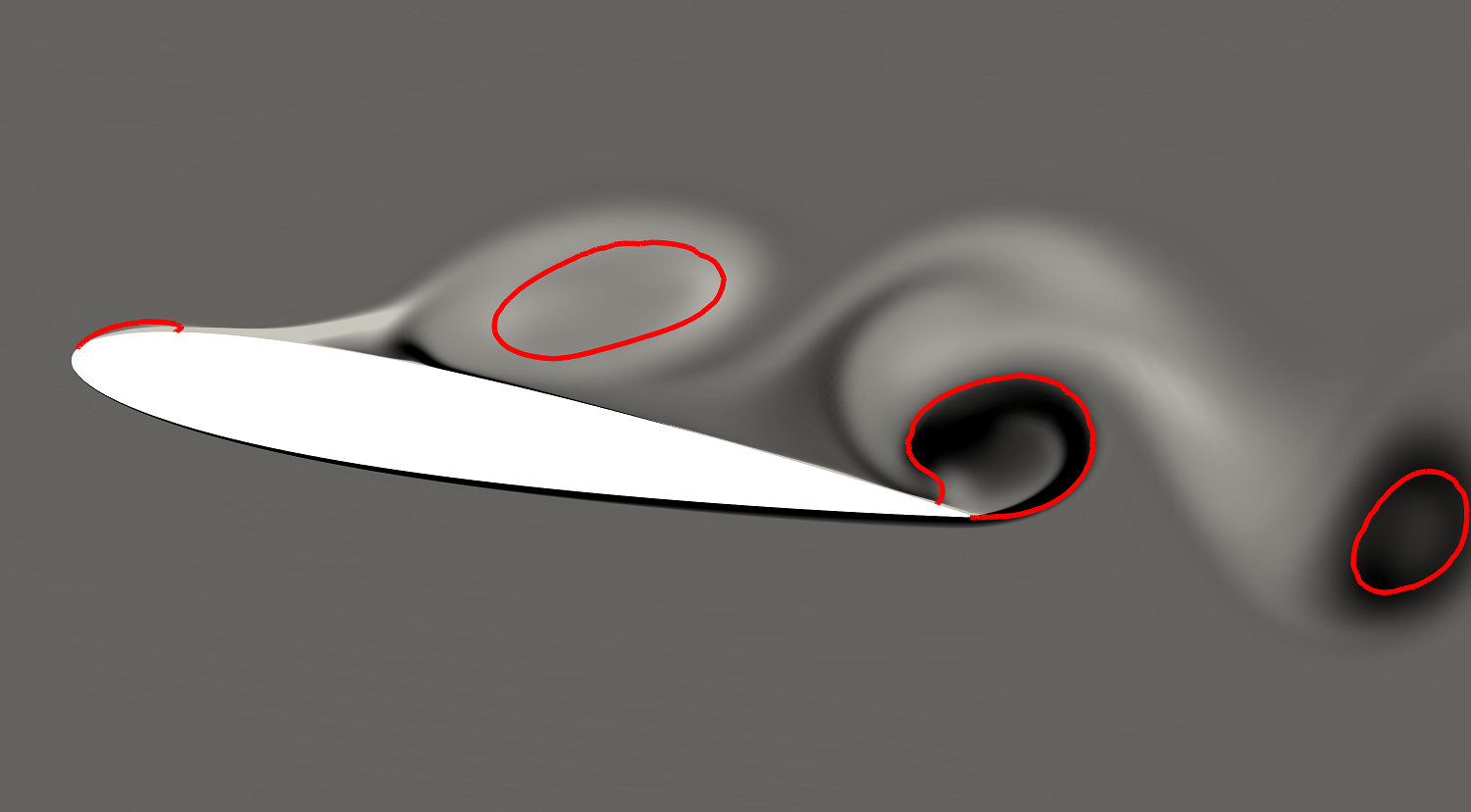}  
  \caption{$t_0$}
  \label{fig:cavShed1}
\end{subfigure}
\begin{subfigure}{.5\textwidth}
  \centering
  \includegraphics[width=.9\linewidth]{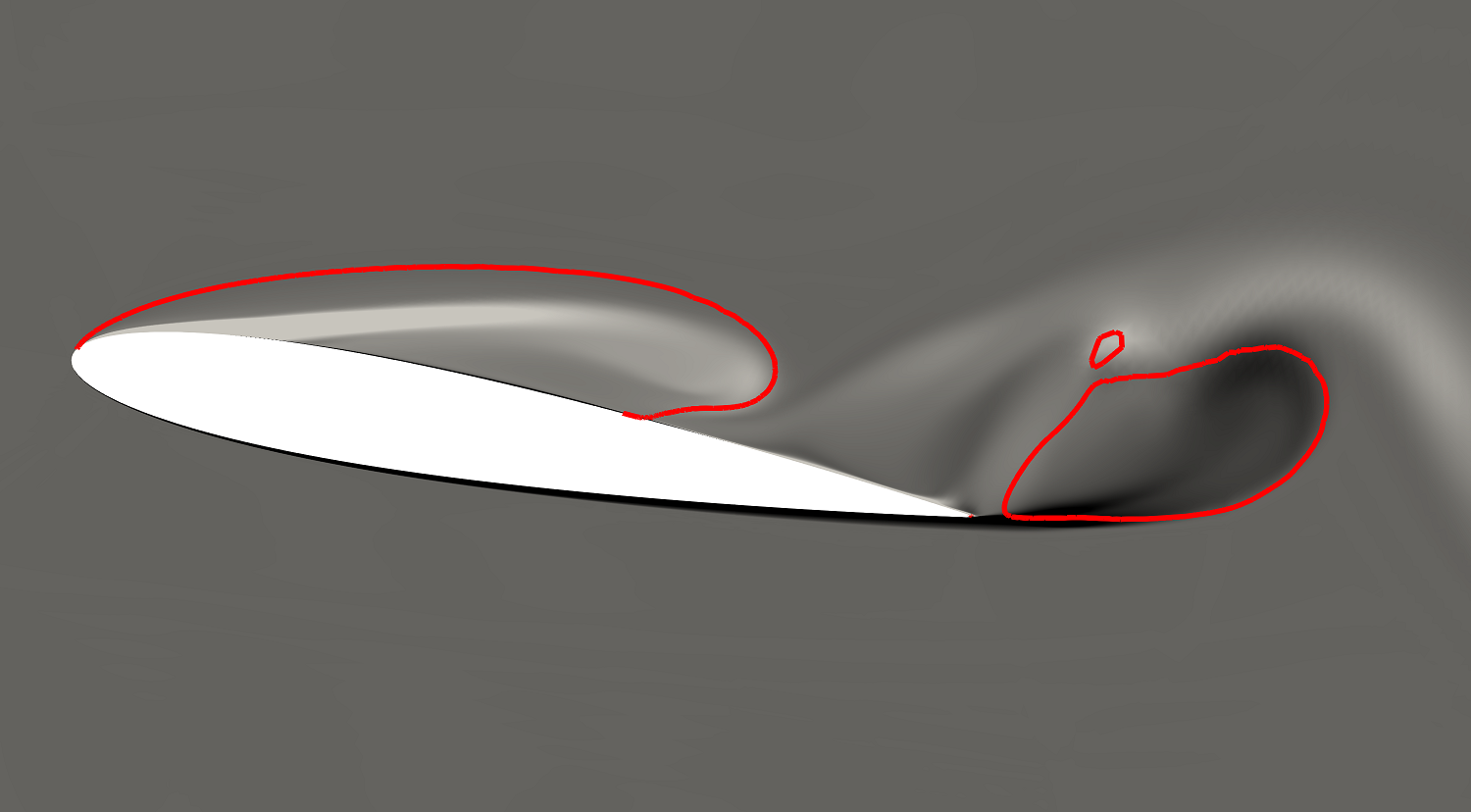}  
  \caption{$t_0 + 0.11t_\infty$}
  \label{fig:cavShed2}
\end{subfigure}


\begin{subfigure}{.5\textwidth}
  \centering
  \includegraphics[width=.9\linewidth]{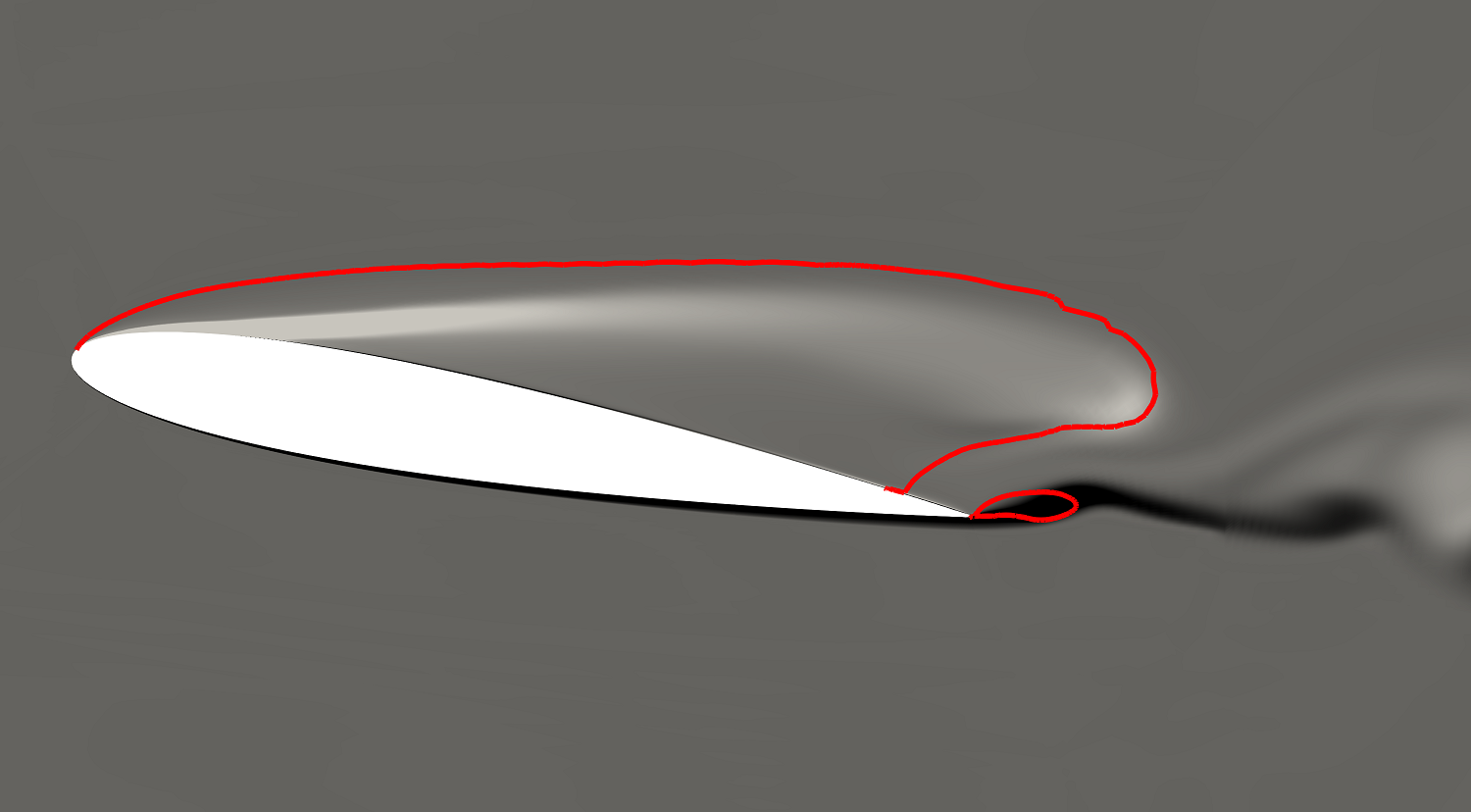}  
  \caption{$t_0 + 0.17t_\infty$}
  \label{fig:cavShed3}
\end{subfigure}
\begin{subfigure}{.5\textwidth}
  \centering
  \includegraphics[width=.9\linewidth]{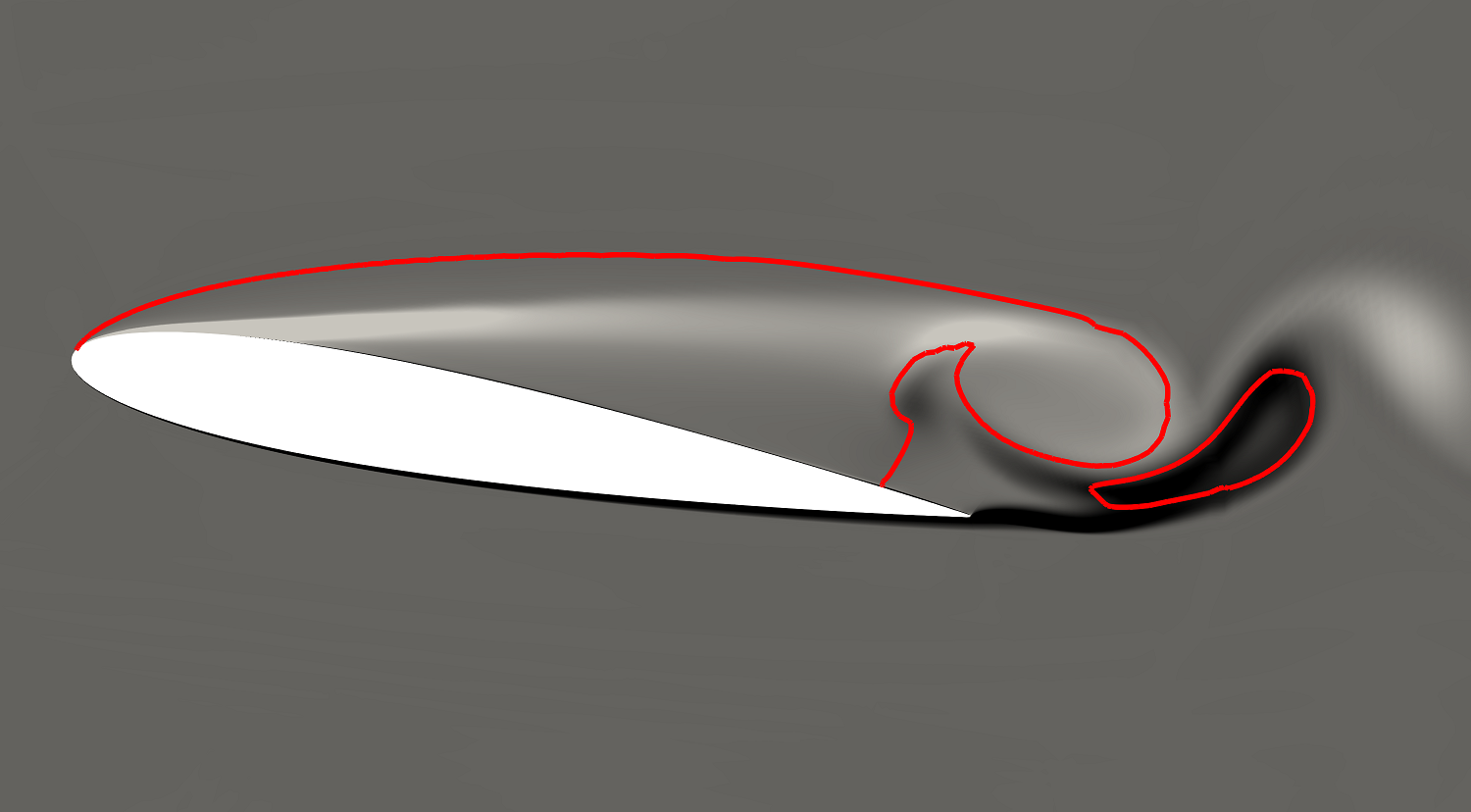}  
  \caption{$t_0 + 0.24t_\infty$}
  \label{fig:cavShed4}
\end{subfigure}


\begin{subfigure}{.5\textwidth}
  \centering
  \includegraphics[width=.9\linewidth]{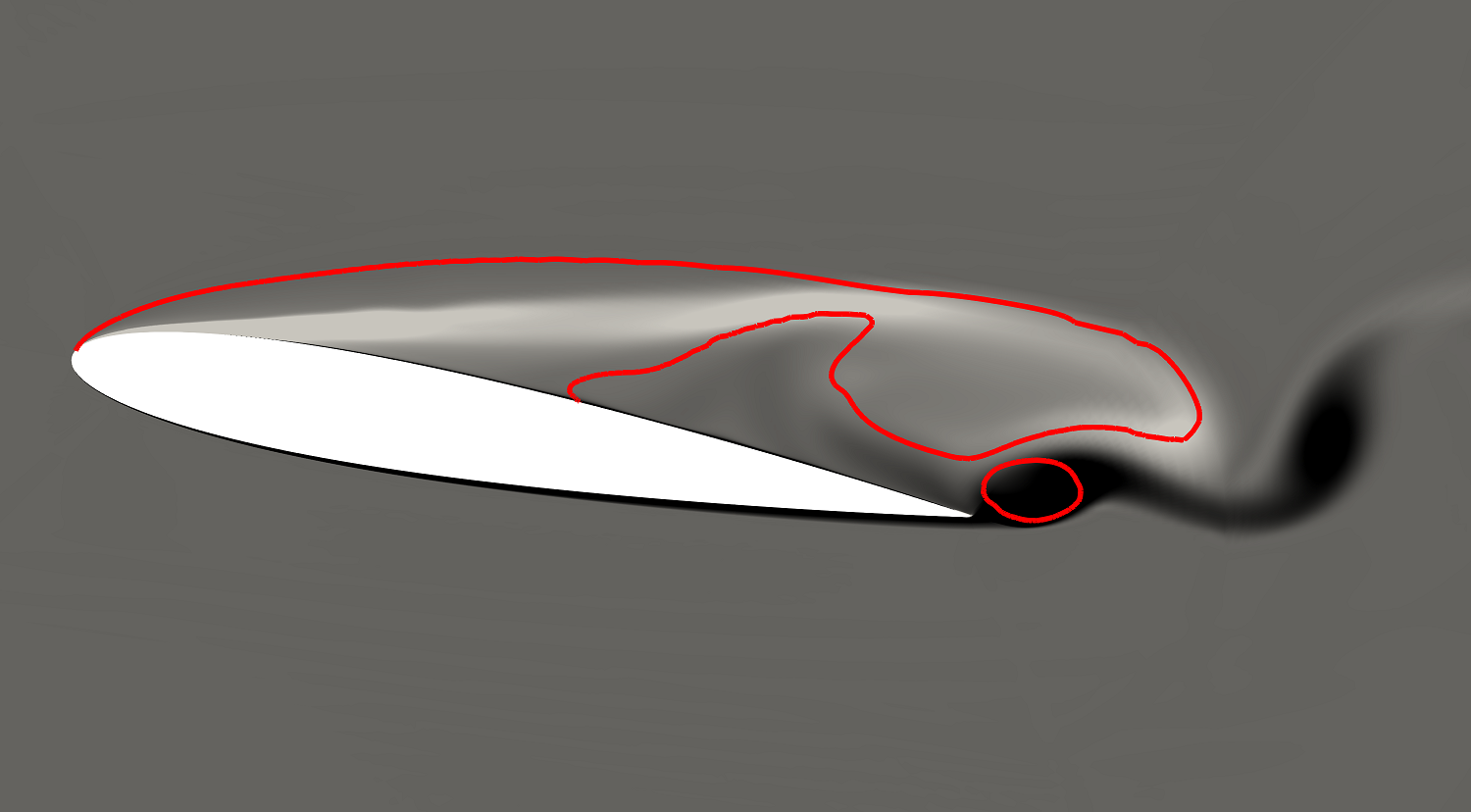}  
  \caption{$t_0 + 0.39t_\infty$}
  \label{fig:cavShed5}
\end{subfigure}
\begin{subfigure}{.5\textwidth}
  \centering
  \includegraphics[width=.9\linewidth]{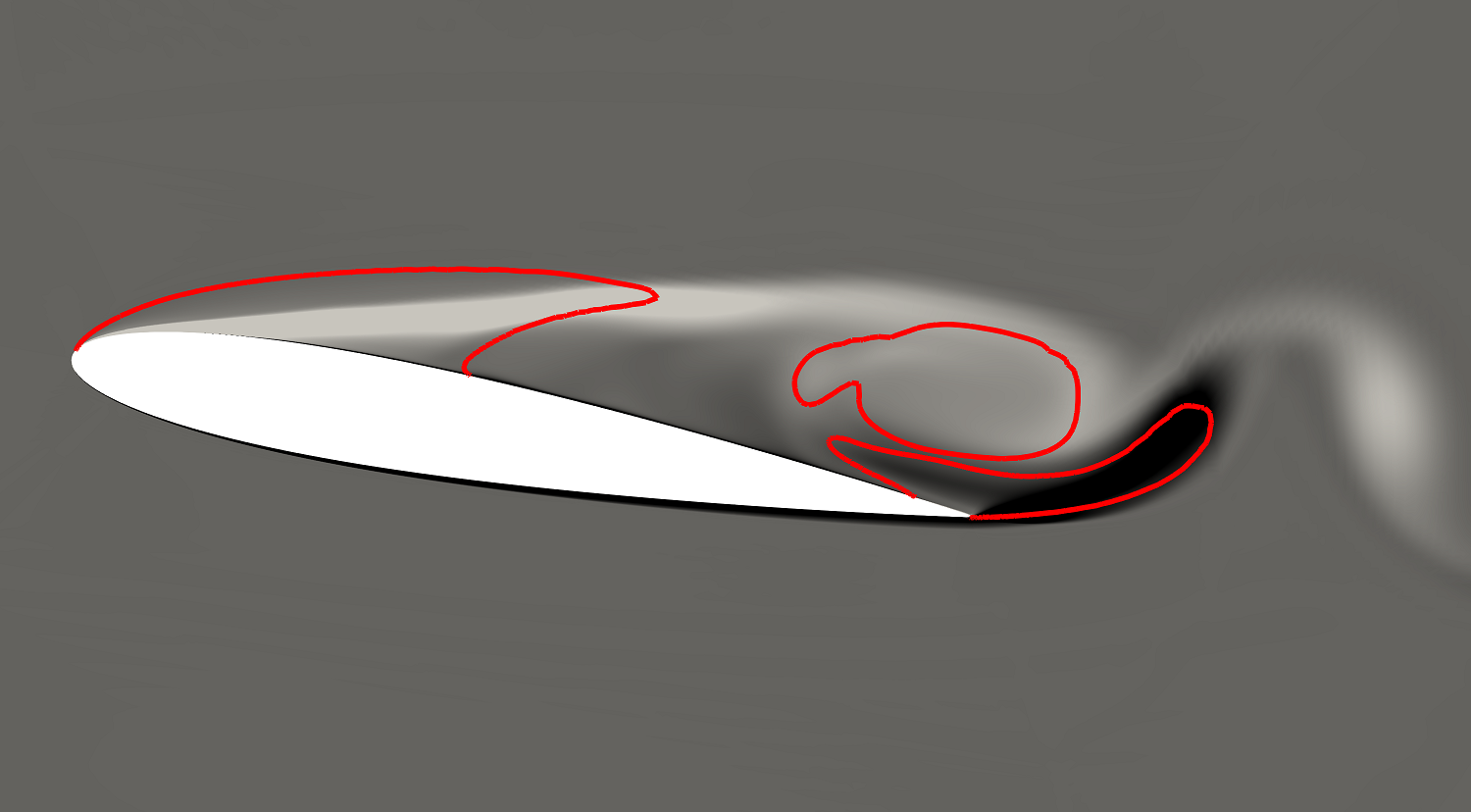}  
  \caption{$t_0 + 0.44t_\infty$}
  \label{fig:cavShed6}
\end{subfigure}


\begin{subfigure}{.5\textwidth}
  \centering
  \includegraphics[width=.9\linewidth]{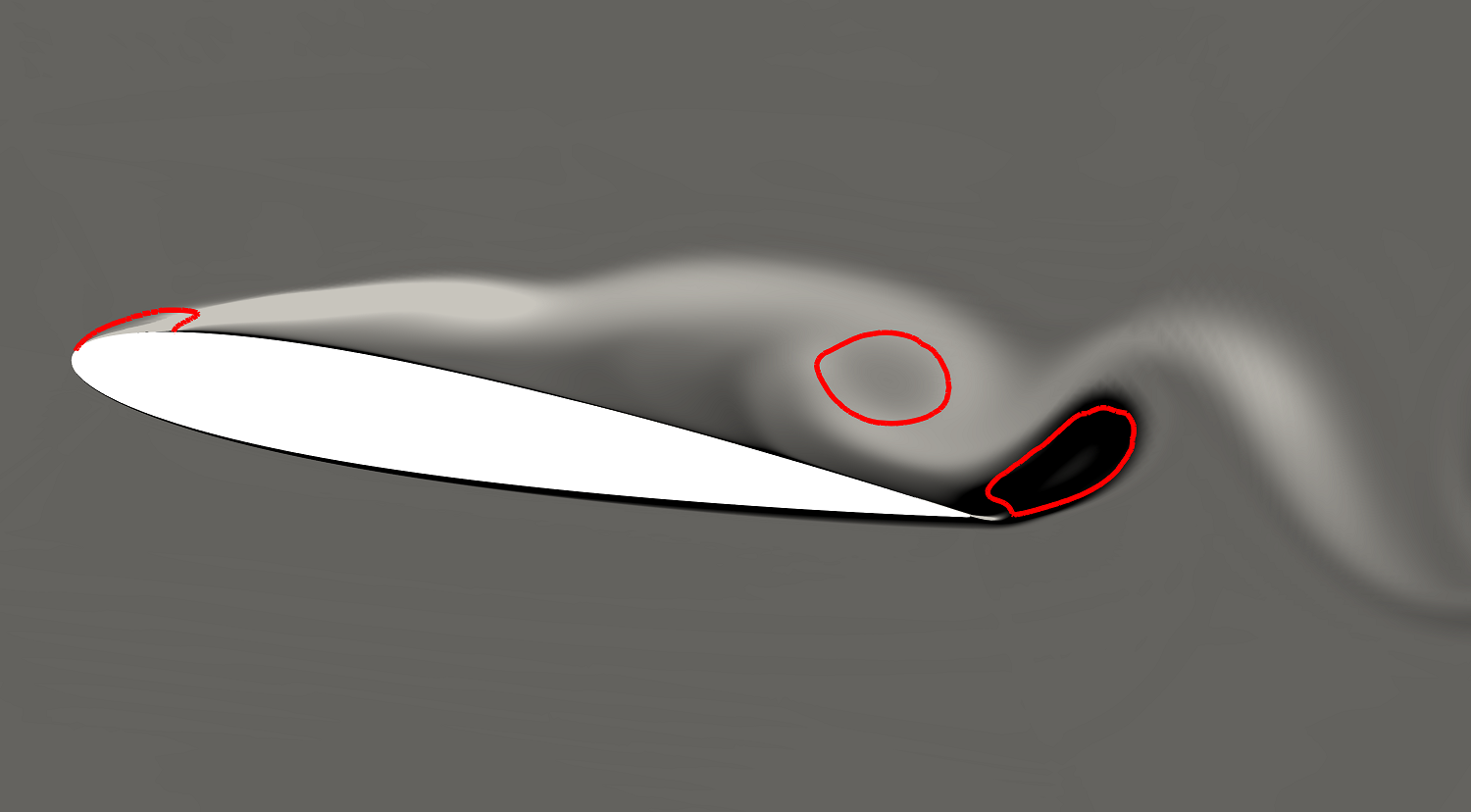}  
  \caption{$t_0 + 0.49t_\infty$}
  \label{fig:cavShed7}
\end{subfigure}
\begin{subfigure}{.5\textwidth}
  \centering
  \includegraphics[width=.9\linewidth]{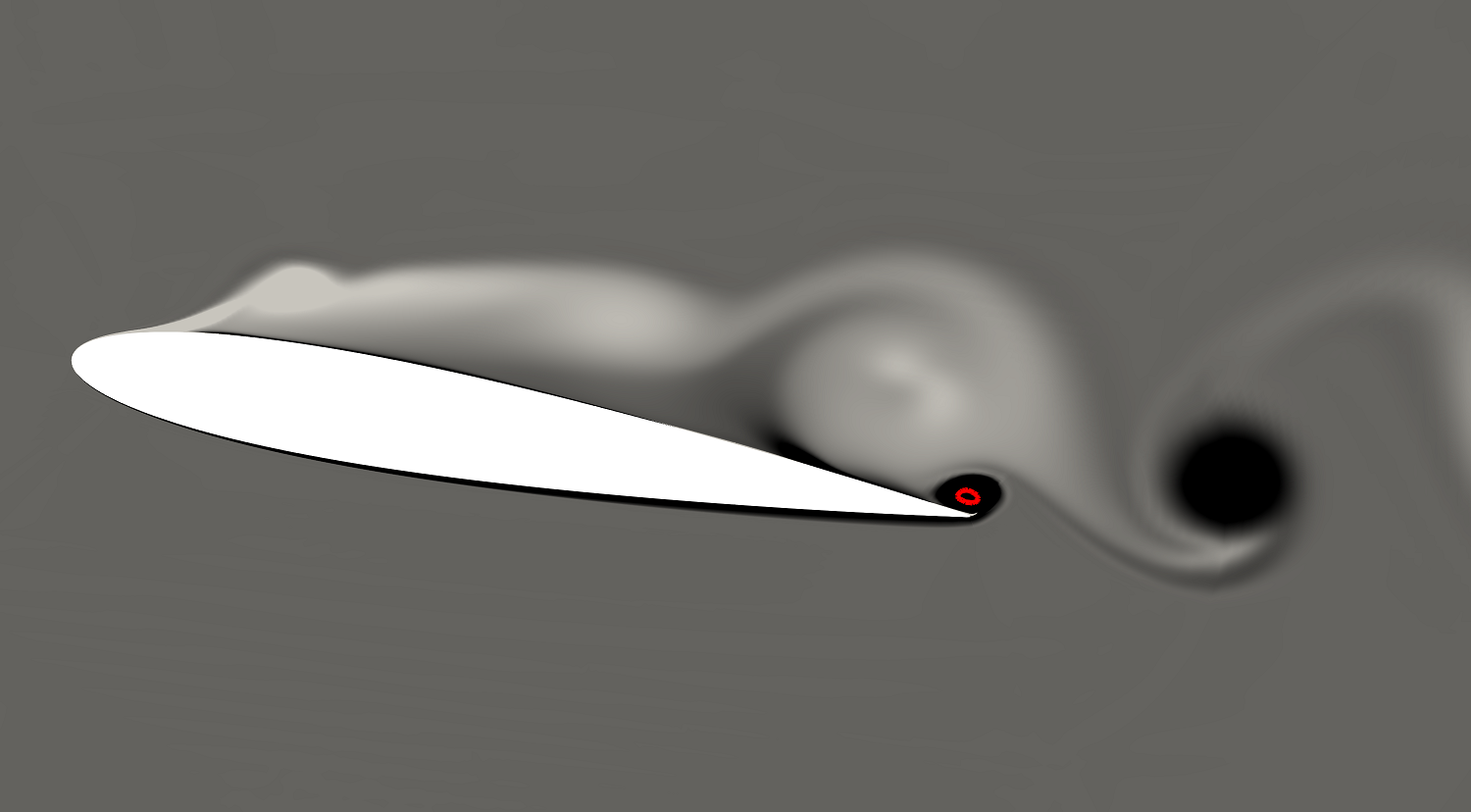}  
  \caption{$t_0 + 0.53t_\infty$}
  \label{fig:cavShed8}
\end{subfigure}


\begin{subfigure}{1\textwidth}
  \centering
  \includegraphics[width=.35\linewidth]{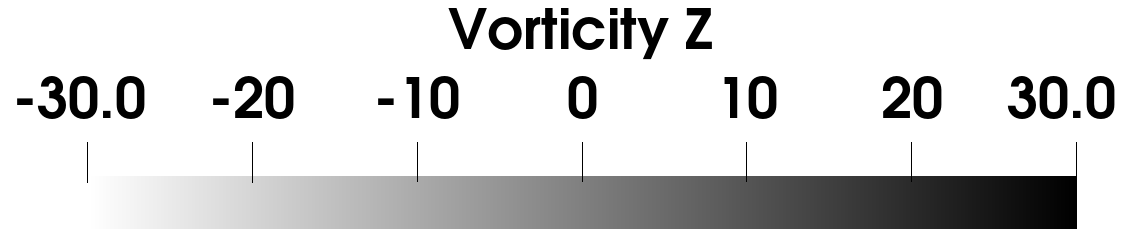}  
\end{subfigure}
\caption{Contours of Z-vorticity (positive is clockwise, negative is anti-clockwise) during one shedding cycle. Cavity (in red) marked by iso-contour of $\phi^{\mathrm{f}}=0.95$. First figure marked by time $t_0$ for reference. Subsequent figures marked in terms of $t_0$ and $t_\infty = C/U_\infty$}
\label{fig:cavShedding}
\end{figure}

\begin{figure}[!h]
\centering
\frame{\includegraphics[width=0.77\columnwidth]{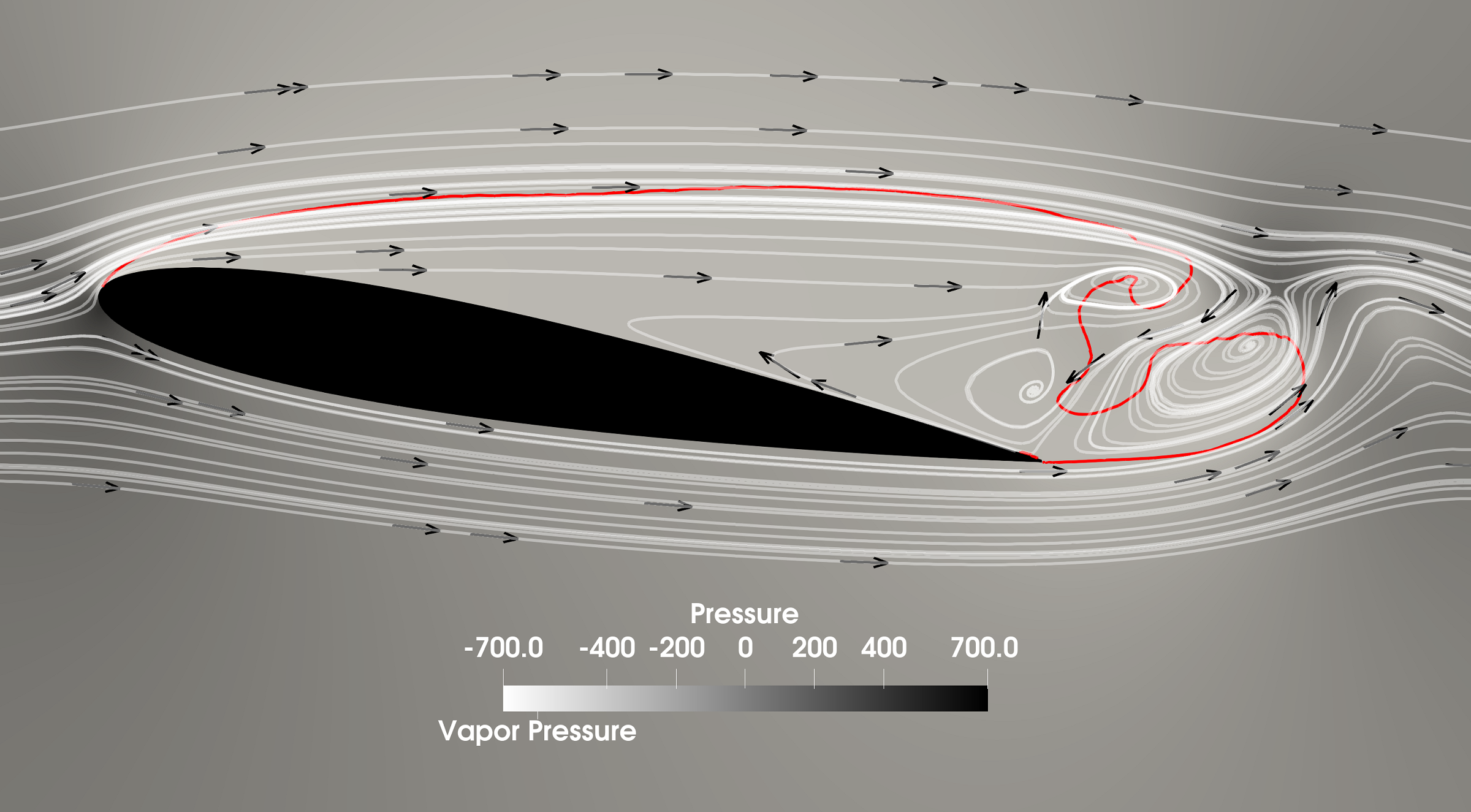}}
\caption{Vortex interaction and formation of re-entrant jet at trailing edge of hydrofoil. Cavity (in red) marked at iso-contour of $\phi^{\mathrm{f}}=0.95$.}
\label{fig:reEntrantJet}        
\end{figure}

 Figure~(\ref{fig:cavShedding}) shows the results of the study at different instances during one cavity shedding cycle. An iso-contour of $\boldsymbol{\phi}^{\mathrm{f}} = 0.95$ (in red) is used to represent the cavity surface. Also shown are the contours of the vorticity in the direction $z$ out of the plane of the figure. Fig.~(\ref{fig:cavShed1}) shows the inception of a leading-edge cavity. The cavity is observed to grow, primarily collocated with a leading-edge vortex (LEV), to the extent of the hydrofoil chord. In Fig.~(\ref{fig:cavShed4}) the clockwise rotating LEV interacts with a counter-clockwise trailing edge vortex (TEV). We observe that the interaction, along with a reverse pressure gradient, leads to the formation of a re-entrant jet along the hydrofoil suction surface. This detaches the cavity which is shed in the form of pockets (clouds) and is convected with the mean flow. Fig.~(\ref{fig:reEntrantJet}) shows the streamlines in the domain at the beginning of the shedding process. The formation of the re-entrant jet can be observed to originate at the intersection of the LEV and TEV. 

\begin{remark}
We note the ability of the present implementation to capture some select physics of interest in turbulent cavitating flow over hydrofoils. Only preliminary results are presented for demonstration. The accuracy of the shedding frequency needs to be investigated, and can require careful calibration of coefficients in the cavitation model and the modified turbulent viscosity. Detailed investigation is beyond the scope of the current work, and will be explored in future studies. 
\end{remark}

\subsection{Pitching hydrofoil}

\begin{figure}[!h]
\centering
\includegraphics[width=0.7\columnwidth]{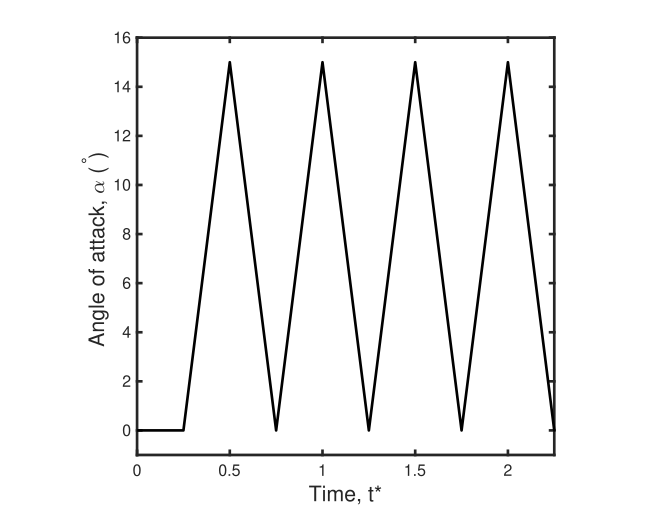}
\caption{Representative prescribed pitching motion }
\label{fig:pitchingMotion}        
\end{figure}

\begin{figure}
\begin{subfigure}{.5\textwidth}
  \centering
  \includegraphics[width=.9\linewidth]{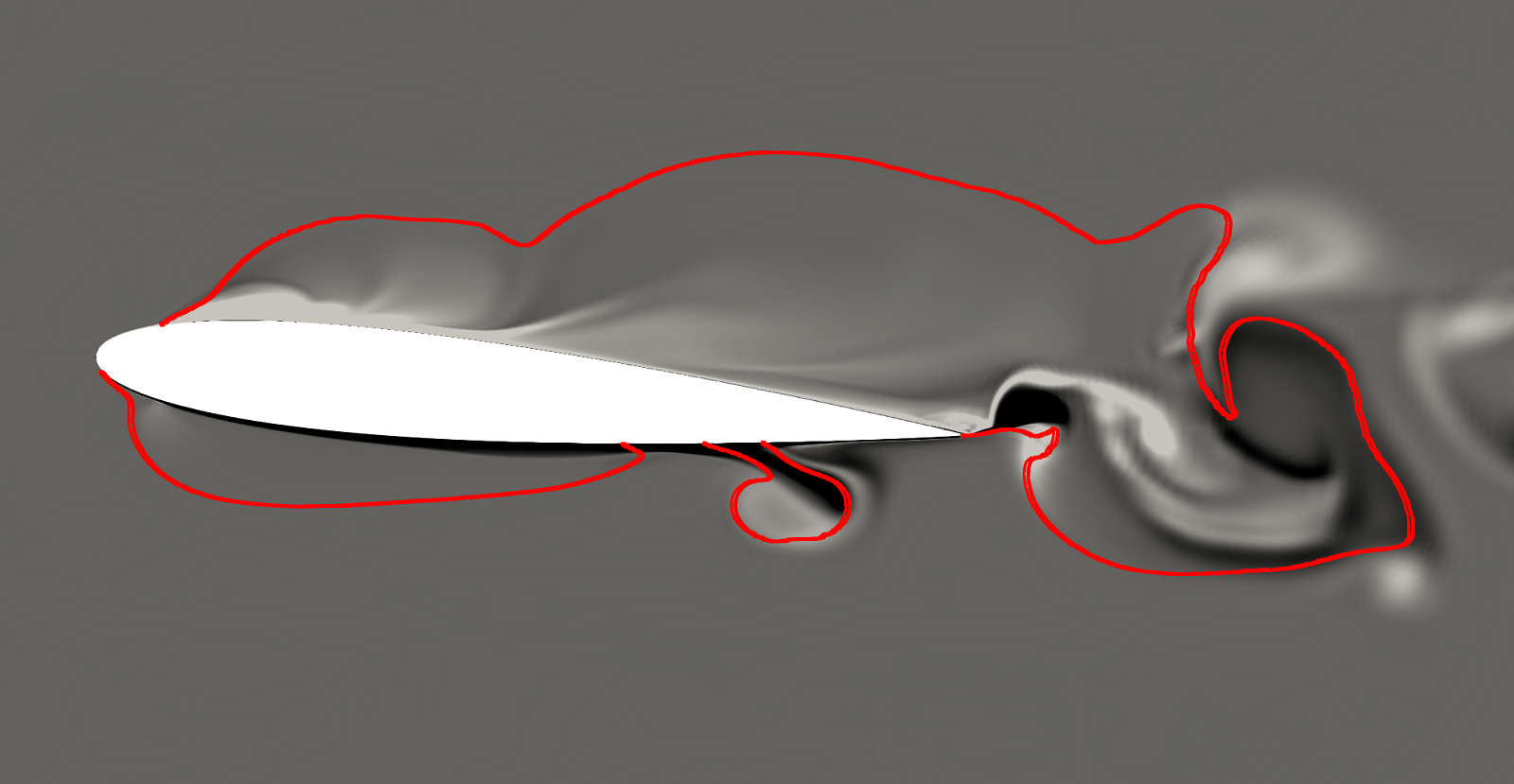}
  \setcounter{subfigure}{0}
  \caption{$(+)5^{\circ}$}
  \label{fig:cavPitch1}
\end{subfigure}
\begin{subfigure}{.5\textwidth}
  \centering
  \includegraphics[width=.9\linewidth]{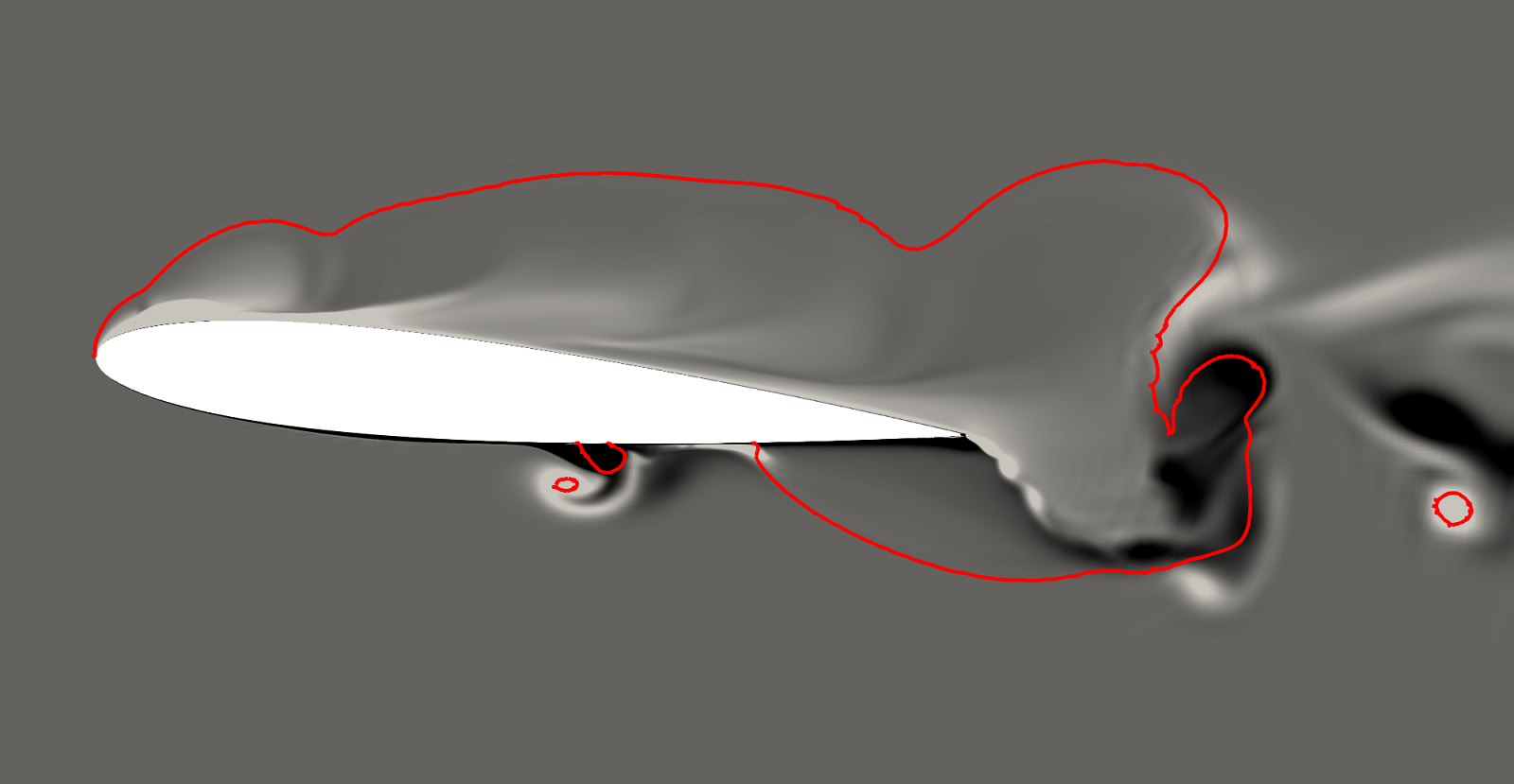}
  \setcounter{subfigure}{7}
  \caption{$(-)5^{\circ}$}
  \label{fig:cavPitch2}
\end{subfigure}


\begin{subfigure}{.5\textwidth}
  \centering
  \includegraphics[width=.9\linewidth]{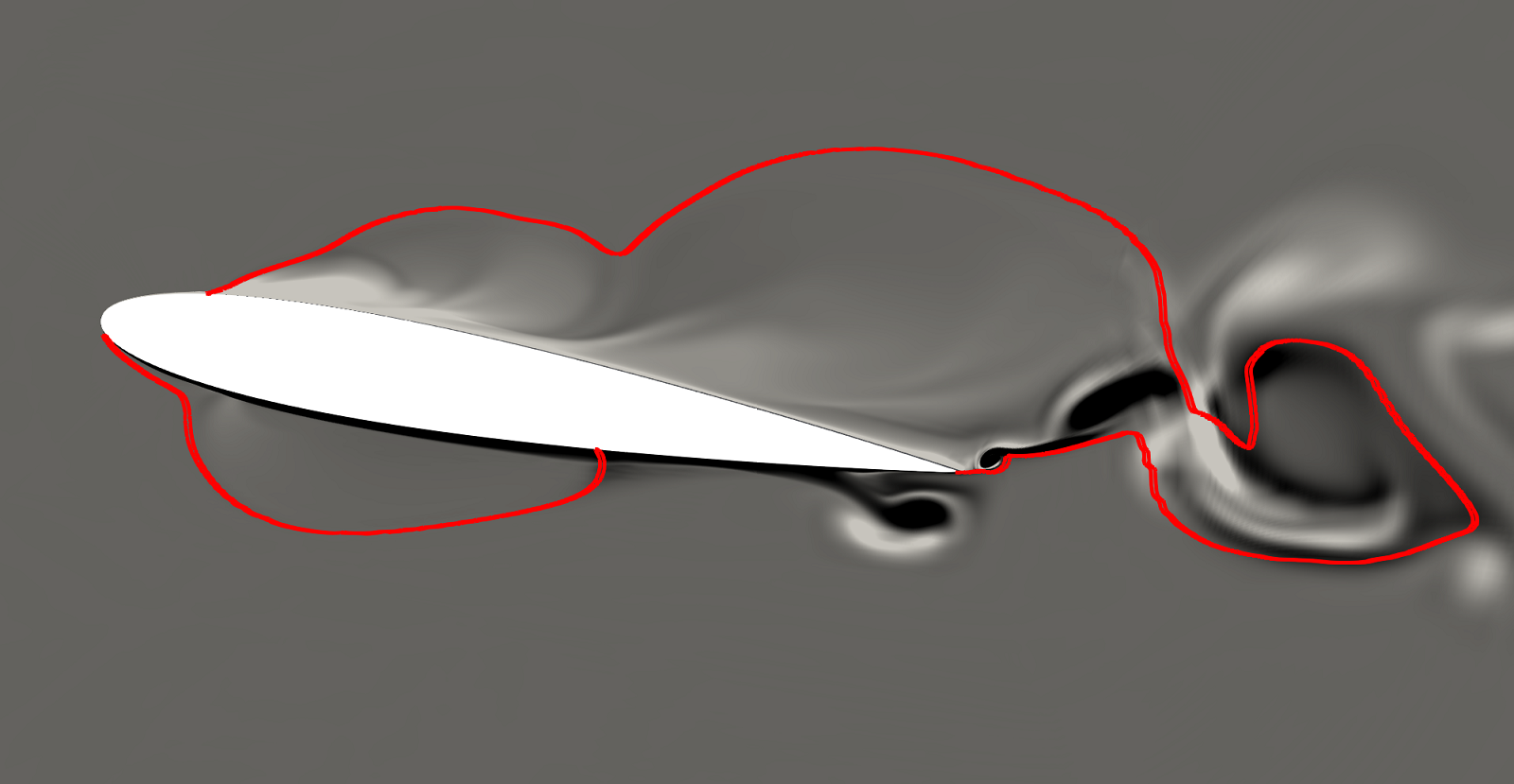}
  \setcounter{subfigure}{1}
  \caption{$(+)10^{\circ}$}
  \label{fig:cavPitch3}
\end{subfigure}
\begin{subfigure}{.5\textwidth}
  \centering
  \includegraphics[width=.9\linewidth]{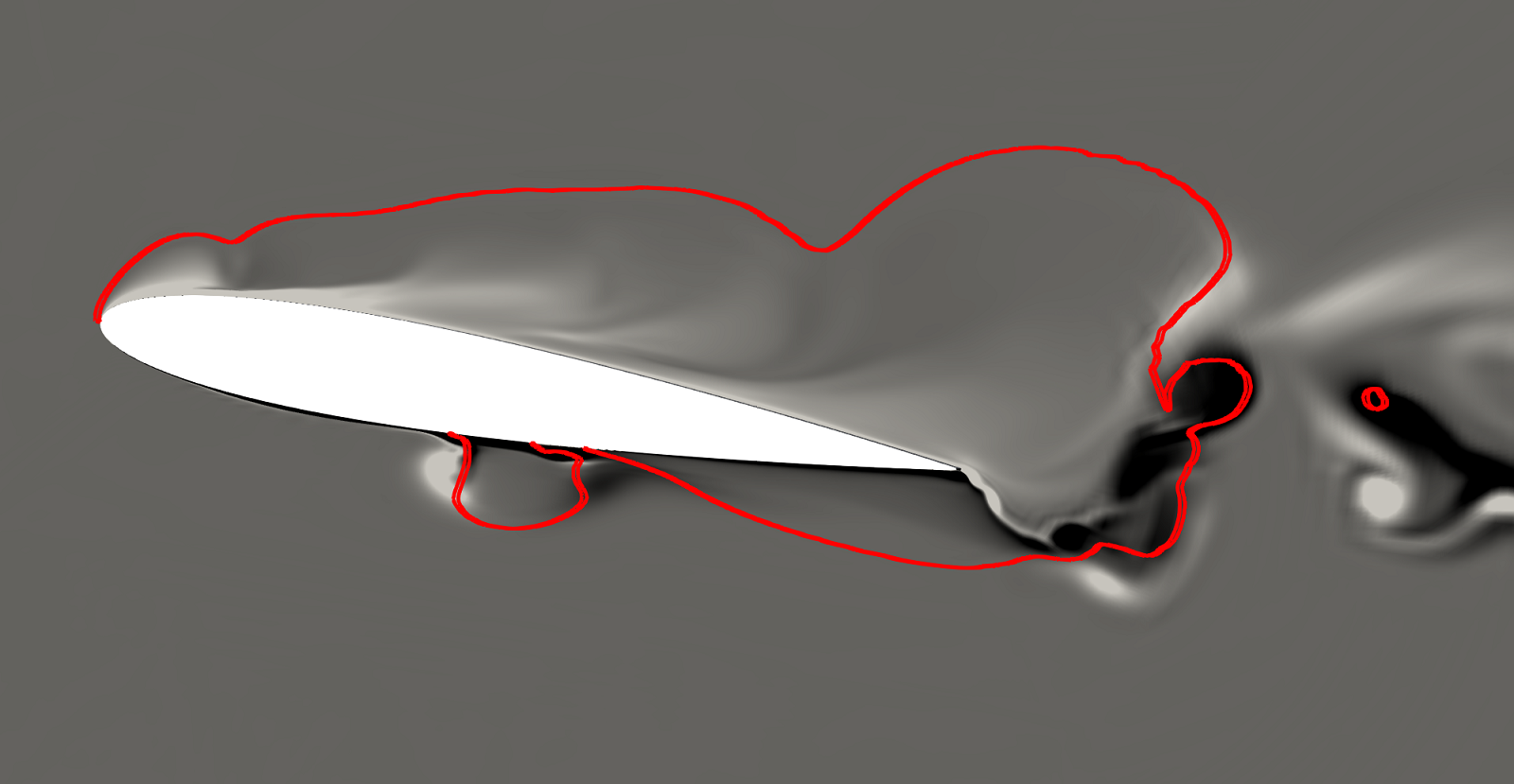}
  \setcounter{subfigure}{6}
  \caption{$(-)10^{\circ}$}
  \label{fig:cavPitch4}
\end{subfigure}


\begin{subfigure}{.5\textwidth}
  \centering
  \includegraphics[width=.9\linewidth]{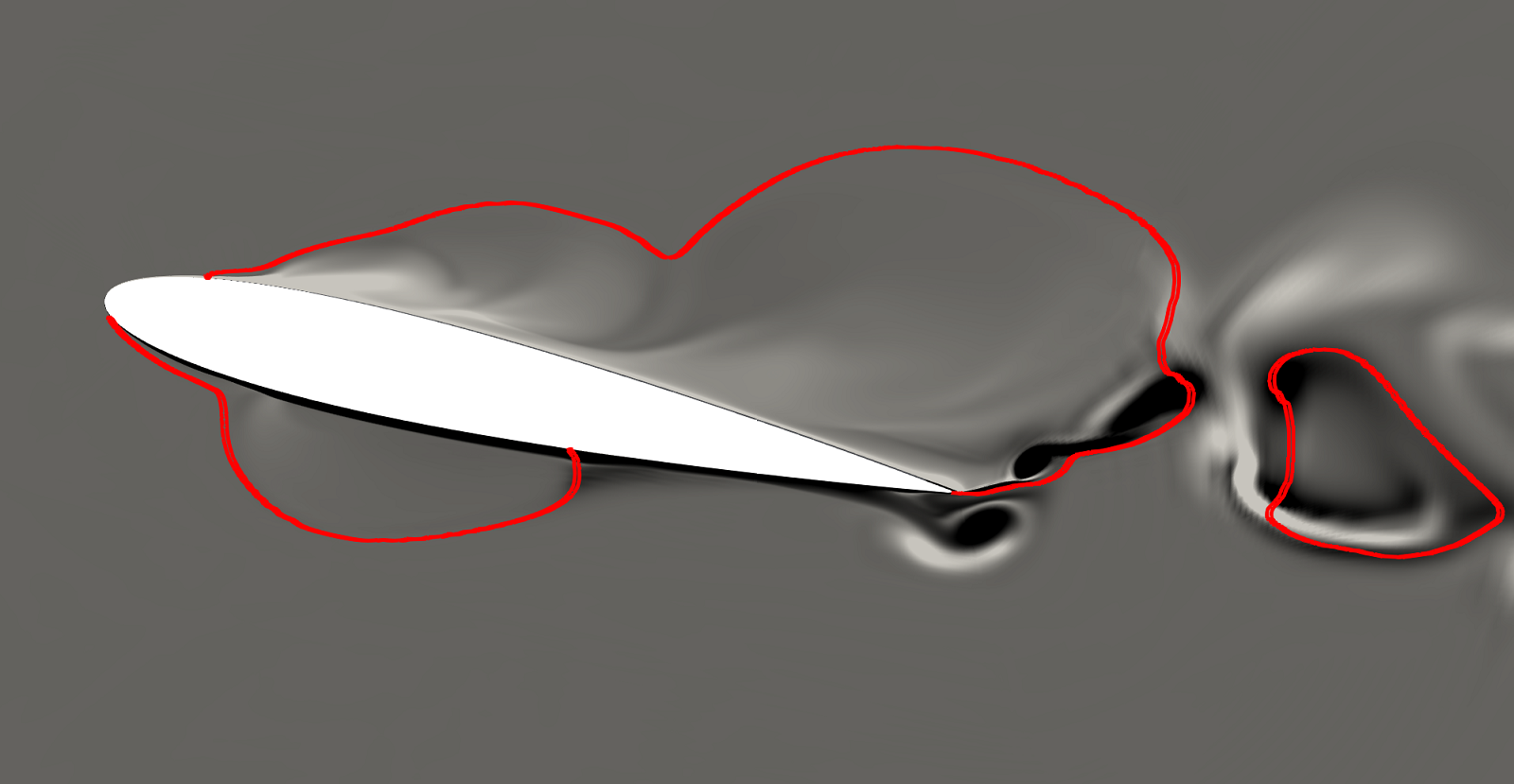}
  \setcounter{subfigure}{2}
  \caption{$(+)12.5^{\circ}$}
  \label{fig:cavPitch5}
\end{subfigure}
\begin{subfigure}{.5\textwidth}
  \centering
  \includegraphics[width=.9\linewidth]{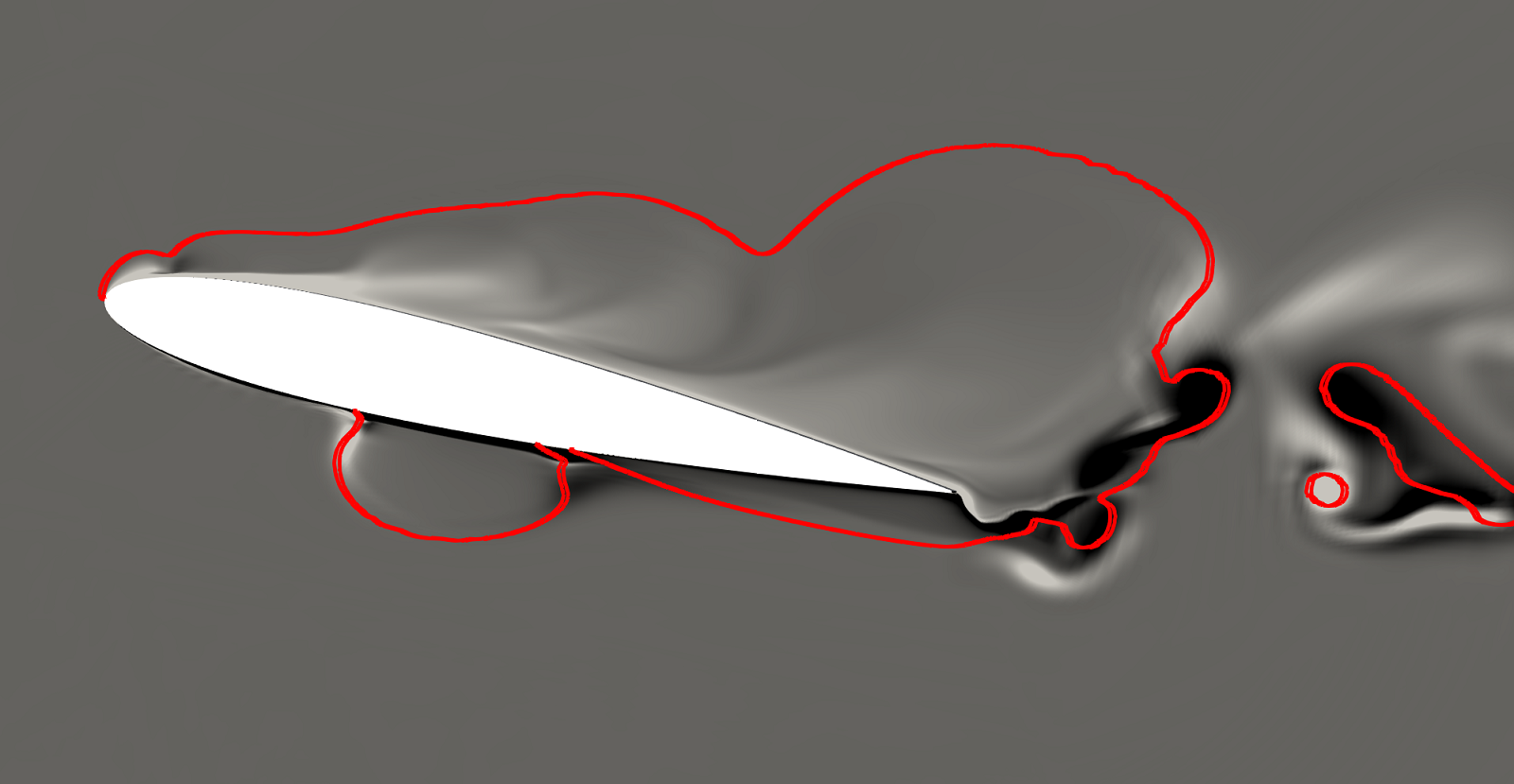}  
  \setcounter{subfigure}{5}
  \caption{$(-)12.5^{\circ}$}
  \label{fig:cavPitch6}
\end{subfigure}


\begin{subfigure}{.5\textwidth}
  \centering
  \includegraphics[width=.9\linewidth]{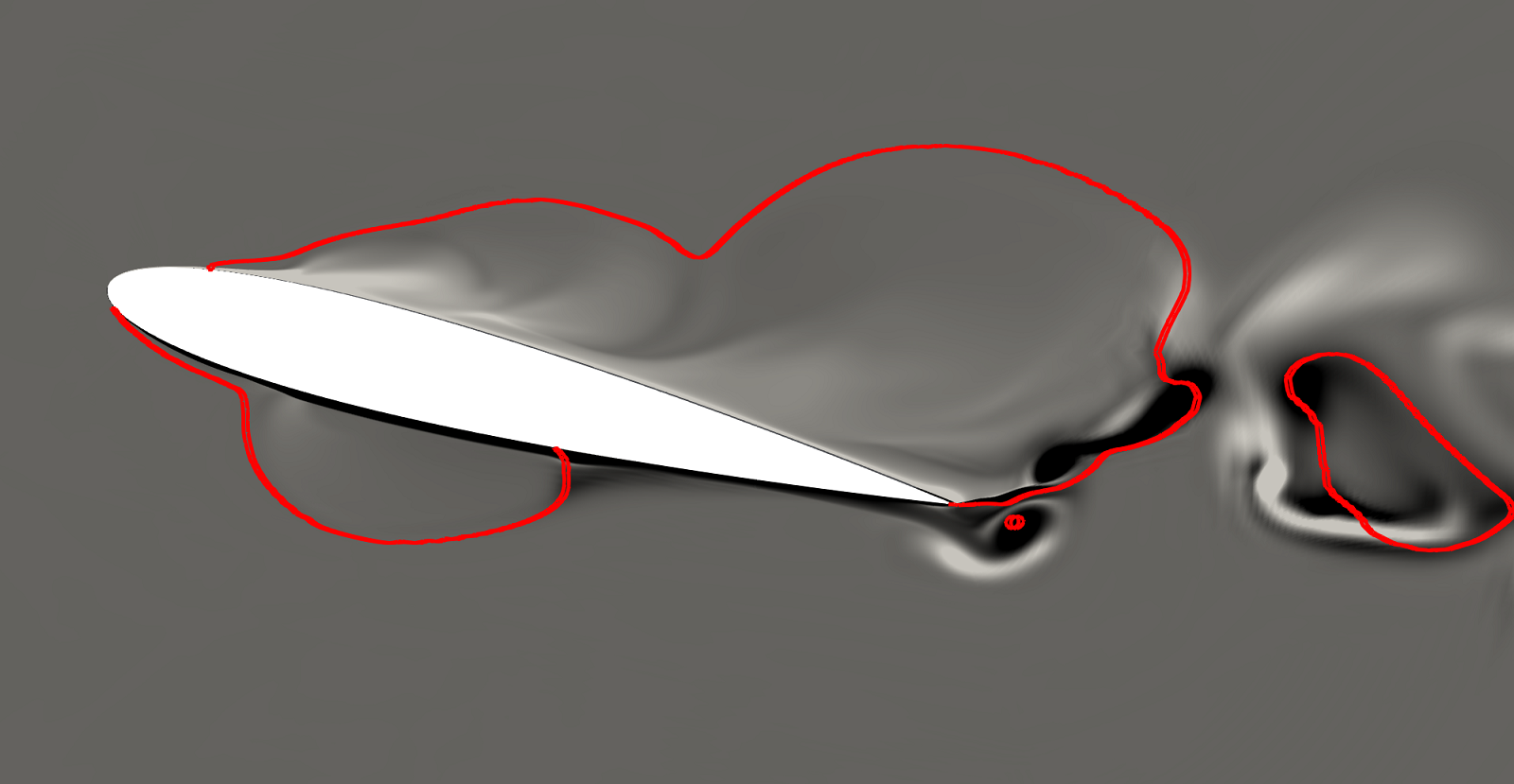}  
  \setcounter{subfigure}{3}
  \caption{$(+)14^{\circ}$}
  \label{fig:cavPitch7}
\end{subfigure}
\begin{subfigure}{.5\textwidth}
  \centering
  \includegraphics[width=.9\linewidth]{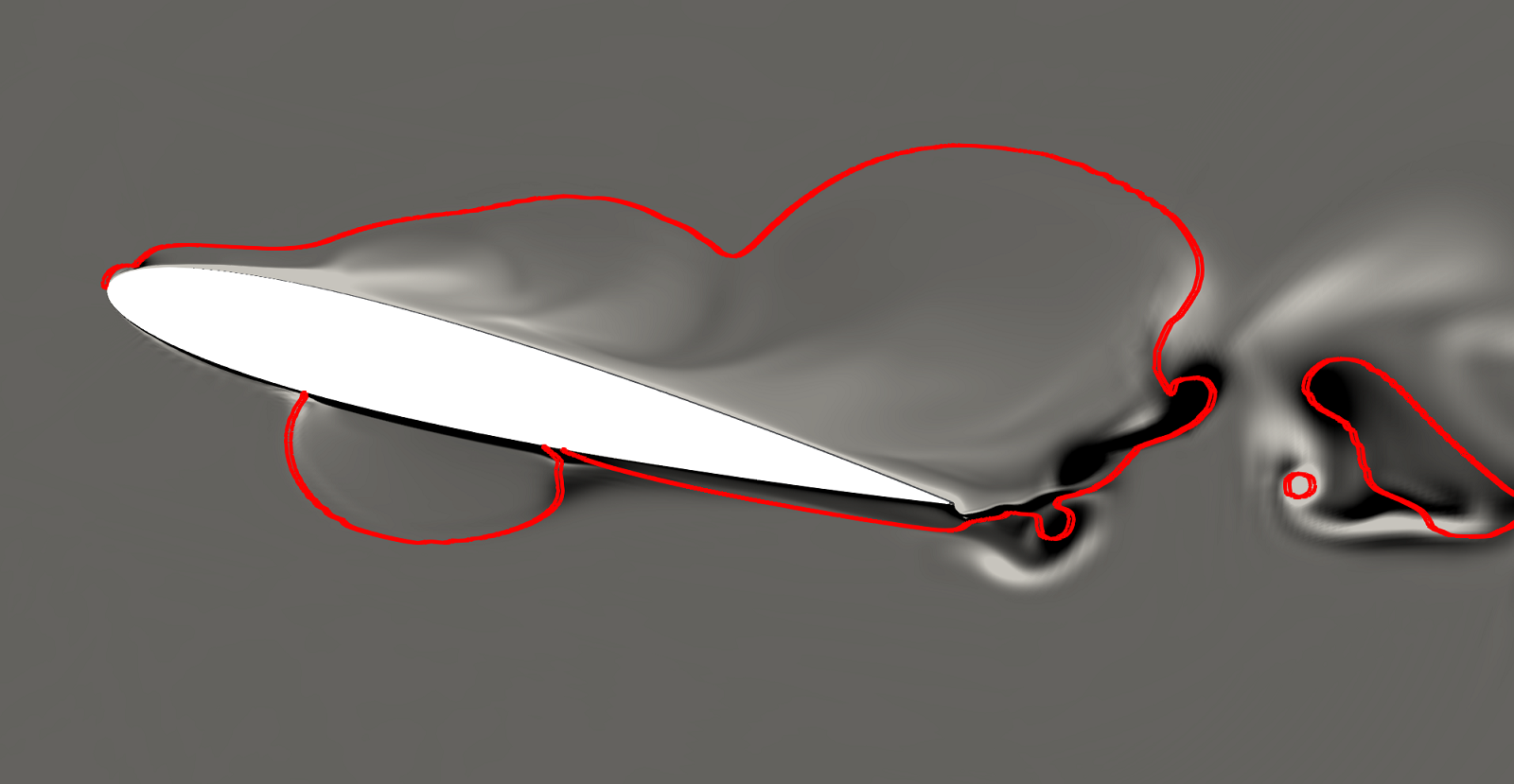}  
  \setcounter{subfigure}{4}
  \caption{$(-)14^{\circ}$}
  \label{fig:cavPitch8}
\end{subfigure}


\begin{subfigure}{1\textwidth}
  \centering
  \includegraphics[width=.35\linewidth]{colorbar.png}  
\end{subfigure}
\caption{Contours of Z-vorticity (positive is clockwise, negative is anti-clockwise) during one pitching cycle. Cavity (in red) marked by iso-contour of $\phi^{\mathrm{f}}=0.95$. (+) marks the upward pitching stroke and (-) marks the downward stroke. }
\label{fig:cavPitching}
\end{figure}

We next investigate turbulent cavitating flow over the NACA0012 section subject to a prescribed periodic pitching motion. By using our ALE-based FSI framework, we linearly ramp the angle of attack of the hydrofoil between $0^{\circ}$ and $15^{\circ}$. Fig.~(\ref{fig:pitchingMotion}) shows the first four cycles of the prescribed motion. The frequency $f_{pitch}$ of the motion is $2~\si{Hz}$. The rest of the study parameters are kept the same as in Section~(\ref{sec:statHydrofoil}).
 Figure~(\ref{fig:cavPitching}) shows the cavities in the domain marked by the iso-contour $\phi^{\mathrm{f}}=0.95$. Also plotted are the contours of z-vorticity.   At the high pitching frequency $f_{pitch}=2~\si{Hz}$ and the associated hydrofoil acceleration, cavities are observed to originate on both surfaces because of the low pressures during the pitching motion. The primary cavity generation is at the hydrofoil leading edge. The cavity shedding frequency is low compared to the pitching frequency and the hydrofoil suction surface is seen to be perennially covered by cavities that are continuously being shed and convected with the mean flow. The collocation between the vortices and the cavities can be discerned. Detailed investigations on the influence of the pitching frequency on the cavity and vortex shedding frequency are reserved for future work. 

We note the compatibility of the cavitation and the flow solvers with structural deformation. The versatility provided by the partitioned coupling was exploited to couple the additional solvers for turbulence modeling and ALE mesh update. The numerical solution obtained is stable and $2-3$ non-linear predictor-corrector iterations are seen to give a converged solution at each time-level. This sets the stage for large-scale fully coupled FSI studies in cavitating flows, and will be of interest for future work.

\section{Conclusion}
A robust and accurate variational finite element formulation for the numerical study of cavitating flows has been presented for stationary and moving hydrofoils. We introduced novel stabilized linearizations of two cavitation transport equation models based on the two-phase homogeneous mixture theory. The numerical implementation has been employed to study two cavitating flow configurations with vastly different temporal and spatial scales. An initial verification study on the micro-scale collapse of a spherical vaporous bubble has been shown to maintain stability across a range of  time steps. Accurate solutions of the pressure field were obtained, devoid of spurious numerical oscillations. The solution of the phase indicator is demonstrated to be bounded, and the solver is numerically stable at large density ratios. The implementation has been validated on fully turbulent cavitating flow over a hydrofoil with good agreement with previous numerical and experimental studies. We also explored the ability of the implementation to predict select characteristic features of macro-scale cavitating flows, including re-entrant jets, periodic cavity shedding and cavity-vortex interaction. Further, we examined the versatility of the implementation to be coupled with solvers for studying fluid-structure interaction, demonstrating the case of a pitching hydrofoil. In future work, the authors plan to extend the implementation to study fully-coupled FSI studies in cavitating flows. One potential application is cavitating flow-induced vibrations taking into account the hydroelastic response of structures. Another potential application is the study of material erosion resulting from cavitation bubble collapse using fully-Eulerian fluid-solid formulations.

\section*{Acknowledgements}
The authors would like to acknowledge the Natural Sciences and Engineering Research Council of Canada (NSERC) for the funding. This research was enabled in part through computational resources and services  provided by (WestGrid) (https://www.westgrid.ca/), Compute Canada (www.computecanada.ca) and the Advanced Research Computing facility at the University of British Columbia.

%
%


\bibliography{mybibfile}   

\end{document}